\documentclass[10pt,aps,prb,reprint,superscriptaddress]{revtex4-2}
\usepackage[pages=all, color=black, position={current page.south}, placement=bottom, scale=1, opacity=1, vshift=5mm]{background}
\SetBgContents{
	\tt 
}    
\usepackage{amsmath}
\usepackage{amsthm}
\usepackage{amssymb}
\usepackage{times}
\usepackage{braket}
\usepackage{ulem}
\usepackage[utf8]{inputenc}
\usepackage{svg}
\usepackage{hyperref}
\usepackage[english]{babel}

\usepackage{graphicx, color}
\graphicspath{{fig/}}
\usepackage{algorithm, algpseudocode} 
\usepackage{mathrsfs} 
\usepackage{lipsum}
\usepackage{caption}
\usepackage{subcaption}
\usepackage{placeins}
\usepackage{etoolbox}
\usepackage{caption}
\let\origsection\section
\let\origsubsection\subsection

\preto\section{\FloatBarrier}
\preto\subsection{\FloatBarrier}

\newcommand{\beginsupplement}{%
  \setcounter{figure}{0}%
  \renewcommand{\thefigure}{S\arabic{figure}}%
  \setcounter{table}{0}%
  \renewcommand{\thetable}{S\arabic{table}}%
  \setcounter{section}{0}%
  \renewcommand{\thesection}{S\Roman{section}}%
  \setcounter{subsection}{0}%
  \renewcommand{\thesubsection}{\thesection.\arabic{subsection}}%
}

\begin{document}

\title{Lattice initialisation and finite-size effects of non-equilibrium molecular dynamics simulations for heat transfer across graphene–copper interfaces}

\author{L.A. van Goor}
\affiliation{Mathematics of Multiscale Modeling and Simulation, Department of Applied Mathematics, Faculty EEMCS, University of Twente, PO Box 217, 7500 AE Enschede, The Netherlands.}

\author{W.N. Edeling}
\affiliation{Mathematics of Multiscale Modeling and Simulation, Department of Applied Mathematics, Faculty EEMCS, University of Twente, PO Box 217, 7500 AE Enschede, The Netherlands.} 
\affiliation{Centrum Wiskunde \& Informatica, Scientific Computing Group, Science Park 123, 1098XG Amsterdam, The Netherlands.} 

\author{D. Jafari}
\affiliation{Faculty of Engineering Technology, University of Twente, PO Box 217, 7500 AE Enschede, The Netherlands.}

\author{H. Lee}
\affiliation{Mathematics of Multiscale Modeling and Simulation, Department of Applied Mathematics, Faculty EEMCS, University of Twente, PO Box 217, 7500 AE Enschede, The Netherlands.} 

\author{E. Luesink}
\affiliation{Korteweg-de Vries Institute for Mathematics, University of Amsterdam, P.O. Box 94248, 1090 GE Amsterdam, The Netherlands.}

\author{W.W. Wits}
\affiliation{Mathematics of Multiscale Modeling and Simulation, Department of Applied Mathematics, Faculty EEMCS, University of Twente, PO Box 217, 7500 AE Enschede, The Netherlands.} 
\affiliation{NLR - Royal Netherlands Aerospace Centre, Marknesse, The Netherlands.}

\author{A.V. Lyulin}
\affiliation{Mathematics of Multiscale Modeling and Simulation, Department of Applied Mathematics, Faculty EEMCS, University of Twente, PO Box 217, 7500 AE Enschede, The Netherlands.} 
\affiliation{Soft Matter and Biological Physics Group, Department of Applied Physics and Science Education, Eindhoven University of Technology, 5600 MB, Eindhoven, The Netherlands}

\author{B.J. Geurts}
\affiliation{Mathematics of Multiscale Modeling and Simulation, Department of Applied Mathematics, Faculty EEMCS, University of Twente, PO Box 217, 7500 AE Enschede, The Netherlands.} 
\affiliation{Centrum Wiskunde \& Informatica, Scientific Computing Group, Science Park 123, 1098XG Amsterdam, The Netherlands.} 
\affiliation{Soft Matter and Biological Physics Group, Department of Applied Physics and Science Education, Eindhoven University of Technology, 5600 MB, Eindhoven, The Netherlands}

\date{
	\today
}

\begin{abstract}
We study thermal transport across copper–graphene–copper interfaces using Non-Equilibrium Molecular Dynamics (NEMD), and focus on the influence of (i) finite domain length and (ii) domain configuration, including lattice initialisation and associated graphene wrinkling, on the predicted thermal conductivity and Kapitza resistance. NEMD simulations have been employed in the literature to identify trends in the Kapitza resistance of graphene-copper interfaces. However, the outcomes and thus reliability of these simulations may depend heavily on configuration choices, which are currently underexplored in the literature. Here, we identify a strong sensitivity of the Kapitza resistance to domain configuration choices that affect the lattice constants and atomic density. We show that two conventional lattice initialisation strategies yield a difference of nearly a factor of two in the Kapitza resistance, despite differences of only a few per cent in the lattice parameters. This behaviour is accompanied by strain-dependent shifts in the graphene and copper phonon spectra, and by increased phonon overlap at lower strain. Counter to conventional expectations, however, greater phonon-mode overlap coincides with higher Kapitza resistance, demonstrating that spectral overlap alone is insufficient to capture the interfacial heat-transfer dynamics in this system. We suggest that in lattices initialised with lower residual strain, a damping boundary layer develops near the interface, increasing thermal resistance, as indicated by increased local structural disorder and local spectral broadening over a wider interfacial region. Beyond strain- and density-related effects associated with the lattice constants, we find no significant dependence of Kapitza resistance on domain length or on the temperature enforcement at the boundaries as implemented in this study. By contrast, the copper lattice conductivity exhibits clear domain-size and temperature dependence, consistent with phonon mean-free-path limitations and supported by phonon spectral analysis. Our findings call for further exploration of NEMD sensitivity to uncertainties in controlled, continuously varied input parameters to determine reliable thermal transport properties.
\end{abstract}

\maketitle
\section{Introduction}
State-of-the-art manufacturing techniques have enabled the development of copper-graphene composites for thermal-management applications. With the unprecedented demand for data storage and the ever-decreasing size of devices, we are facing challenges in electronic cooling, e.g., in data centres and on-application computing \cite{MOORE2014163}\cite{KHOSRAVI2024114834}. Recent advances in manufacturing technologies, including additive manufacturing, have enabled the production of copper matrix composites containing nanoscale graphene inclusions \cite{Lee2025_RSER}. These composites are promising since graphene, composed of two-dimensional sp$^2$-bonded carbon layers, combines exceptionally high in-plane thermal conductivity of 5000 W/mK with favourable mechanical properties \cite{grapheneproperties}\cite{qadirreview}\cite{corona_graphene_nanocompostie}. The thermal performance of these composites is not determined solely by constituent properties and mass fractions. Literature reports a strong dependence on the morphology, orientation, and distribution of graphene within the metal matrix, as well as on the efficiency of heat transfer across the large number of graphene-copper interfaces \cite{Lee2025_RSER}.\\
\\
Experimentally reported thermal conductivities of copper-graphene composites span a wide range, from $200$ to $800$ W/mK with varying degrees of anisotropy \cite{Lee2026_inprep}, \cite{WANG2025}. This illustrates that both significant increases and decreases in conductivity can be achieved relative to the reference value of approximately $400$ W/mK for pure copper \cite{Thermalconductivities}. The spread in measured thermal conductivities reflects the combined influence of the strong anisotropy of graphene, its sensitivity to substrate interactions, internal defects, thermal resistance across interfaces, and the size and distribution of graphene inclusions in the copper matrix \cite{Boden2014Nanoplatelet} \cite{Pop2012}\cite{substrate_graphene_metal_Kelly}. To gain deeper insight into the interfacial thermal transport, Non-Equilibrium Molecular Dynamics (NEMD) simulations have been widely used to study the interfacial thermal conductance (ITC) and Kapitza resistance of the graphene-metal interface by resolving atomistic dynamics, lattice vibrations, phonon transport, anharmonic effects, and interfacial scattering processes\cite{Zhu2022}\cite{Li2022}\cite{Fang2025}. These studies have identified trends in thermal transport as a function of physical parameters, such as density, strain, defects, and the number of graphene layers.\\
\\
Although NEMD simulations provide atomistic insight into thermal transport, their predictions depend not only on physical interface properties but also on computational domain construction and boundary treatment. In nanoscale systems, thermal transport is inherently sensitive to the finite size of the computational domain, because phonon mean free paths, the length over which coherent lattice vibrations decay, and boundary scattering affect thermal conductivity \cite{CASIMIR1938495}\cite{Park2013}. Length-dependent thermal conductivity for graphene has been reported up to micrometres, and likewise for copper up to nanometers \cite{Xu2014}, which corresponds to reported (collective) phonon mean free paths of $100\,\mu$m and $10$nm, respectively \cite{Mei2014}\cite{Fugallo2014}\cite{Tong2019}\cite{Saether2022}. Furthermore, thermal transport in graphene is strongly affected by substrate interactions, which can reduce the effective thermal conductivity through phonon scattering, phonon leakage, and thermal boundary resistance \cite{Pumarol2012}\cite{Cai2010}. In metal-supported graphene, additional electronic effects such as charge transfer, Fermi-level shifts, and metal-dependent graphene--metal bonding may also occur, which can likewise impact phonon-electron contributions at the interface \cite{substrate_graphene_metal_Kelly}.
EMD simulations of graphene have been shown to match qualitative trends in the length dependence of graphene's thermal conductivity to experimental results \cite{Xu2014}. These considerations raise the possibility that the thermal properties obtained from MD simulations may depend on computational choices such as the definition of conductance, domain length, boundary treatment, and thermostat implementation, rather than reflecting only intrinsic interfacial physics. This challenge is the main motivation behind the current study, in which we focus on two main sources of uncertainty, i.e., (1) the computational domain size and (2) the disparate length scales of the atomic lattices that come together at the interface, where lattice mismatch and variations in lattice constants can alter the interfacial dynamics:
\begin{enumerate}
    \item The effect of domain length on Kapitza resistance in MD simulations has received some attention in the literature. Domain length has been shown to influence the resolved phonon spectrum, as well as the computed phonon-mode overlap of the phonon spectra of two adjacent material domains, and consequently, the resulting thermal interface resistance \cite{MD_interface_length}\cite{Zhan2015}. Thermostat choices have also been reported to disrupt long-range mode-mode correlations even when heat baths are located far from the interface \cite{Gordiz2017}. However, the literature reports are inconsistent regarding the order of magnitude of the characteristic computational domain length over which the thermal interface resistance varies, ranging from a few to at least 100 nanometers \cite{Chalopin2012}\cite{Landry2009}. These contradictions have been attributed to differences in simulation setup, such as domain size, boundary treatment, and thermostat implementation \cite{Stanley2023}. When the computed phonon spectrum is biased due to boundary or thermostat artefacts that are unable to decay within the domain or over the simulation time, the robustness and accuracy of the extracted thermal properties become questionable. Contradictory reports on the length dependence of thermal interface resistance from MD simulations may arise from methodological and numerical artefacts, and a systematic study is therefore needed.
    \item A second source of uncertainty arises from the initialisation of the atomic lattices. The equilibrium densities and lattice constants required for lattice initialisation are not always known for the specific composite system. Residual strain introduced during the initialisation can affect thermal properties, including the thermal interface resistance \cite{Li2022}. To reduce residual strain, MD simulations can be preceded by an equilibration procedure involving domain relaxation \cite{Braun2019BestPractices}. In composites, however, residual strain due to lattice mismatch cannot be eliminated entirely, and full structural relaxation remains unattainable. In this study, we compare two initialisation strategies: one based on lattice constants and interfacial distances consistent with experimental observations \cite{Fang2025}, and one based on an equilibration procedure that minimises strain through domain relaxation and minimal mismatch. We assess how these choices affect the geometry of the interfacial region, the ITC, and the domain-length dependence of the ITC.
\end{enumerate} 
In this paper, we examine the relative contributions of finite-size effects, lattice initialisation, and temperature-enforced boundary conditions to thermal transport across a single layer of graphene embedded in copper in NEMD simulations. To isolate these effects, we restrict the study to a single choice of force field, thermostat, and MD method, and quantify the influence on the thermal properties due to variations in the domain length, initialisation strategy, and temperature enforcement at the boundaries. Here, we focus on the sensitivity of the thermal conductivity and thermal interface resistance to these choices, and use phonon spectra to analyse boundary and initialisation effects. We show that the ITC is highly sensitive to initialisation choices that affect residual strain. Beyond these residual-strain effects, which depend on the initialisation approach and can be enhanced in smaller domains where residual stress has less opportunity to redistribute across the system, the ITC shows no significant dependence on domain length. By contrast, the copper conductivity of the bulk volumes adjacent to the graphene layer exhibits clear temperature and domain-size dependence. These findings highlight the importance of distinguishing genuine interfacial physics from artefacts introduced by the simulation setup. To the best of our knowledge, no systematic study has yet examined this sensitivity of MD simulations for single-layer graphene embedded in copper.  \\
\\ 
The remainder of this paper is organised as follows. In Section \ref{sec:Methods}, we introduce the NEMD method and simulation details, including the two lattice-initialisation approaches and the two temperature-enforcement approaches used when varying the domain length. Section \ref{sec:resul} presents the results in four parts: first, validation of the NEMD implementation against a literature case \cite{Zhu2022}; second,  the interfacial thermal conductance for various domain lengths; third, the differences between the two initialised cases; and fourth, the computed copper conductivities. Concluding remarks are provided in Section \ref{sec:concl}.

\section{Modelling and Simulation Details}\label{sec:Methods}
In this section, we describe the NEMD method employed in this study, the simulation details and post-processing choices, the lattice initialisation and temperature enforcement for varying domain lengths. Through NEMD, we calculate the bulk conductivities and the thermal interfacial resistance, known as the Kapitza resistance, for setups of the graphene-copper interface of varying domain length $L_z$, keeping the cross-sectional area constant. We describe the NEMD method in subsection \ref{sec:NEMDMETHOD}. We introduce the force fields, domain setup and post-processing steps in subsection \ref{subsec:simdetail}. In subsection \ref{subsec:initialisation}, the initialisation choices of the lattices are described, and in subsection \ref{subsec:temp} we formulate how temperature gradients are imposed as a function of domain lengths in this paper. 

\subsection{The Non-Equilibrium Molecular Dynamics method}\label{sec:NEMDMETHOD}
The NEMD method is used to calculate conductivities and interfacial conductance from a driven non-equilibrium steady state, in which an imposed thermal perturbation generates a heat flux and temperature profile that can be related to material properties through Fourier's law. Such driving may be imposed by direct temperature enforcement at the boundaries (Direct NEMD), momentum exchange at the boundary particles (Müller-Plathe \cite{MuellerPlathe1997}), or other types of perturbations. The NEMD method relies on a set of assumptions that resemble the statistical derivations of equilibrium methods, such as the Green-Kubo method\cite{Kubo1957}\cite{Green1954}, which is founded in linear response theory. In this paper, the direct NEMD method is employed by imposing hot and cold temperatures at the domain boundaries and thereby generating a stationary heat flux in an emerging statistically steady-state. The non-equilibrium steady state is interpreted within the standard local-equilibrium framework of linear irreversible thermodynamics; although the system is globally out of equilibrium, in each point a local region may be defined small enough to admit an equivalent local temperature, through equipartition. This local-equilibrium framework, together with  Fourier-type constitutive behavior and Onsager reciprocity, provides the basis for extracting thermal conductivity and interfacial thermal resistance from the simulated statistically steady-state temperature profile and heat flux.\cite{Onsager1931}\cite{deGrootMazur1984}\\
\\
There is some debate in the literature over the equivalence of NEMD methods and equilibrium-state MD methods (EMD), such as the Green-Kubo approach, \cite{Dong2018} \cite{Nejatolahi2021}\cite{Khadem2013}\cite{Green1954}\cite{Kubo1957}\cite{Matsubara2017}. From a fundamental point of view, bulk NEMD and EMD methods should correspond closely within the linear response regime, and discrepancies should be sought in domain, thermostat, or sampling choices, \cite{Chen2022} \cite{Ghatage2020} \cite{Kubo1991_StatPhysII} \cite{Onsager1931}. This is not the case when interface properties are concerned. EMD and NEMD simulations measure inherently different conductances \cite{Chen2022}. NEMD simulations of graphene were found to match experimentally observed qualitative trends in the length dependence of graphene's thermal conductivity, whereas the Green-Kubo method failed \cite{Xu2014}. Therefore, with a focus on domain configuration effects on interfacial resistance, we will employ the direct NEMD approach.\\
\\
From the NEMD simulations, we derive two material properties: the lattice thermal conductivity within the homogeneous copper domains and the interfacial thermal resistance:
\begin{enumerate}
    \item The thermal conductivity tensor $\mathbf{K}$ is defined through Fourier's constitutive relation between the gradient of the temperature $T$ and the heat flux $\mathbf{J}$: 
\begin{equation}\label{eq:Fourier}
    \mathbf{J} = -\mathbf{K} \nabla T
\end{equation}
For a material region that is homogeneous along the temperature-gradient direction and remains within the linear-response regime, a one-dimensional linear temperature profile is expected to emerge in the statistically steady state of the Direct NEMD simulation. Denoting the direction of the temperature gradient by unit vector $\mathbf{e}$, the corresponding conductivity $\kappa_\mathbf{e}$ is given by:
\begin{equation}
   \kappa_\mathbf{e}= \mathbf{e}\cdot \mathbf{K}\mathbf{e} =-\frac{\mathbf{J}\cdot \mathbf{e}}{a} , \quad T(\mathbf{x}) = a (\mathbf{x\cdot e})+b
\end{equation} 
\item The interfacial thermal resistance is defined in terms of an effective discontinuity in the linear temperature profile at the location of the interface. From the normal heat flux and this discontinuity in the temperature profile, we define  the interfacial thermal resistance, known as the Kapitza resistance, as \cite{Kapitza1941}:
 \begin{equation} \label{eq:Kapitza}
        R_K = \frac{1}{\text{ITC}} = \frac{\delta T}{J_n}, \quad \delta T = |T^+-T^-|
    \end{equation}
where $\delta T$ denotes the magnitude of the temperature discontinuity at the interface, $T^\pm$ denote the one-sided interfacial limits of the local temperature profile at the interface $T^{\pm} = \lim_{\epsilon\to0^+} T(\mathbf{x_I}\pm \epsilon \mathbf{n})$ where $\mathbf{x}_I$ is a point on the interface, and $J_n = |\mathbf{J\cdot n}|$ is the magnitude of the heat flux normal to the interface. In this work the temperature gradient is perpendicular to the average interfacial plane $\mathbf{n}=\mathbf{e}$.
\end{enumerate} 
\begin{figure*}
\centering 
    \includegraphics[width=0.7\linewidth]{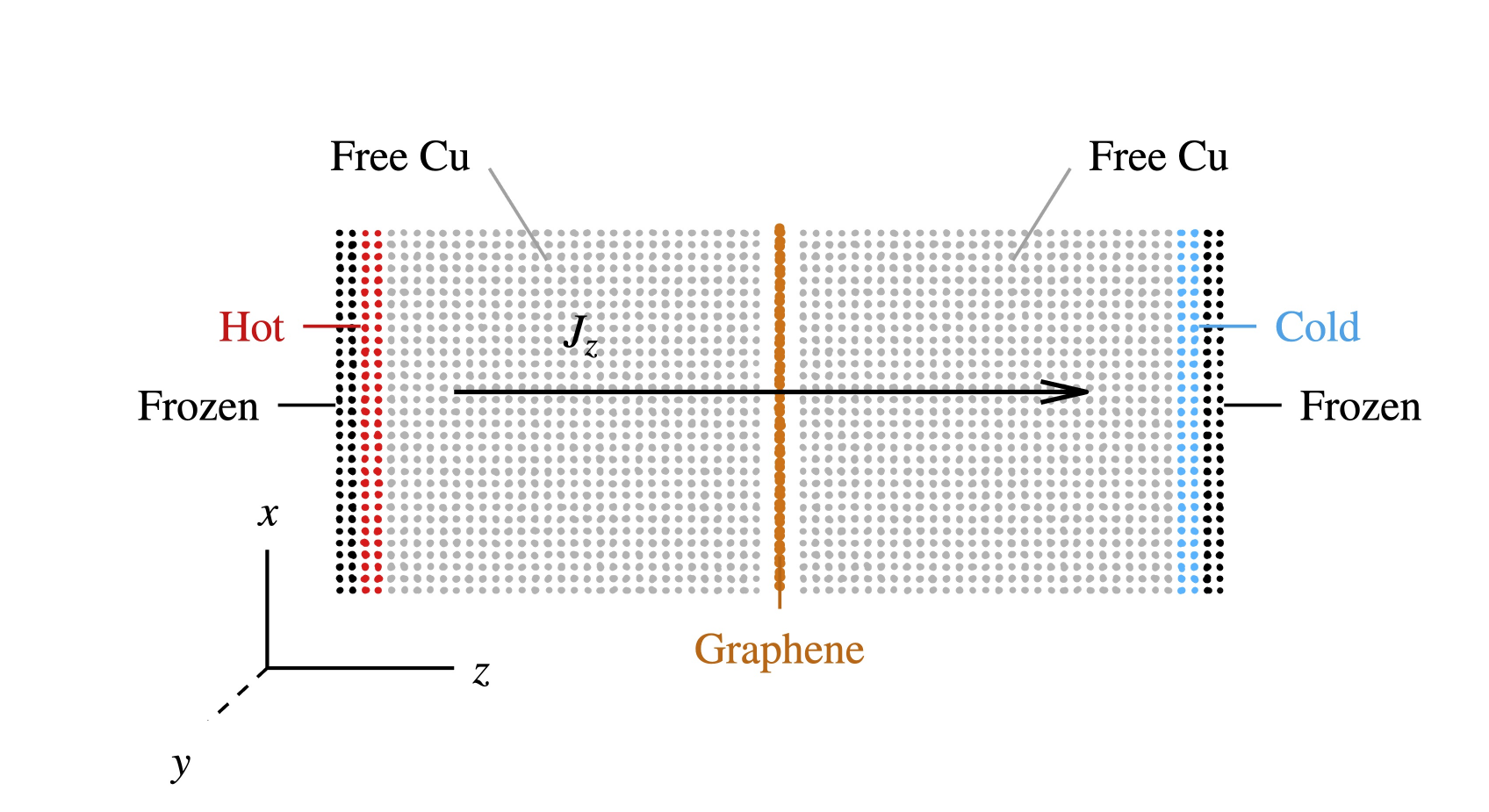}
        \caption{Direct NEMD setup domain with fixed boundaries along $z$ and periodic in $x,y$. A temperature gradient is imposed using Langevin thermostats in slabs of $4$\AA\, at $T_{hot}$ (red), and at $T_{cold}$ (blue) adjacent to slabs of $3$\AA\, with fixed atoms (black), over the Cu-G-Cu interface, viewed in the x-z plane. A stationary flux is achieved in the statistically steady state $J_z = \mathbf{J\cdot e}_z$ through the free copper atoms (grey) and the graphene layer (orange).}
        \label{fig:NEMDsetup}
    \end{figure*}
To derive the thermal conductivity and Kapitza resistance from the Direct NEMD simulations, we need to define local temperatures and the heat flux from the simulation. In a non-equilibrium system, the local temperature should be understood as an equivalent equilibrium temperature, that is, as a measure of the local average energy density rather than corresponding to a uniquely defined underlying distribution \cite{chen2005nanoscale}. There are no universally accepted rigorous criteria for local equilibrium on which to base the region's size within an NEMD system, which forms a compromise between statistical accuracy and adhering to the equilibrium assumption. Accordingly, the equivalent equilibrium temperature in some local region $\Omega_k$ centered around point $\mathbf{x}_k$ is defined through the equipartition theorem as:
 \begin{equation}
       {T}(\mathbf{x}_k) = \frac{2}{3N_kk_B} \left<\sum_{i\in \Omega_k} \frac{1}{2}m_i|\mathbf{v}_i|^2\right> \label{eq:temperature}
    \end{equation}
\\
where $T$ defines the ensemble average, which is the instantaneous estimator at sampling time $t$, with $m_i$ and $\mathbf{v}_i$ being the mass and velocity of particle $i$, and $k_B$ is the Boltzmann constant, $N_k$ is the number of atoms in $\Omega_k$. The local region $\Omega_k$ around $\mathbf{x}_k$, can be for instance a ball or radius $r_\Omega$, $\Omega_k = \left\{\mathbf{x}: \|\mathbf{x}-\mathbf{x}_k\| \leq r_\Omega \right\}.$ \\
\\
We estimate the local temperature $T$ by a converged time average, $\bar{T}(\mathbf{x}_k)$. Assuming ergodicity, time averages over a sufficiently long statistically steady trajectory are equivalent to ensemble averages. The resulting $\bar{T}$ is therefore used as the statistically steady temperature profile associated with the equilibrated configuration.
\begin{equation}
   \bar{T}(\mathbf{x}_k) = \lim_{N_t \to \infty} \frac{1}{N_t}\sum_{j=1}^{N_t} \frac{2}{3N_kk_B} \sum_{i\in \Omega_k} \frac{1}{2}m_i|\mathbf{v}_i|^2 \label{eq:temperature_average}
\end{equation}

\subsection{Simulation details and post-processing}\label{subsec:simdetail}
\begin{figure}
    \centering
    \includegraphics[width=0.9\linewidth]{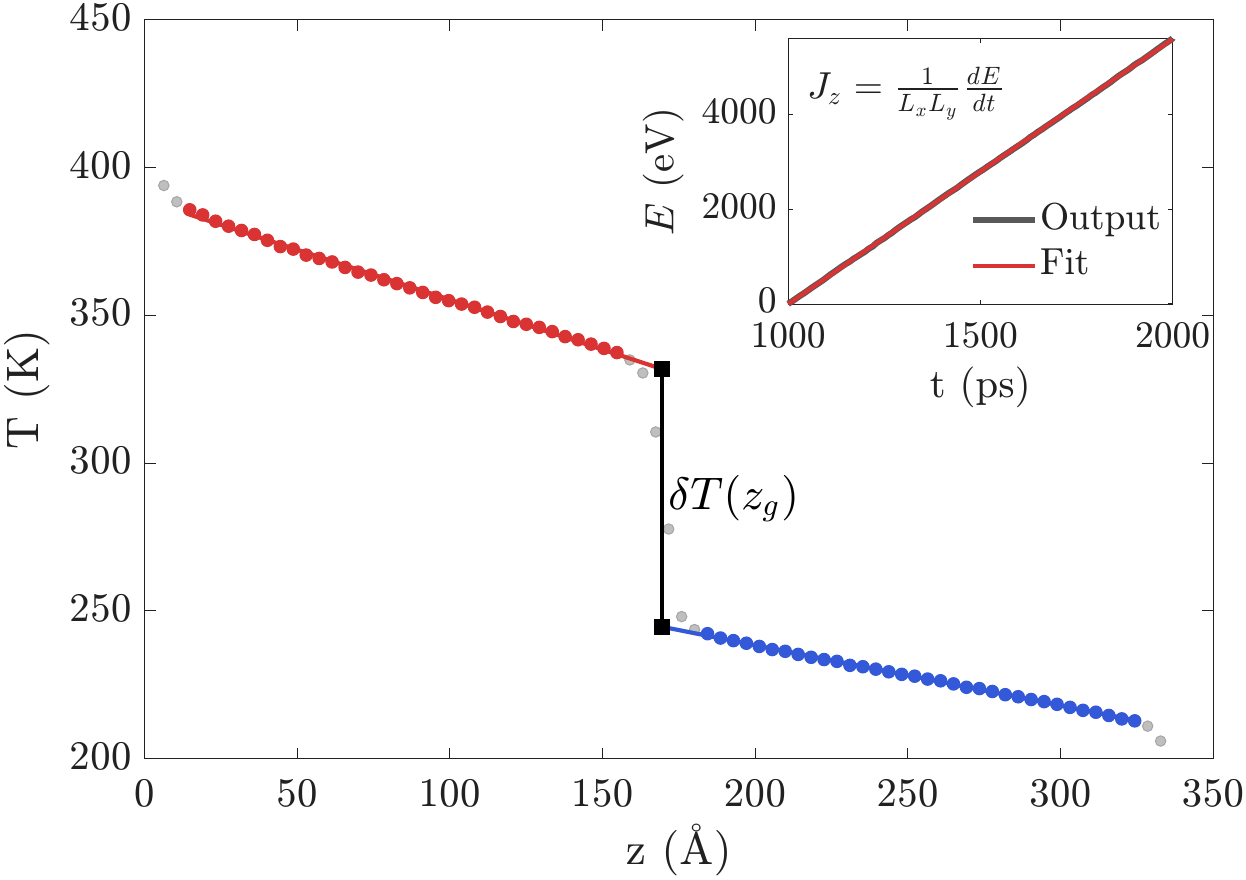}
    \caption{Temperature profile as a function of $z$ during the NEMD run. The intervals included in the fits are colored (red, blue) while the discarded intervals are grey. The plotted lines are the least-squares fits $ (a_i z + b_i)$ (red, blue), and the calculated jump $\delta T(z_g)$ (black). Both Cu and G atoms are included in the average temperature, accounting for the centre points due to an oscillating graphene layer along $z$. Cumulative energy from the thermostats $E = \sum \frac{f_{hot}-f_{cold}}{2}$ during the production run and linear fit from which $J_z = \frac{1}{L_xL_y}
\frac{dE}{dt}$ is derived (top right).}
    \label{fig:tmp_relax}
\end{figure}
In this subsection, the simulation method, force fields and thermostats are described. Subsequently, the domain and regions of temperature enforcement are discussed. Then, the temperature sampling and derivation of the interfacial temperature difference from the simulation output are described, and lastly, the formulae describing the phonon spectra are shown.\\
\\
\textbf{Dynamics and Force Fields}\\
Direct NEMD simulations were performed using the Large-scale Atomic/Molecular Massively Parallel Simulator (LAMMPS) \cite{Thompson2022_LAMMPS}. Here, atoms are represented as point particles whose motions are dictated by Newton's equation under forces arising from prescribed interatomic potentials and thermostats. The velocities and positions of the atoms are updated through the symplectic St\o rmer-Verlet integration scheme \cite{Verlet1967}\cite{Hairer1993}. The atomic masses were assigned from isotope-averaged standard atomic weights, using 63.546 amu for copper and 12.011 amu for carbon \cite{weights_report}. Interatomic forces are prescribed through a combination of potentials following previous literature approaches, since a single force field for a graphene-copper embedding is currently unavailable \cite{Zhu2022} \cite{Li2022}\cite{Fang2025}. The interactions between copper atoms are calculated following the Embedded Atom Method (EAM) potential for copper atom interactions as developed by Mishin et al. (2001) using ab initio calculations \cite{Mishin2001}. The EAM potential considers both a pairwise potential function and an embedding contribution from the electron density within a cutoff radius of $\sim 5.5$ \AA\,, which provides a reliable representation of structural and thermal properties of copper \cite{DawBaskes1984} \cite{Mishin2001}. Carbon interactions are modelled with the Adaptive Intermolecular Reactive Bond Order  (AIREBO) potential, known for representing both chemical and van der Waals bonding, enabling bond formation and breaking dynamics \cite{Stuart2001}. This is particularly relevant for graphene, where the short-range intralayer $sp^2$ bonds are much stronger and chemically bonded compared to interlayer bonds, which are held only by van der Waals forces and are prone to wrinkling, folding and defects \cite{Pop2012}. The copper-carbon bonds are assumed to arise solely from van der Waals forces and are represented with a Lennard-Jones (LJ) potential with mixing-rule coefficients of $\epsilon=0.02578$ eV, $\sigma=3.0825$\AA\,  and a cutoff radius of $r_c=10.0$ \AA\, \cite{Guo2006}. Temperature-controlled regions are subjected to additional stochastic and dissipative forcing terms through a Langevin thermostat with a damping time of 0.1 ps \cite{Schneider1978}. The stochastic component aids phase-space sampling and helps promote ergodic behaviour. For additional information and full expressions of the Langevin thermostat and force fields, the reader is referred to Section\ref{sec:add_theory}.\\
\\
\textbf{NEMD domain}\\
In this study, we consider a domain of size $(L_x,L_y,L_z)$, periodic in the $x,y$ plane and fixed along $z$. The simulation domain contains two regions of copper atoms, separated by a graphene layer. In slabs of $3.0$\AA\, at the fixed boundaries along $z$, atoms are fixed with zero initial velocity and zero forces, such that their positions and velocities are fixed throughout the simulation. A temperature gradient over the domain is enforced along $z$ by $\Delta T(0,L_z) = |T(0)-T(L_z)|=T_{hot}-T_{cold}$. A slab of $4.0$\AA\, adjacent to the fixed slab at one boundary is forced to $T_{hot}$, and a slab of $4.0$\AA\, next to the fixed slab at the opposite boundary is forced to $T_{cold}$. Different seeds are assigned to the thermostat in each slab to avoid correlations in the stochastic noise. The NEMD setup of the domain is
visualised in Figure \ref{fig:NEMDsetup}. \\
\\
\textbf{NEMD simulation and post-processing}\\
The lattice initialisation is described in section \ref{subsec:initialisation}. After lattice initialisation, each direct NEMD simulation is equilibrated under the temperature gradient for 1 ns, with the domain dimensions held constant. After the NEMD equilibration run, the system is assumed to be in a non-equilibrium statistically steady state, and a production simulation is performed for 1 ns unless otherwise specified. The timestep is 1 fs. \\
\\
During the NEMD production run of 1 ns, local temperatures are sampled every ps from intervals of width $\Delta z$ along $z$ under symmetry in the $x,y$ plane, e.g. in slab regions $\Omega_k = \left\{ \mathbf{x} \in \Omega : z_k-\frac{\Delta z}{2} \leq \mathbf{x}\cdot\mathbf{e}_z < z_k+\frac{\Delta z}{2} \right\}$. These samples are assumed to be approximately decorrelated, in line with the argument that time averaging and ensemble averaging are equivalent. From the decorrelated samples, the temperature in each interval is averaged, and the temperature profile in the copper domains is approximated by an ordinary least-squares linear fit, shown in Figure \ref{fig:tmp_relax}.
\begin{equation}\label{eq:T_Cu}
        T_{Cu}(z) = \begin{cases}
            a_1z+b_1, \quad \text{ for }z\in [0, z_g)\\
            a_2z+b_2, \quad \text{ for }z\in (z_g, L]
        \end{cases},    
    \end{equation}
    \[(a_i,b_i) = \arg\min_{a_i,b_i}\left[
    \sum_{k; z_k \in \Omega_{Cu_i}}
    \left[ \bar{T}(z_k) - (a_i z_k + b_i) \right]^2 \right]\]
    where $\bar{T}$ the average temperature as defined in equation \ref{eq:temperature_average}, $\sigma_{\bar{T}_k}$ is the standard error of the time-averaged temperature in slab $k$, and $z_g$ is the position of the graphene sheet, assuming the temperature gradient is small enough to ensure the linear response regime in copper, $a \approx \left(\frac{d T}{dz}\right)_{Cu}$.\\
\\
The temperature difference across the interface is derived by evaluating equation \ref{eq:T_Cu} at the approximate position of the graphene layer. By using the temperature profiles within these regions, rather than local temperatures adjacent to the interface, we avoid the ambiguity in the definition of local temperature near the interface \cite{Chen2022},\cite{Stanley2023}. 
   \begin{equation}
      \delta  T(z_g) = |T_{Cu}(z_g)^+- T_{Cu}(z_g)^-| \label{eq:T_Cu2}
    \end{equation}
The magnitude of the heat flux along $z$,  $J_z = \mathbf{J\cdot e}_z$, is obtained from a linear ordinary least squares fit to the energy input from the thermostats at the boundaries and shown in Figure \ref{fig:tmp_relax}. Using the flux, temperatures, and temperature difference across the interface, the conductivity in the copper domains is calculated using equation \ref{eq:Fourier}, and the Kapitza resistance is calculated using equation \ref{eq:Kapitza}. \\
\\
Additionally, from the NEMD simulation we can derive the coherent lattice vibrations and construct the phonon spectrum. First, the normalised velocity autocorrelation function (VACF) is calculated as
\begin{equation} \label{eq:VACF}
    \text{VACF}(t)= \frac{\sum \left< v_i(0)v_i(t)\right>}{\sum \left< v_i(0)v_i(0)\right>}
\end{equation}
which decays within 1 ps to $|\phi_{VACF}|<0.1$, as shown in Figure \ref{fig:VACF1}.

Then, using a Fourier transform, the vibrational density of states (VDOS) is derived as:
\begin{equation}
    \text{VDOS}(\omega) = \int_0^\infty \text{VACF}(t) \exp(-2\pi i\omega t)\, \mathrm{d}t
\end{equation} 
For details on error propagation in the post-processing steps, please see section \ref{sec:uncertainties}. For details on the equilibration and convergence of temperatures $\bar{T}(z)$ within an interval along $z$, the VACF and velocity distributions, the reader is referred to section \ref{sec:details_velocities}.

\subsection{Lattice initialisation}\label{subsec:initialisation}\begin{figure}
    \centering
      \begin{subfigure}{\linewidth}
\includegraphics[width=\linewidth]{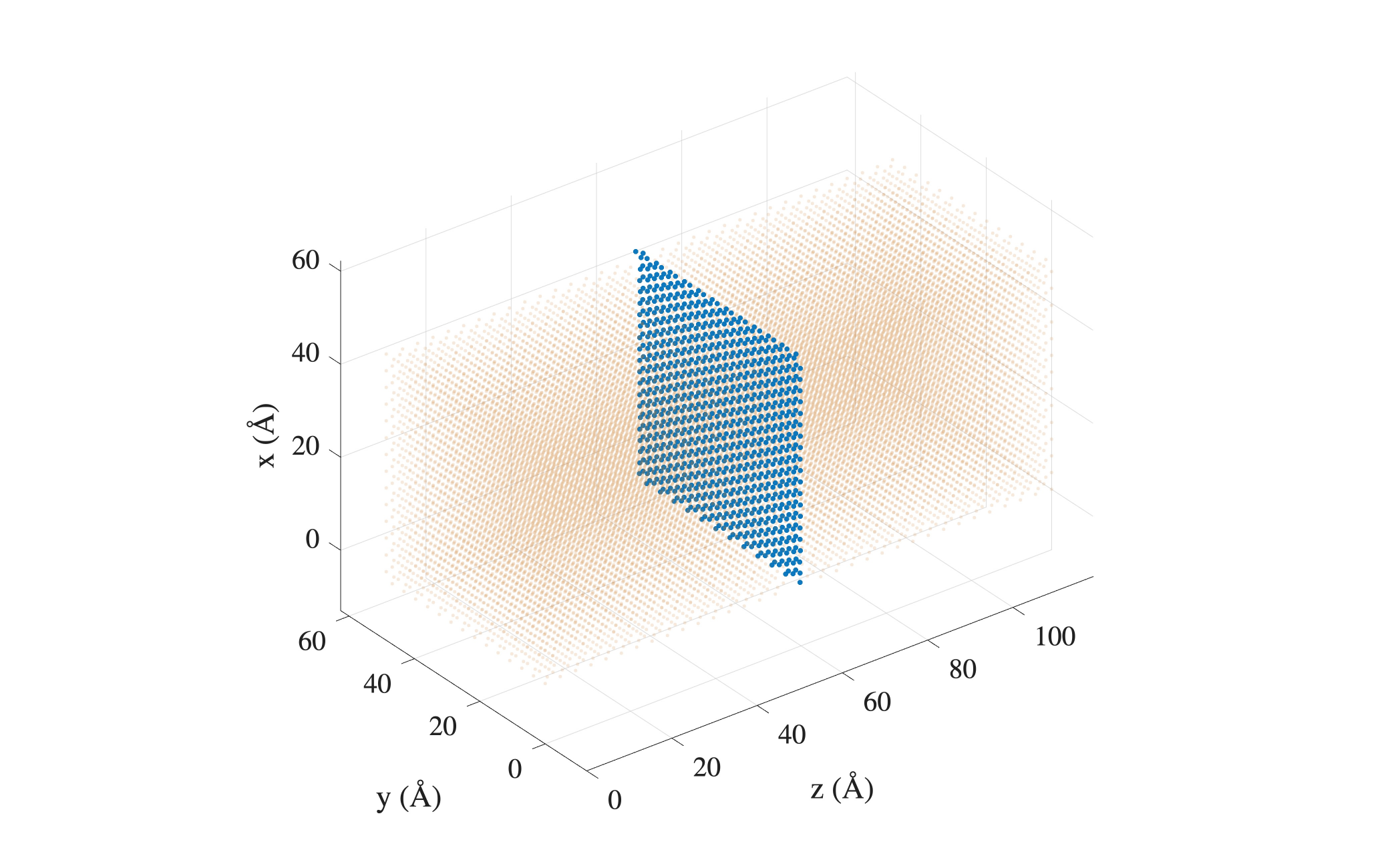}
    \end{subfigure}
  \begin{subfigure}{\linewidth}
\includegraphics[width=\linewidth]{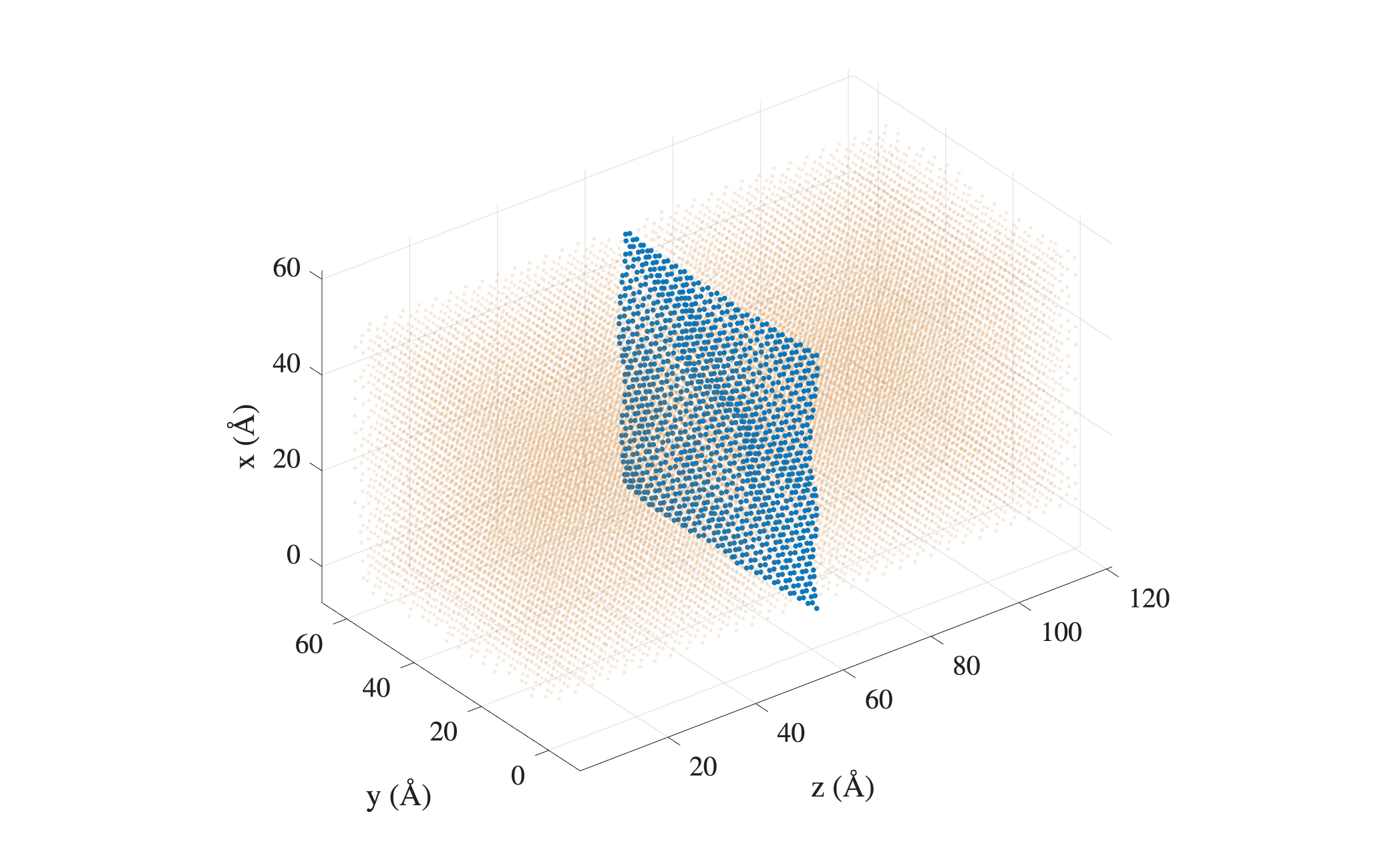}
  \end{subfigure}
    \caption{Graphene-copper composite with periodic boundaries in the $xy$ plane and fixed boundaries for $z$. Case I, $(L_x,L_y,L_z) = (50, 50 ,120)$\AA\,  with $a_G=2.46$\AA\,, $a_{Cu}=3.6$ \AA\,, after an NVT run of 200 ps at 300K without gradient (top). Case II equilibrated to $(L_x,L_y,L_z) =(54.3, 57.9, 121.2)$\AA\, at 300K using anisotropic NPT relaxation to $a_G\approx 2.41 $\AA, $a_{Cu}\approx 3.61$ (bottom).}
    \label{fig:initialisse_strained}
\end{figure}
In this paper, we will employ two approaches to initialise the copper-graphene-copper composite for varying domain lengths, which we will refer to as Case I and Case II. Case I relies on experimentally derived material properties, while Case II relies on domain relaxation. The lattices are constructed in the following two ways:
\begin{enumerate}
    \item \textbf{\textit{(Case I):}} A domain is initialised of size $L = (L_x,L_y,L_z)$ with two domains filled with an FCC Cu lattice with given lattice parameter $a_{Cu}=3.6$\AA. The two copper domains are separated by a graphene layer at $L_z/2$ with an initial separation distance, $d_{Cu-G}$, of $3$\AA\, from each copper domain. The graphene layer is initialised as a honeycomb structure with lattice parameter $a_{G}=2.46$\AA \,\cite{Fang2025}\cite{Zhu2022}\cite{BokdamGraphenedistance}. For varying domain sizes, this construction yields approximately constant lattice strain and mismatch strain, up to finite-size variations, of the domain size modulo lattice periodicity, i.e. $[L_\alpha \bmod \lambda_\alpha]$, where $\lambda_\alpha$ is the smallest translational repeat length along the unit vector $\mathbf{e}_\alpha$, with $\alpha\in\{x,y,z\}$. The variation in strain can be roughly proportional to the variation in the atomic density in the domain. Furthermore, $(L_x,L_y) = (50,50)$ \AA\, unless otherwise specified. 
    \item  \textbf{\textit{(Case II):}} A domain is initialised by a composition of two relaxed fcc Cu lattices and a relaxed honeycomb graphene layer. These structures were separately relaxed at 0K and equilibrated at 300K from which the lattice parameters are derived of $a_{Cu}=3.632$\AA\, and $a_G=2.422$\AA. The composite is constructed with minimal mismatch strain by matching the number of unit cells based on the derived lattice parameters. An initial average separation distance of $d=3$\AA\, is used between the equilibrated graphene layer and the equilibrated copper lattices along $z$, but this distance is not leading for the initialised structure since the domain dimensions are allowed to equilibrate. The constructed composite is again allowed to equilibrate to the resulting domain dimensions and atomic positions, thereby minimising the system pressure. The equilibrated system is assumed to approach a strain-free system up to mismatch strain. The final domain is within $0.2\%$ of the equilibrated domain dimension $(L_x, L_y,L_z)$ as can be seen in Figure \ref{fig:relaxation300Kgraphene_cu__lattice}. For different domain sizes, the fraction of copper atoms relative to graphene atoms determines the effective lattice constant. Obtained mismatch strains and lattice dimensions are given in Table \ref{tab:mismatchlattice}. The full equilibration procedure is described in section \ref{sec:lattice_initilization_caseII}.
\end{enumerate}
The main differences in lattices obtained through initialisation of Case I and Case II are found to be deviations of a few per cent in the lattice parameters and the apparent order of the graphene layer and copper lattice in the interfacial region. After an NVT run of 200 ps at 300K, the graphene layer of Case I simulations remains highly structured, while in Case II the graphene layer as well as the neighbouring copper atoms display wrinkling, as can be seen in Figure \ref{fig:initialisse_strained}. This is also seen in atomic positions in NEMD simulation visualised in Figures \ref{fig:average_Z} and \ref{fig:VDOS_interf_regions}, as viewed in the $z-x$ plane. Since the total separation between the copper domains is approximately $6$\AA\, in both Case I and Case II, there is no direct indication that any observed differences arise from a change in the overall inter-domain separation or from altered force magnitudes due solely to the distance dependence of the potential.

\subsection{Temperature enforcement}\label{subsec:temp}
To identify the domain length dependence of the thermal transport, we employ two means of temperature enforcement for varying domain lengths. The approaches keep different thermal quantities constant and introduce distinct biases. Both approaches keep the average temperature $(T(L)+T(0))/2$, at $300$K, and the temperatures are enforced in regions of the same width with a Langevin thermostat with the same damping time. The temperatures at the boundaries as a function of domain length are enforced in the following two ways:

\begin{enumerate}
    \item \textit{\textbf{(\boldmath$\Delta T = $const.):}} The first approach keeps the temperature difference across the whole domain constant, thus keeping the boundary temperatures constant on average: 
    \begin{equation}
        \Delta T  = |T(0) - T(L_z)| = \text{ constant}
    \end{equation}
    where $L_z$ is the domain length.\\
 The advantage of this method is that the boundary conditions are identical for all domain lengths. The main limitation lies in the decrease of the temperature gradient as the domain length increases. Thermal transport becomes difficult to distinguish from statistical and physical fluctuations and noise in the limit of large domain lengths. Additionally, the magnitude of the temperature difference across the interface will not remain constant. If the ITC depends on the magnitude of this difference, then varying the domain length while keeping $\Delta T$ constant will reproduce the effects of finite domain size as well as the influence of the interfacial temperature difference.
    \item \textit{\textbf{(\boldmath$\delta T = $const.):}} The second approach keeps the temperature difference across the interface $\delta T$ as well as the temperature gradient within each of the copper domains $\nabla T_{Cu}$ approximately constant. 
    \begin{equation}
        \delta T = |T(z_g)^+ -T(z_g)^-|  = \text{ constant},    \end{equation}\[ \nabla T_{Cu} = \Delta T_{Cu} / L_{Cu} = \text{constant}\]

    where $z_g$ is the position of the graphene layer.\\
    This is achieved by extrapolating the boundary temperatures required to maintain the interfacial temperature difference, under the assumption that dynamics within the copper domains remain in the linear response regime following the Fourier law with constant conductivity within each domain. The main limitation of this approach is the non-constant boundary temperatures for various domain lengths. Different enforced boundary temperatures have a different phonon population enforced by the Langevin thermostat. Therefore, boundary artefacts are not controlled in this approach. Moreover, for sufficiently large domain lengths, the required boundary temperatures may drive the system outside the linear-response regime.
\end{enumerate}
\begin{figure}
  \centering
  \begin{subfigure}[c]{\linewidth}
    \centering
    \includegraphics[width=0.9\linewidth]{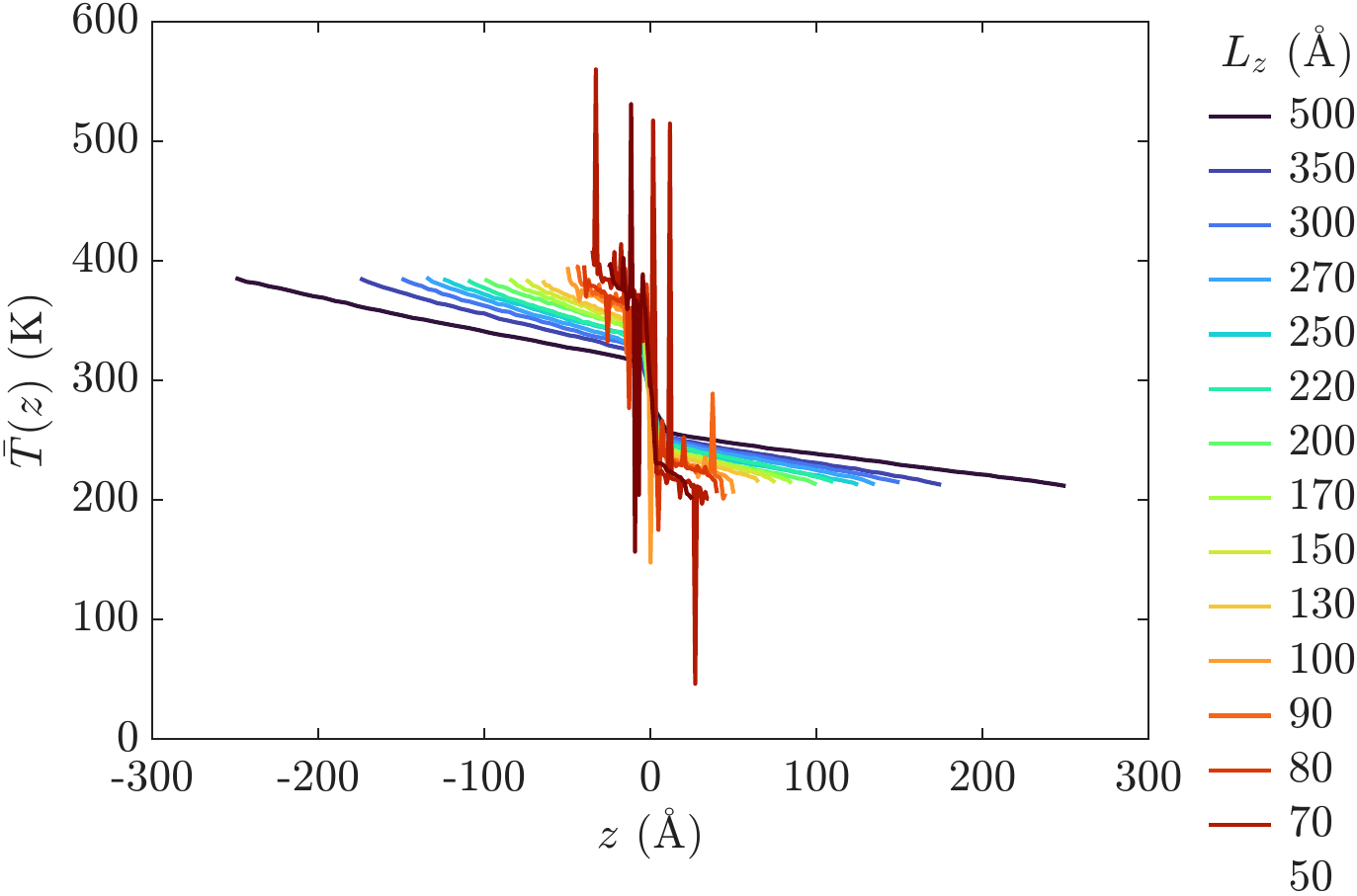}
  \end{subfigure}\hfill
  \begin{subfigure}[c]{\linewidth}
    \centering
\includegraphics[width=0.9\linewidth]{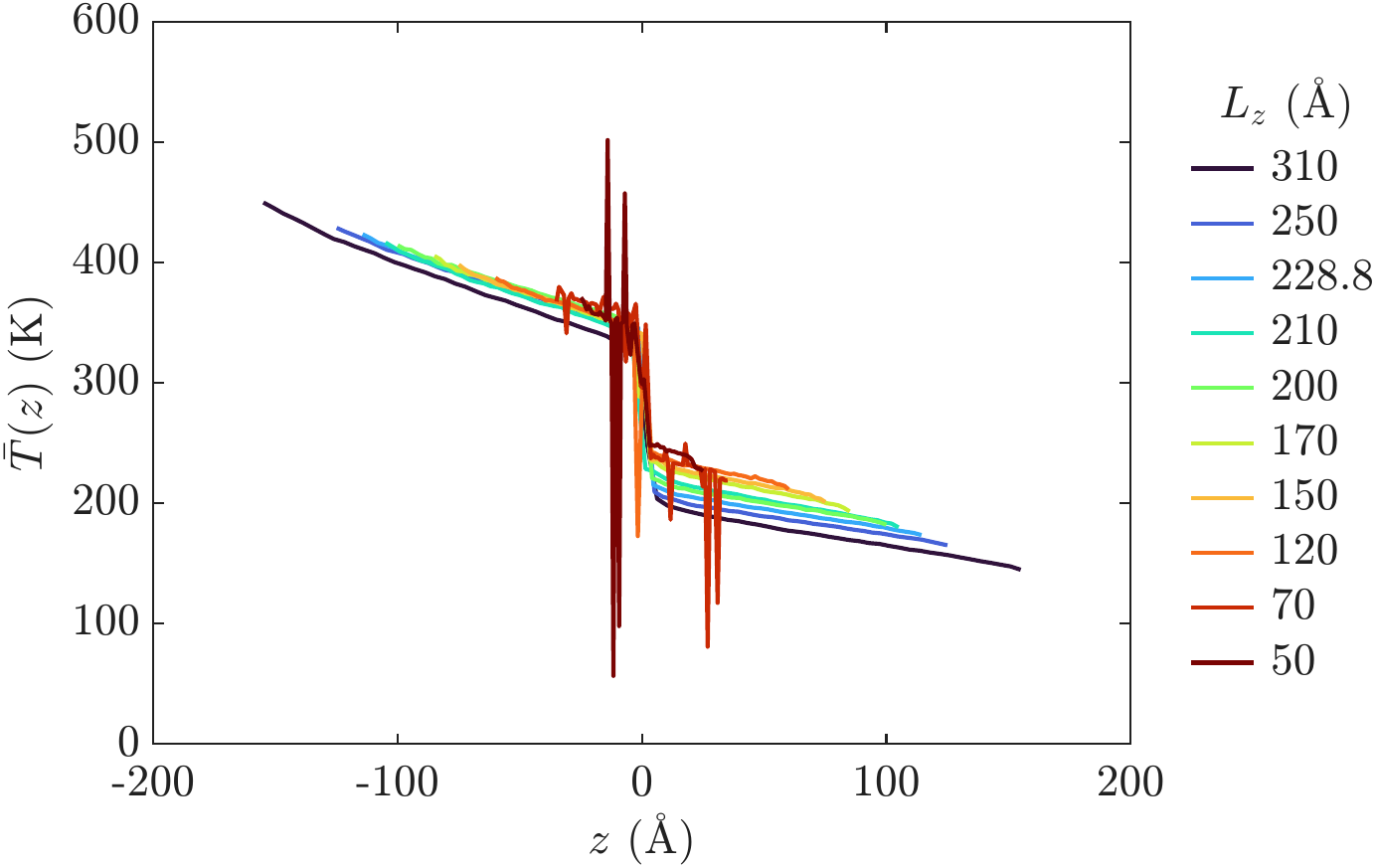}
  \end{subfigure}
\caption{Temperature profiles for different domain length $L_z$, with $L/2$ centered at $z=0$ with constant $\Delta T$ (top) and constant $\delta T$ (bottom) for Case I. For temperature profiles of Case II, see Figure \ref{fig:temp_profiles_case2}.}
\label{fig:temp_profiles}
\end{figure}
The temperature profiles obtained using the two approaches for Case I are shown in Figure \ref{fig:temp_profiles}. The temperatures are obtained from the recorded atomic velocities using a fixed number of spatial bins across varying domain lengths. As a result, smaller domain lengths yield seemingly less smooth temperature profiles in Figure \ref{fig:temp_profiles}. This does not affect the calculated outcomes, since the full temperature profile is used in Equations \ref{eq:T_Cu} and \ref{eq:T_Cu2}. Trends that are consistent between the two temperature-enforcement strategies appear to be more reliable and show the robustness of the ITC in NEMD to boundary artefacts.

\section{Results}\label{sec:resul}
\begin{figure}
        \centering
        \includegraphics[width=\linewidth]{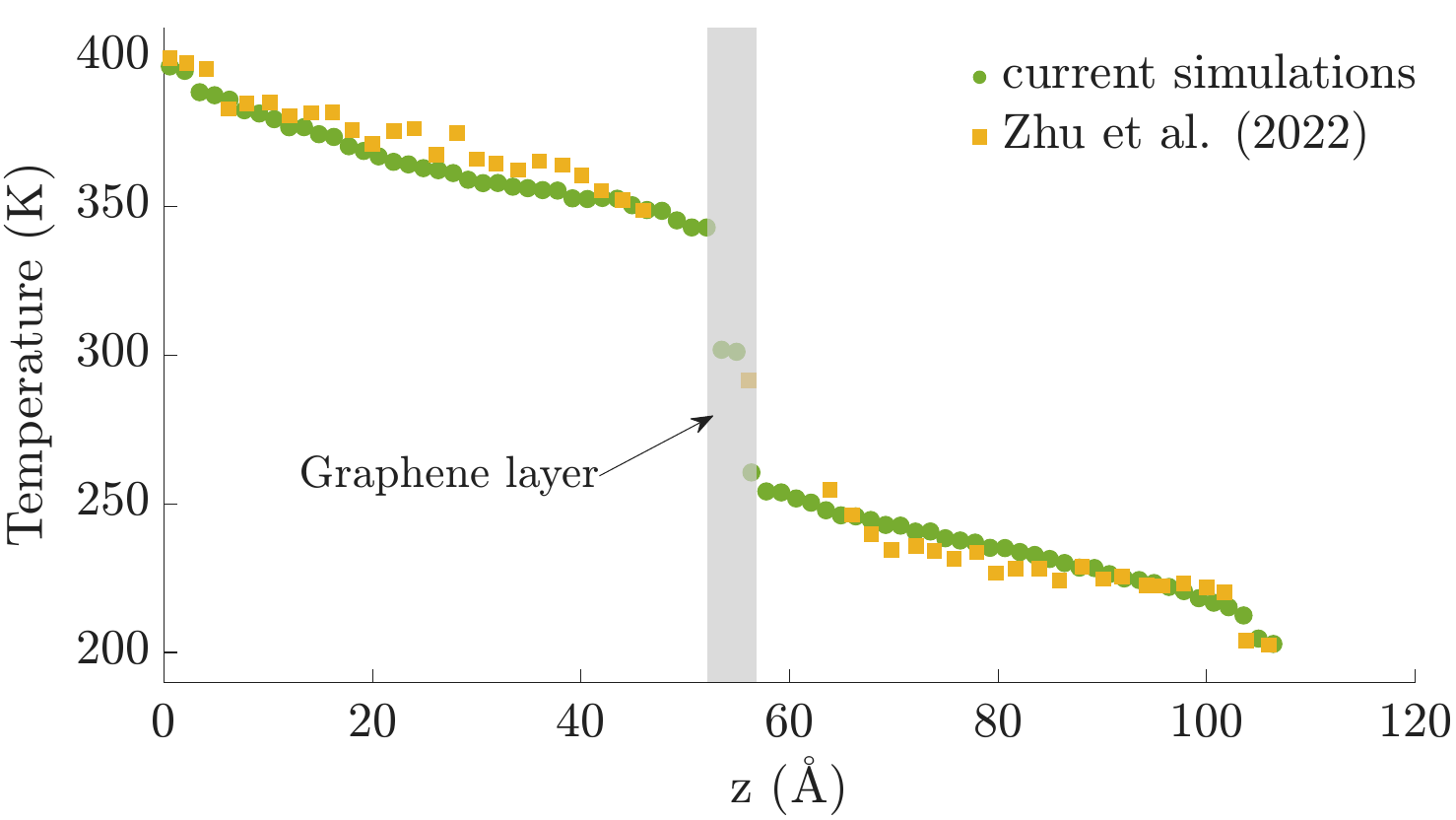}
        \caption{Temperature profile $\bar{T}(z_k)$ and linear fit $T(z)$ from the NEMD simulation of Case I at $(L_x,L_y,L_z)=(46.9, 46.9,114)$ \AA\, for which ITC$= 630\pm7$MW/m$^2$K is found, and compared to the temperature profile from Zhu et al. (2022) \cite{Zhu2022}, of reported ITC$= 640.2$MW/m$^2$K.}
        \label{fig:temp_prof_ref_zhu_comp} \end{figure}
In this section, the simulation results are discussed. First, the implementation of the computational framework is validated by reproducing a case reported in the literature in subsection \ref{subsec:Ref}. Then, in subsection \ref{subsec:domainlength}, the effect of the domain length on the ITC is discussed, showing that the fluctuations in ITC are mainly related to density and strain-induced variations, rather than other domain length relations, such as boundary artefacts. The observed difference in the ITC due to the initialisation procedures, Case I and Case II, is further studied in subsection \ref{subsec:InitChoice} by comparing local interfacial features of two simulations with comparable domain sizes and inspecting local interfacial features. Lastly, in subsection \ref{subsec:copperconductivity}, the copper conductivity retrieved from the temperature profiles on either side of the interface is discussed, which demonstrates domain-length dependence beyond strain-related effects, specifically through differences in the mean free path due to boundary scattering at different temperatures. 

\subsection{Interfacial Thermal Conductance: reference to literature}\label{subsec:Ref} 
As a validation step, we reproduced a graphene-copper MD system reported in the literature and obtained a comparable ITC. For the Case I initialisation with $(L_x,L_y,L_z)=(46.9,46.9,114)$\AA\, and boundary temperatures $T(0) = 400K$, $T(114)=200K$, corresponding closely to Zhu et al. (2022) \cite{Zhu2022}, we obtained an ITC of $630\pm 7$ [MW/m$^2$K] compared with the reported value of $640.2$ [MW/m$^2$K]. This small deviation is likely due to differences in the definition of the slabs for which temperatures are averaged and thermostatted regions, which do not fully overlap between the two setups (Figure \ref{fig:temp_prof_ref_zhu_comp}). The uncertainty reported here is obtained by standard propagation of the noise in the averaging and fitting of the temperature and flux for this specific simulation setup and does not include uncertainties in the setup's definition; please see section \ref{sec:uncertainties} for details. For comparison, Fang et al. reported an ITC of $664.5$  [MW/m$^2$K] for a similar domain with lattice coefficients set to experimental values \cite{Fang2025}. These comparisons show that the present framework convincingly reproduces the literature ITC values for graphene-copper interfaces.

\subsection{Interfacial Thermal Conductance: Domain length}\label{subsec:domainlength}
 \begin{figure*}
        \centering \includegraphics[width=0.8\linewidth]{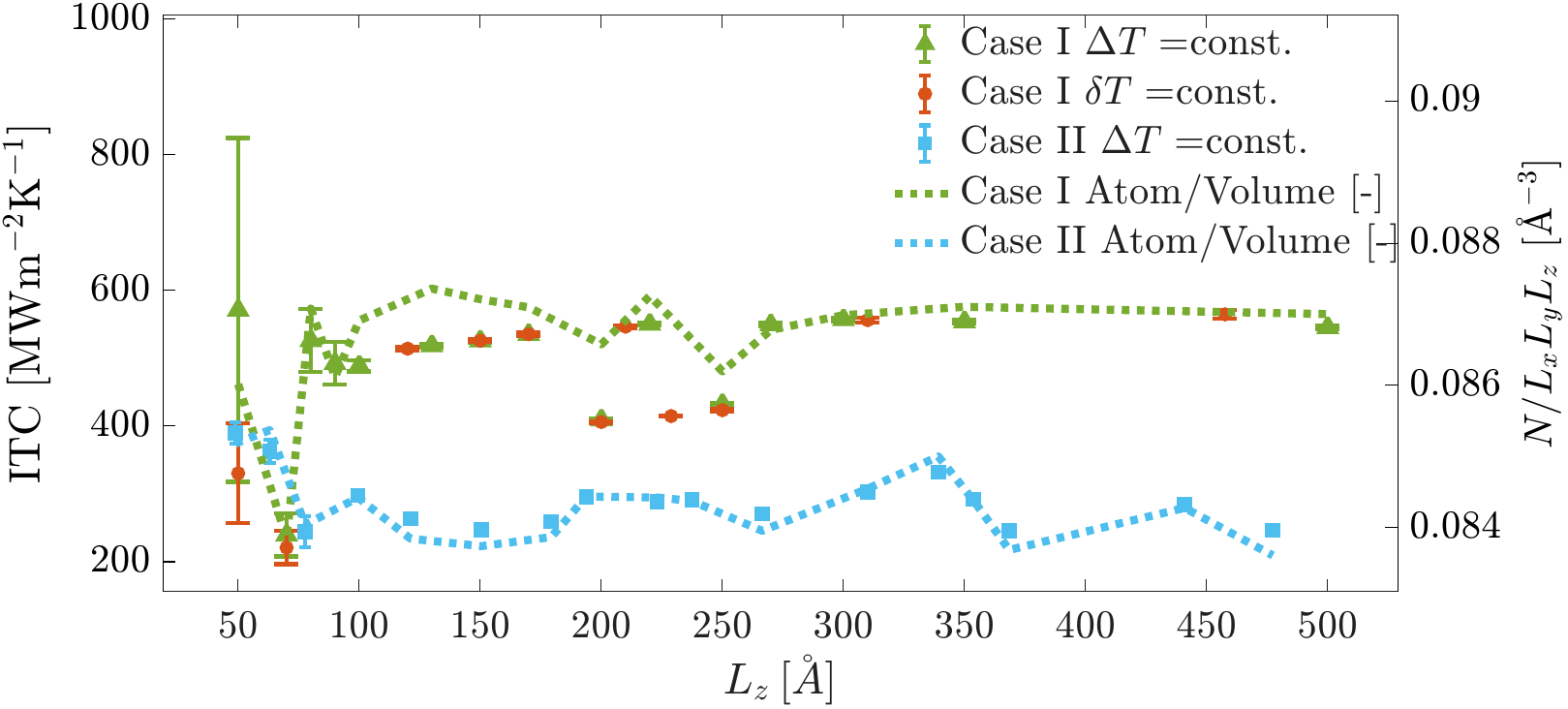}
        \caption{Interfacial thermal conductance (ITC) as a function of domain length ($L_z$) for Cases I and II with a constant temperature difference over the domain length $\Delta T(0,L) = T(z=0)-T(z=L)$ (green triangle, blue square) and with constant temperature gradient across the interface $\delta T(L/2) = T^+(L/2)-T^-(L/2)$ (red circle). Right axis for the number density (the number of atoms per unit volume), of Case I (dashed green) and Case II (dashed blue).}
        \label{fig:strain_conductance_length} \end{figure*}

This subsection examines how the ITC of the graphene-copper interface varies with domain length and how this relates to atomic density. We first describe the overall behavior of the ITC as a function of domain length in both Cases I and II, as well as the two temperature definitions, and then discuss each case separately. We argue that the atomic density is a measure of the residual strain in the system and is the main source of variations in the ITC.\\
\\
For both Cases I and II, Figure \ref{fig:strain_conductance_length} shows that the ITC follows the atomic density across the full range of domain length considered here, indicating that it is the primary parameter impacting the ITC of the graphene-copper interface. A higher density is an indicator of a higher strain, which corresponds within this study, as observed in Figure \ref{fig:strain_conductance_length}, to a higher ITC. That density is an indicator of residual strain in this study is most apparent in the difference between the Case I and Case II configurations; Case II corresponds to a lower density, a lower ITC and is in Figure \ref{fig:initialisse_strained} visibly perturbed and wrinkled graphene layer as compared to the higher density, strained Case I. We argue that the atomic density differences across different domain lengths in each case similarly indicate changes in the system's stiffness and strain, suggesting that strain is the dominant mechanism governing thermal interface conductance, consistent with the observed ITC values. \\
\\
The ITC values obtained through the two temperature definitions ($\Delta T$, $\delta T$) in Case I overlap within the error bars for the same domain length, indicating that the ITC is not significantly affected by temperature-induced changes in phonon mode populations at the domain boundaries. This is a striking result, which indicates that the ITC is insensitive to the changes in boundary dynamics in the $\delta T$ setting, unlike the bulk vibrations and resulting conductivity within the copper domains, as discussed later in section \ref{subsec:copperconductivity}. The apparent robustness of NEMD for ITC under variations in the boundary conditions and thermostat settings warrants further investigation. Let us discuss the response of the ITC for Cases I and II in more detail separately:
\begin{enumerate}
    \item In Case I, we observe large fluctuations in the ITC for domain lengths below $100$\AA\,, which decrease in size as the domain length increases, and become negligible above $250$\AA. This behaviour is consistent with effective lattice constants obtained from Case I initialisation. Since Case I initialises the lattice with a constant parameter, variations in density arise mainly from a mismatch in the lattice periodicity and the domain size. The mismatch induces a difference in overall density and relative strain spread throughout the entire lattice. Its effect decreases uniformly with domain size. We hypothesise that this is the primary phenomenon driving the small-domain fluctuations in the ITC for Case I, since the variations in ITC follow the density trend. Beyond this effect, no systematic dependence of the ITC on domain length is observed in Case I. For domain lengths above $250$\AA\, the average ITC is $552$ MW/m$^2$K. Consistent with this interpretation, Figure \ref{fig:VDOS_small_vs_big}, shows no clear shift in the phonon spectra for the domain length in Case I.
    \item Case II demonstrates a different fluctuation pattern. The largest deviations of the ITC occur at domain lengths below $70$\AA\,, whereas the remaining domain lengths exhibit much smaller ITC fluctuations than in Case I, despite density fluctuations of similar magnitude. This may indicate a superlinear relation between density and ITC. Unlike in Case I, the fluctuations do not dampen with increasing domain length, since the initialisation procedure of Case II does not exhibit the modulo-periodicity artefact of Case I. Instead, all domains are equilibrated and differences in density may be due to deviations in the equilibrated domain at the end of the equilibration run, as illustrated in Figure \ref{fig:relaxation300Kgraphene_cu__lattice}. For Case II, the average ITC is $289$ MW/m$^2$K, i.e. almost half of the ITC observed for Case I.
\end{enumerate}
Overall, the results suggest that the dominant mechanism affecting the ITC is lattice initialisation via density and resulting strain, rather than domain length via boundary artefacts or mean-free-path arguments. Once these strain-related effects are accounted for, no intrinsic length dependence of the ITC is resolved within the investigated range, as can be inferred from Figure \ref{fig:strain_conductance_length}.

\subsection{Interfacial Thermal Conductance: initialisation choices}\label{subsec:InitChoice}
\begin{figure}
  \centering
  \begin{subfigure}[c]{\linewidth}
    \centering
    \includegraphics[width=0.9\linewidth]{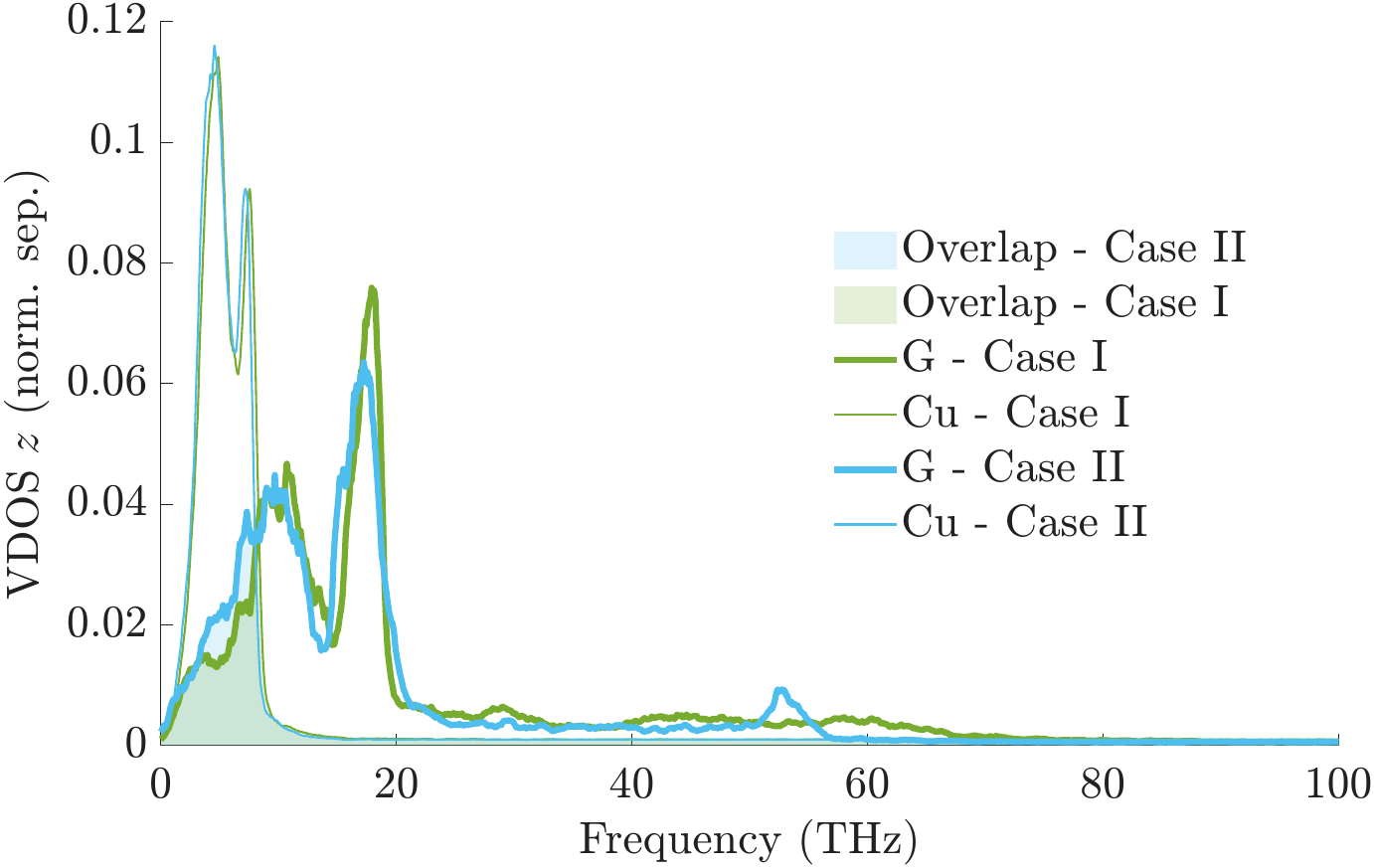}
  \end{subfigure}\hfill
  \begin{subfigure}[c]{\linewidth}
    \centering
\includegraphics[width=0.9\linewidth]{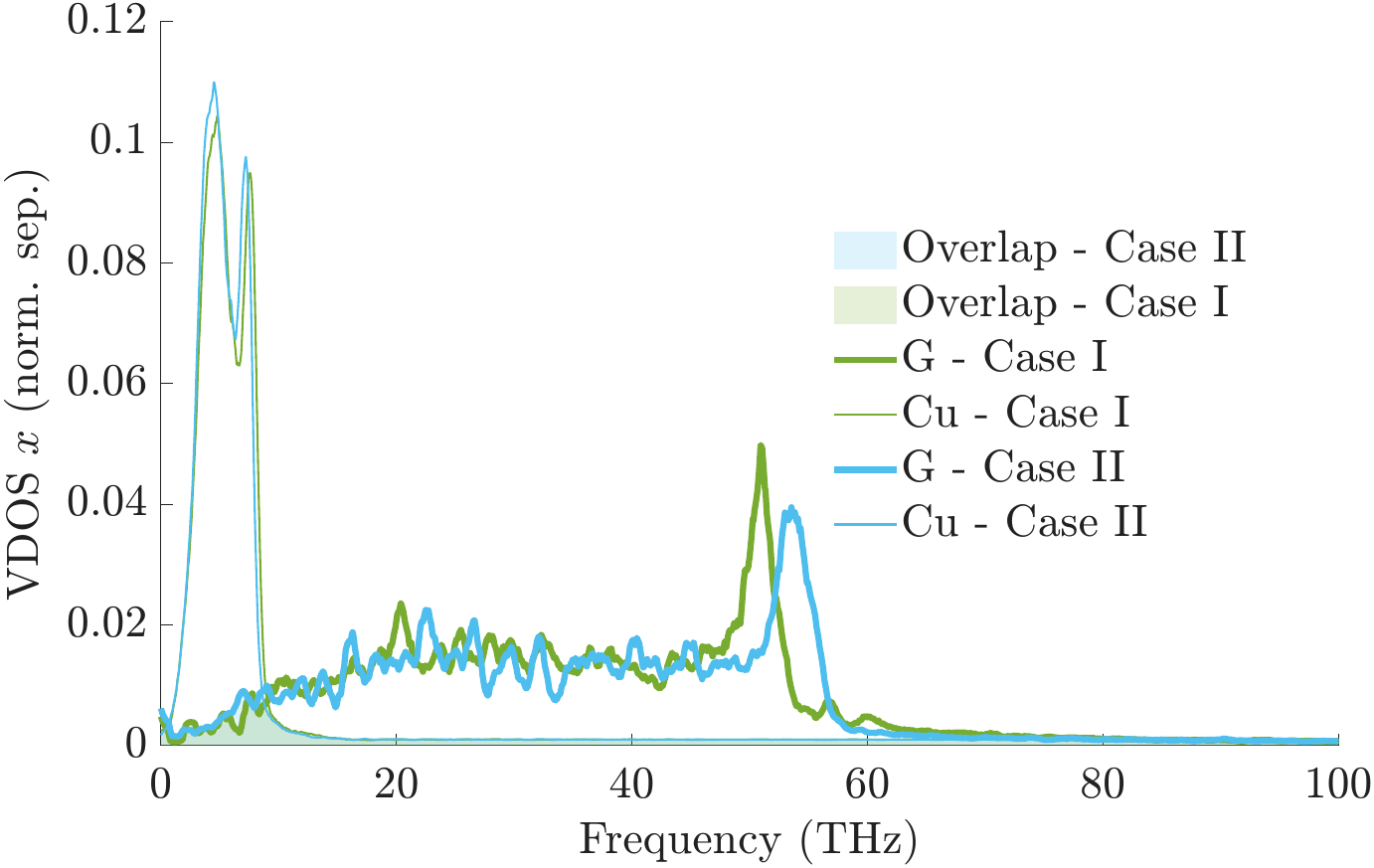}
  \end{subfigure}
  \caption{VDOS of $v_z$ (left) and $v_x$ (right) for Case I with $(L_x,L_y,L_z)=(50.0, 50.0, 300.0) $\AA\,,  ITC$=557.3\pm3$[MW/m$^2$K] and Case II with $(L_x,L_y,L_z)=(54.3,  57.3,  339.1)$\AA\,, ITC$=297.8 \pm 1$[MW/m$^2$K].}
\label{fig:VDOS_set_vs_equil}
\end{figure}
\begin{figure}
  \centering
  \begin{subfigure}[c]{0.55\linewidth}
    \centering
    \includegraphics[width=\linewidth]{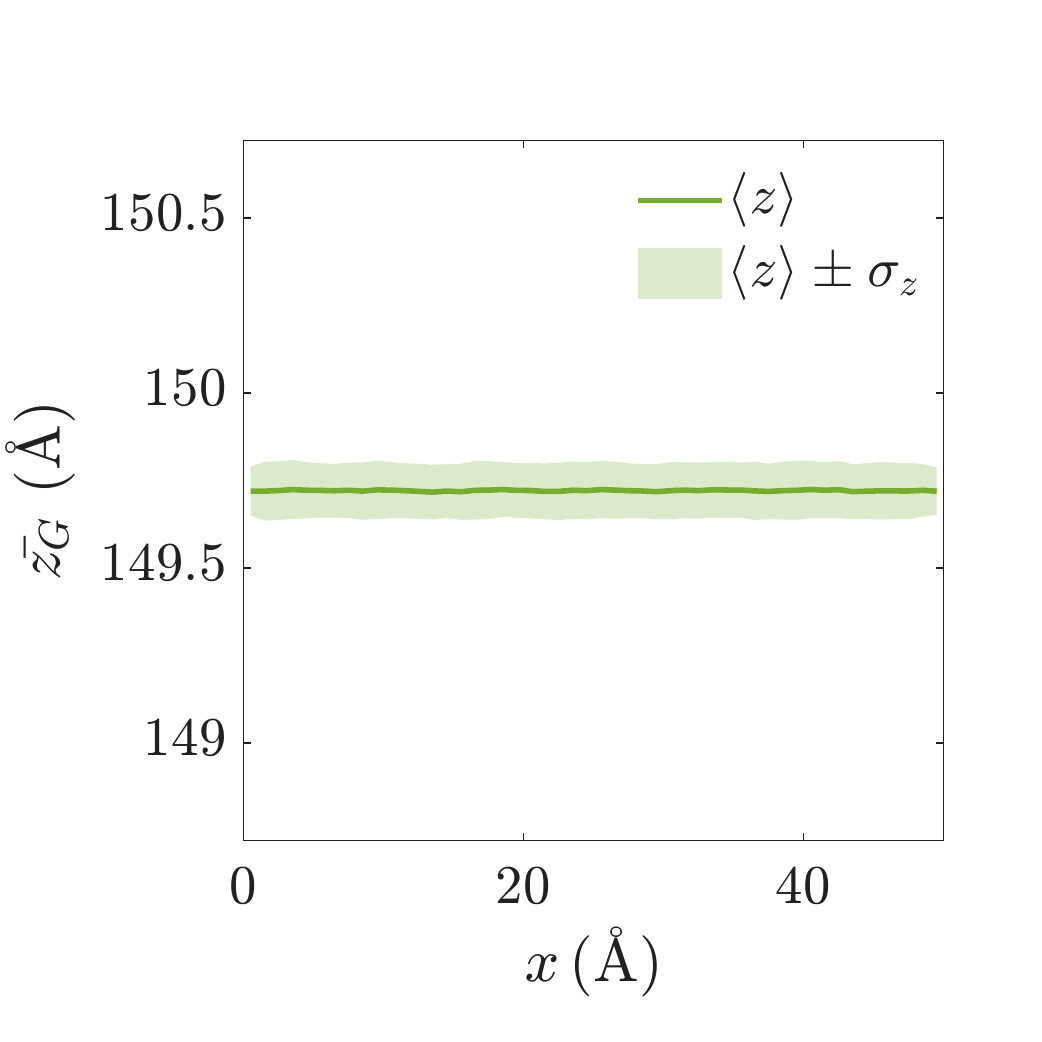}
  \end{subfigure}
  \begin{subfigure}[c]{0.55\linewidth}
    \centering
\includegraphics[width=1\linewidth]{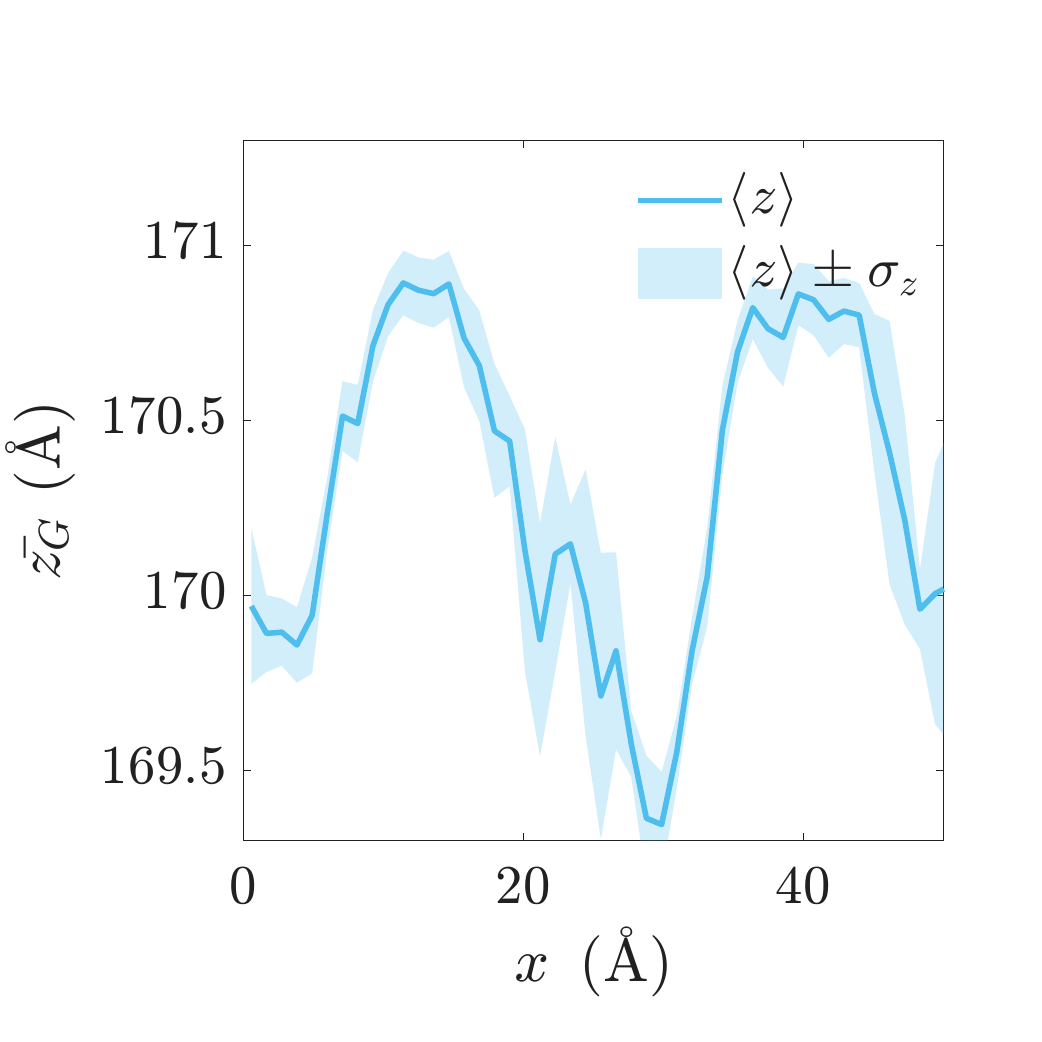}
  \end{subfigure}
  \begin{subfigure}[c]{0.44\linewidth}
    \includegraphics[width=1\linewidth]{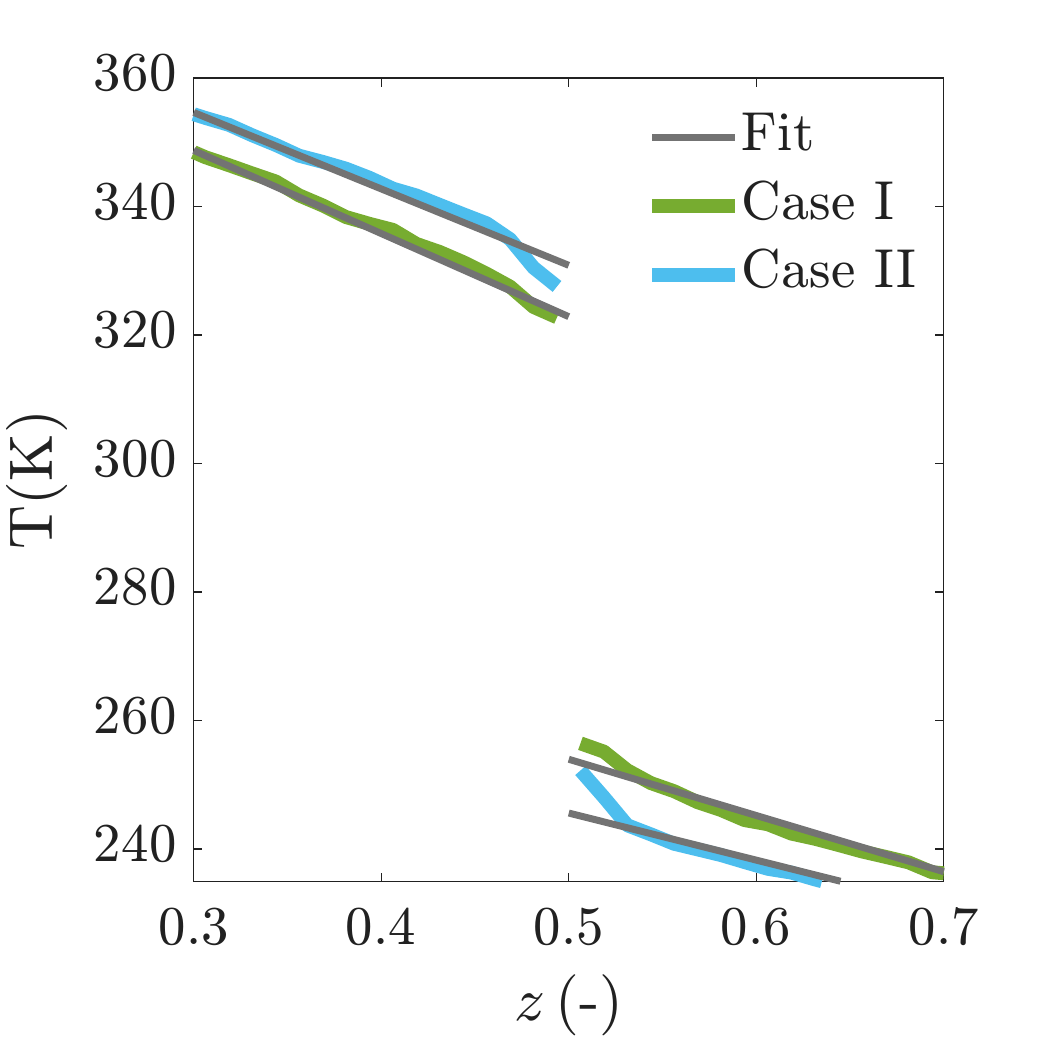}
  \end{subfigure}
  \caption{Average position along $z$ of the graphene atom including the standard deviation $\sigma_z$ during the production run $t\in [1000,2000]$ps for a local slab $y\in[20,23]$\AA\, plotted as a function of their position $x$, for Case I (left) with $(L_x,L_y,L_z)=(50.0, 50.0, 300.0) $\AA\,,  ITC$=557.3\pm3$[MW/m$^2$K] and Case II (middle) with $(L_x,L_y,L_z)=(54.3,  57.3,  339.1)$\AA\,, ITC$=297.8 \pm 1$[MW/m$^2$K]. Temperature from Cu atoms as a function of reduced $z$ coordinates (right) for Case I (green) and Case II (blue), along with the linear fits.}
\label{fig:average_Z}
\end{figure}
\begin{figure*}
  \centering
  \begin{subfigure}[c]{0.45\linewidth}
    \centering
    \includegraphics[width=\linewidth]{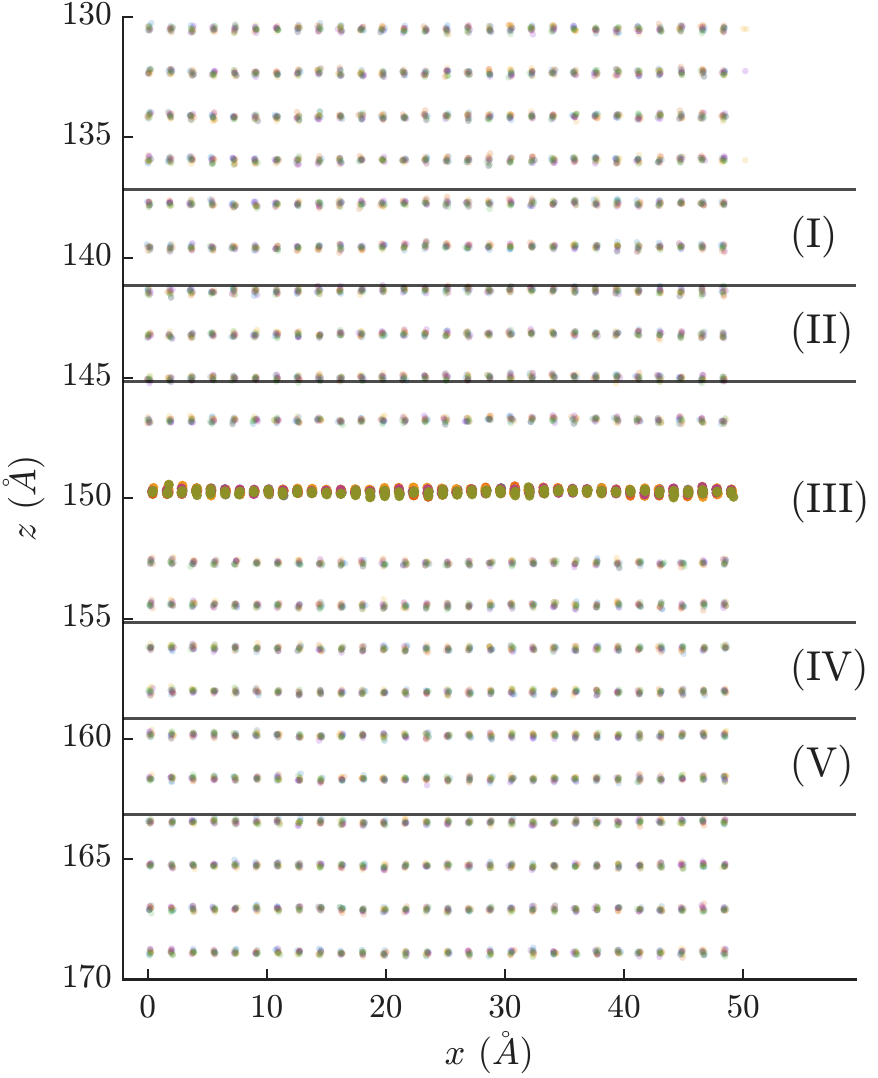}
  \end{subfigure}\hfill
  \begin{subfigure}[c]{0.45\linewidth}
    \centering
\includegraphics[width=1\linewidth]{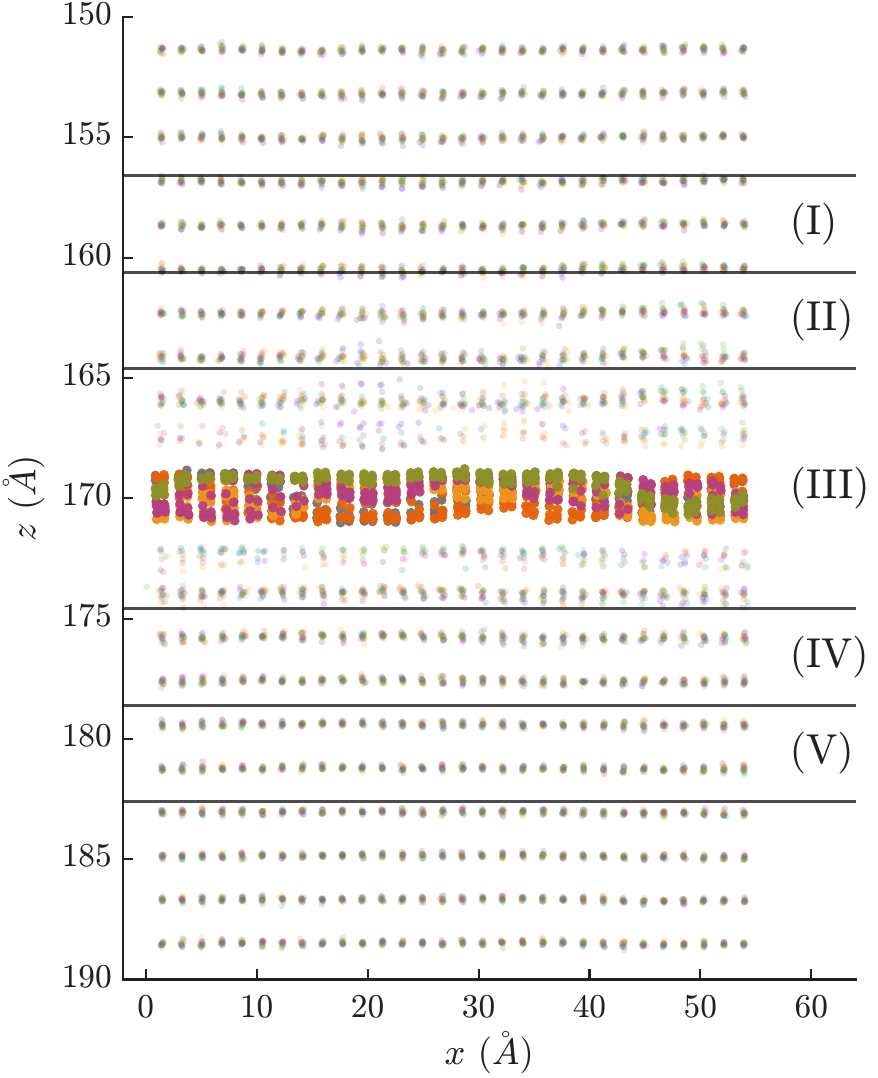}
  \end{subfigure}
  \caption{Interfacial regions for (I), (II), (III), (IV), (V), with widths of $(4, 4, 10, 4, 4)$\AA\, respectively, of Case I (left) with the average distance of the blocks $\langle d_{Cu-Cu}\rangle_{I}= 5.95$\AA\,, and Case II (right) with $\langle d_{Cu-Cu}\rangle_{II}= 6.05$\AA\,.}
\label{fig:VDOS_interf_regions}
  \centering
  \begin{subfigure}[c]{0.45\linewidth}
    \centering
    \includegraphics[width=\linewidth]{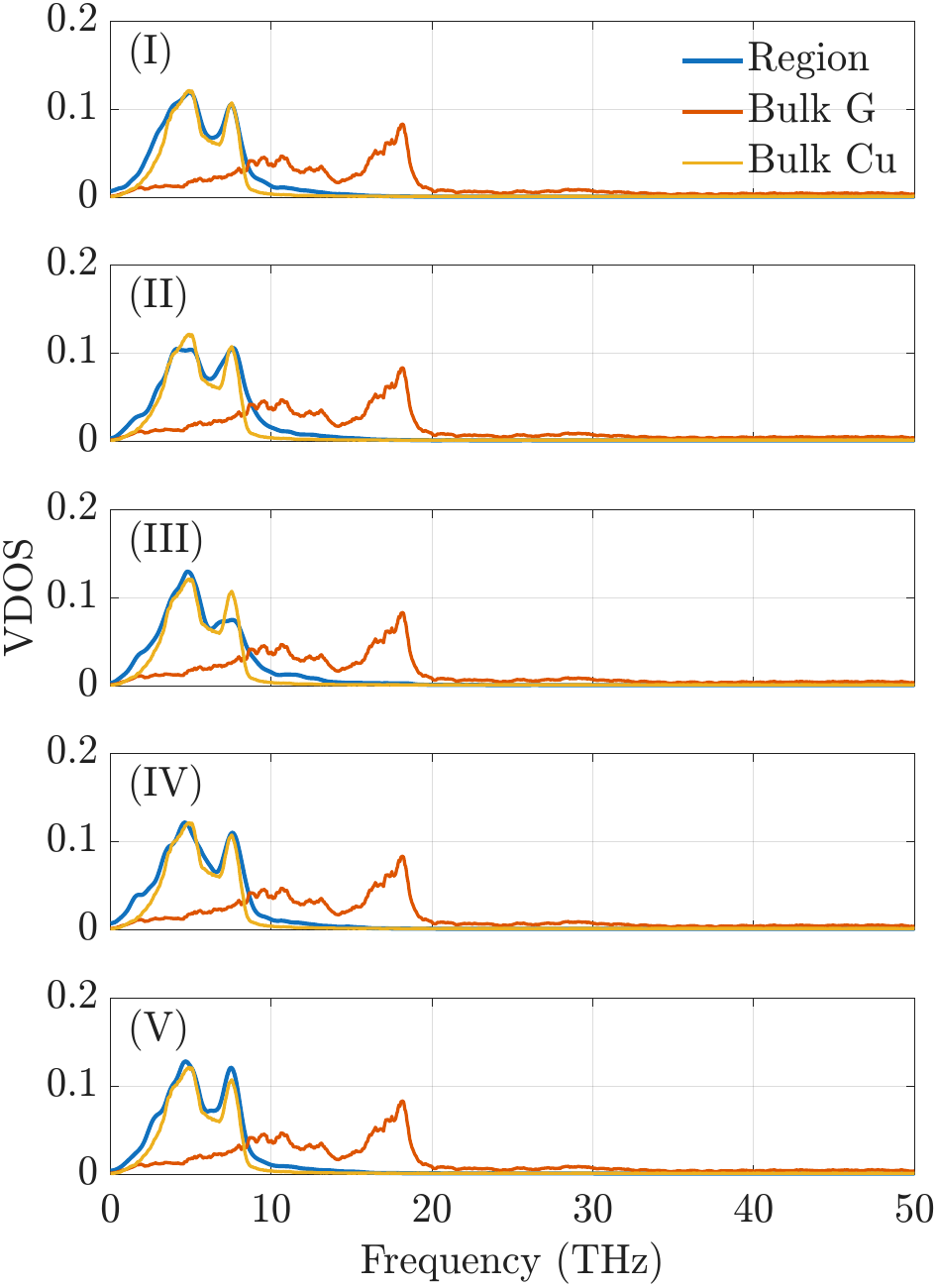}
  \end{subfigure}\hfill
  \begin{subfigure}[c]{0.45\linewidth}
    \centering
\includegraphics[width=1\linewidth]{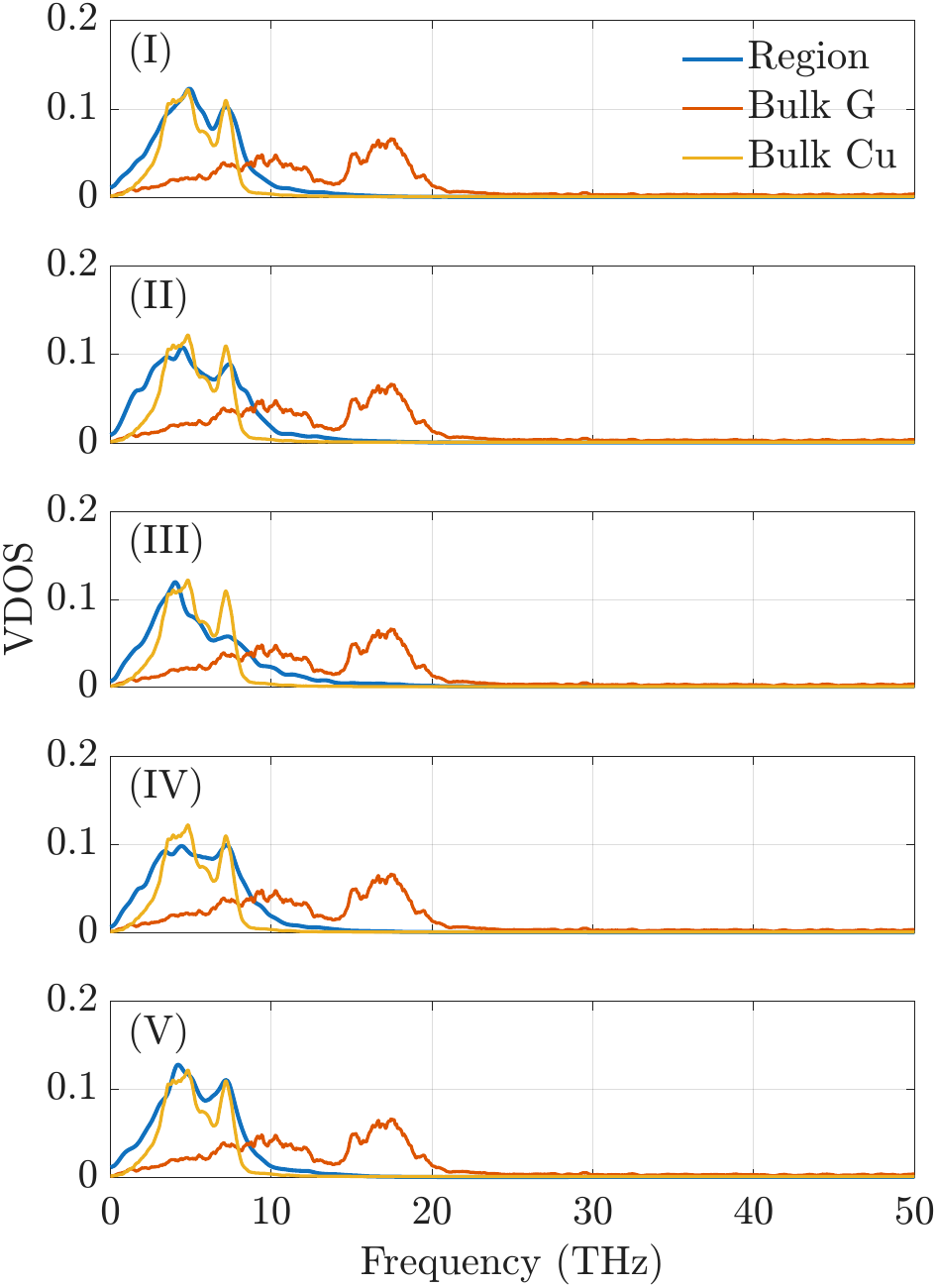}
  \end{subfigure}
  \caption{Bulk VDOS of copper and graphene atoms plotted with interfacial VDOS of copper atoms in regions (I), (II), (III), (IV), (V), of Case I (left), and Case II (right).}
\label{fig:VDOS_interfaces}
\end{figure*}

The two initialisation procedures, Case I and II, differ only by a few per cent in their lattice constants and mismatch strain, yet they produce nearly a twofold difference in interfacial thermal conductance (ITC), as shown in Figure \ref{fig:strain_conductance_length}. To identify the origin of this difference, we compare two systems of similar size: Case I with $(L_x,L_y,L_z)=(50.0, 50.0, 300.0) $\AA\, and Case II with $(L_x,L_y,L_z)=(54.3,  57.3,  339.1)$\AA\,.  The corresponding ITC values are $=557.3\pm3$[MW/m$^2$K]  and $297.8 \pm 1$[MW/m$^2$K], respectively. Despite the modest difference in effective lattice parameters, the interfacial dynamics differ significantly. Visualisation of the lattice in Figure \ref{fig:initialisse_strained} indicates a strain difference in the graphene sheet, corresponding to either a higher-strain, stretched configuration or a lower-strain, wrinkled configuration. We argue that the lower residual strain in Case II allows the formation of a damping interfacial region, which is absent in Case I. This interpretation is supported by the phonon spectra, the copper temperature profiles, and the geometry of the graphene layer, as will be discussed in this subsection.\\
\\
Bulk phonon spectra alone fail to explain the ITC difference. Figure \ref{fig:VDOS_set_vs_equil} shows that relative to Case II, the graphene spectrum in Case I is shifted to higher frequencies in the $xy$ plane and toward lower frequencies in the $z$ plane, consistent with strain-induced spectral shifts reported by Li et al. for embedded graphene in silicon \cite{Li2022}. The shifts are consistent across sufficiently long domains ($L_z>250$~\AA) and are not found in length variations for the same initialisation visualised in Figure \ref{fig:VDOS_small_vs_big}. These shifts slightly modify the normalised graphene-copper spectrum overlap in the $z$-direction, from 0.23 in Case I to 0.25 in Case II. Under Acoustic Mismatch and Diffusive Mismatch reasoning, this would suggest improved interfacial transport in Case II \cite{Chen2022}. Instead, the ITC is substantially lower, indicating that bulk phonon mode overlap alone is not sufficient to explain the thermal transport across this interface.\\
\\
Leading differences are not captured by bulk behaviour, but appear in the interfacial dynamics. The existence of localised interfacial phonon modes has been demonstrated through calculations and simulations and experimentally confirmed for the Si/Ge interface \cite{Cheng2021}. As shown in Figure \ref{fig:VDOS_interfaces}, the local VDOS for interfacial regions I-V defined in Figure \ref{fig:VDOS_interf_regions} are visibly different for Case I and for Case II. In Case II, the regional copper spectra exhibit less pronounced peaks and a broader frequency distribution than the bulk spectrum in all regions I-V. Although the interfacial VDOS in Case I also shows a somewhat broader distribution of modes, this effect is much weaker. Moreover, in Case I, the regional spectra retain clearly distinguishable peaks corresponding to the bulk copper spectrum, and the bulk character is recovered again in regions I and V. This demonstrates a significant difference: the copper dynamics near the interface are more perturbed over a wider region in Case II than in Case I. The broader spectra are consistent with increased scattering and diffusive dampening. This interpretation is further supported by the copper temperature profiles in Figure \ref{fig:average_Z}, where Case II displays a stronger deviation from the linear bulk temperature profile, again indicating that the interfacial region is more strongly perturbed.\\
\\
\begin{figure}
    \centering
    \includegraphics[width=\linewidth]{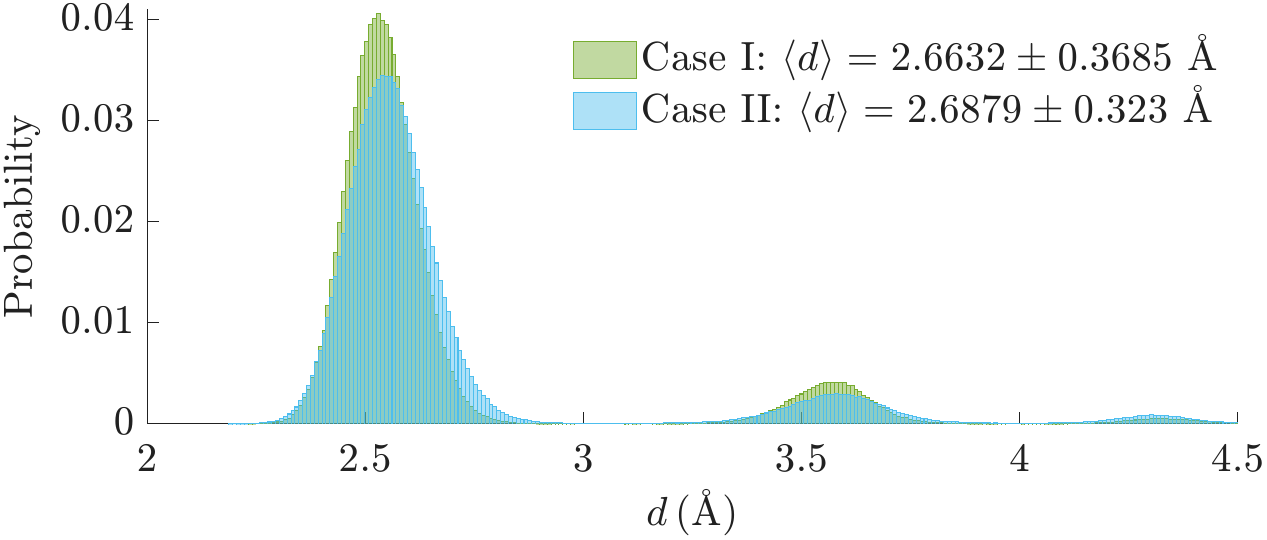}
    \caption{Nearest Neighbour distance Cu-Cu atoms within the center slab $(135<z<185)$ \AA, for Case I (green) and Case II (blue).}
    \label{fig:nn_copper}
\end{figure}

The differences in the interfacial conductance and spectra can be related directly to the geometrical differences in the graphene layer as visualised in Figure \ref{fig:relaxation300Kgraphene_cu__lattice}. In the equilibrated Case II system, the graphene layer is visibly wrinkled, related to higher surface roughness, whereas in Case I it remains essentially flat. In the wrinkled Case II structure, the separation distance over the whole interface is on average larger ($\langle d_{Cu-Cu}\rangle_{II}= 6.05$\AA, $\langle d_{Cu-Cu}\rangle_{I}= 5.95$\AA). Figure \ref{fig:average_Z}, which shows the average position of the atoms for a narrow slab in $y$ during a production run, confirms that the wrinkled morphology in Case II is stable throughout the simulation. The local standard deviation of the graphene position is comparable in both cases, indicating a similar magnitude of the motion of the graphene layer. The reduced ITC in Case II can therefore not be attributed to a change in coherent dynamics in the graphene layer. Consistent with this interpretation, the Cu nearest-neighbour distance distribution in Figure \ref{fig:nn_copper} is broader in Case II, indicating lower crystallinity and a more amorphous local structure, which enhances scattering and diffusion and may act as a damping layer.\\
\\
In conclusion, these results indicate that the lower-strain, wrinkled graphene configuration in Case II contains a structurally and dynamically perturbed boundary layer near the interface. Although the bulk graphene and copper spectra exhibit slightly greater overlap in Case II, the local interfacial dynamics is dominated by increased disorder, spectral broadening, and damping, all of which suppress heat transfer and together reduce the ITC by approximately a factor of two. The ITC is therefore controlled more strongly by the detailed interfacial structure and local scattering processes than by bulk phonon overlap alone.

\subsection{Thermal conductivity: Copper}\label{subsec:copperconductivity}
\begin{figure}
    \centering
    \includegraphics[width=0.8\linewidth]{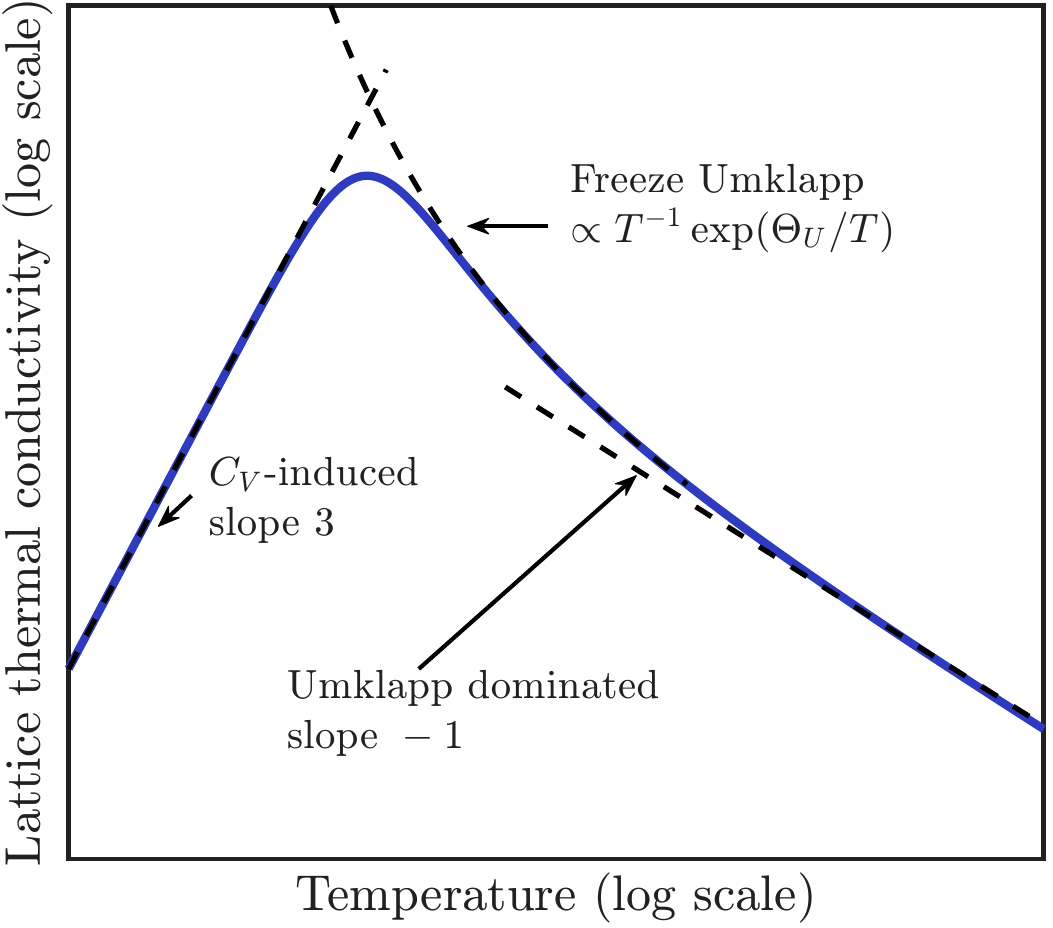}
    \caption{Typical temperature dependence of the lattice thermal conductivity \cite{Boer2023}\cite{ashcroft_mermin_ssp}}
    \label{fig:cond_temp_sketch}
\end{figure}

In this subsection, we examine the domain length and initialisation effects on the thermal conductivity of the copper domains adjacent to the graphene interface, obtained from the NEMD simulations, and explain trends consistent with the established theory of lattice thermal conductivity. The lattice thermal conductivity is proportional to the specific heat $C_V$, the phonon velocity $v_s$, and the phonon mean free path $\lambda$ as  \cite{Boer2023}
\begin{equation}\label{eq:kappa_propto}
    \kappa \propto C_V v_s\lambda
\end{equation}
Therefore, the domain length, which impacts the phonon mean free path through boundary scattering, affects the copper conductivities. Furthermore, the simulations for which the boundary temperatures are altered with domain size ($\delta T$ constant) demonstrate temperature dependence of the lattice conductivity. The thermal conductivity values discussed in this subsection represent only the lattice contribution to heat transport in copper. This distinction is important because thermal transport in copper is dominated by electrons rather than phonons. The lattice conductivity values obtained here are nevertheless consistent with values reported in MD studies of copper \cite{wei2025,Saether2022}.\\ 
\\
\begin{figure}
    \centering
\begin{subfigure}[c]{\linewidth}
    \centering
    \includegraphics[width=\linewidth]{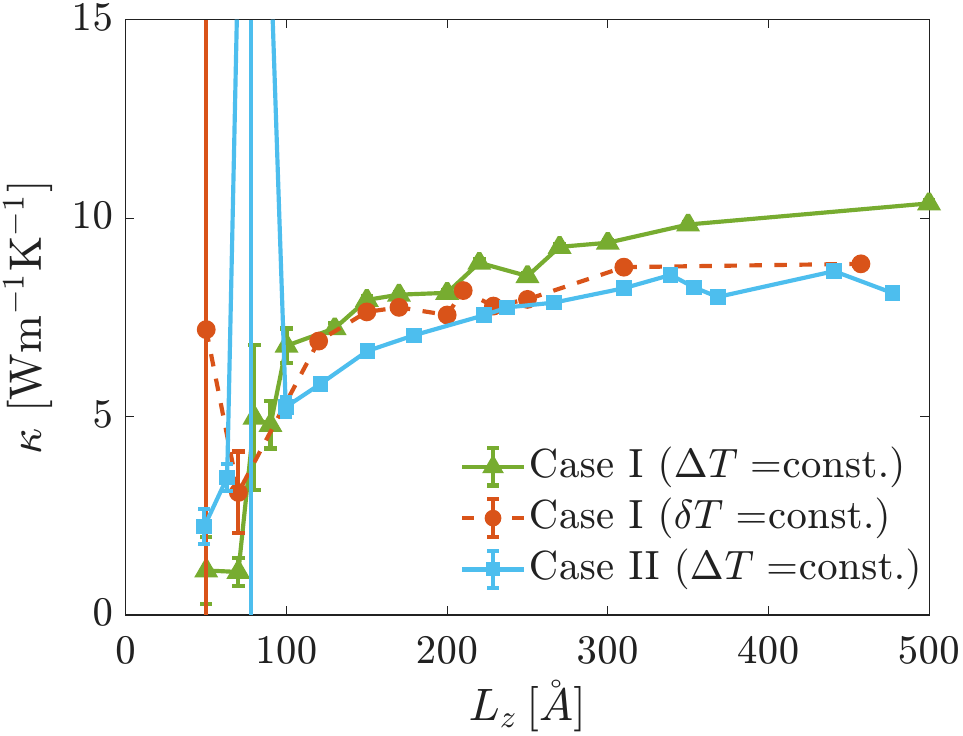}
  \end{subfigure}\hfill
  \begin{subfigure}[c]{\linewidth}
    \centering
\includegraphics[width=1\linewidth]{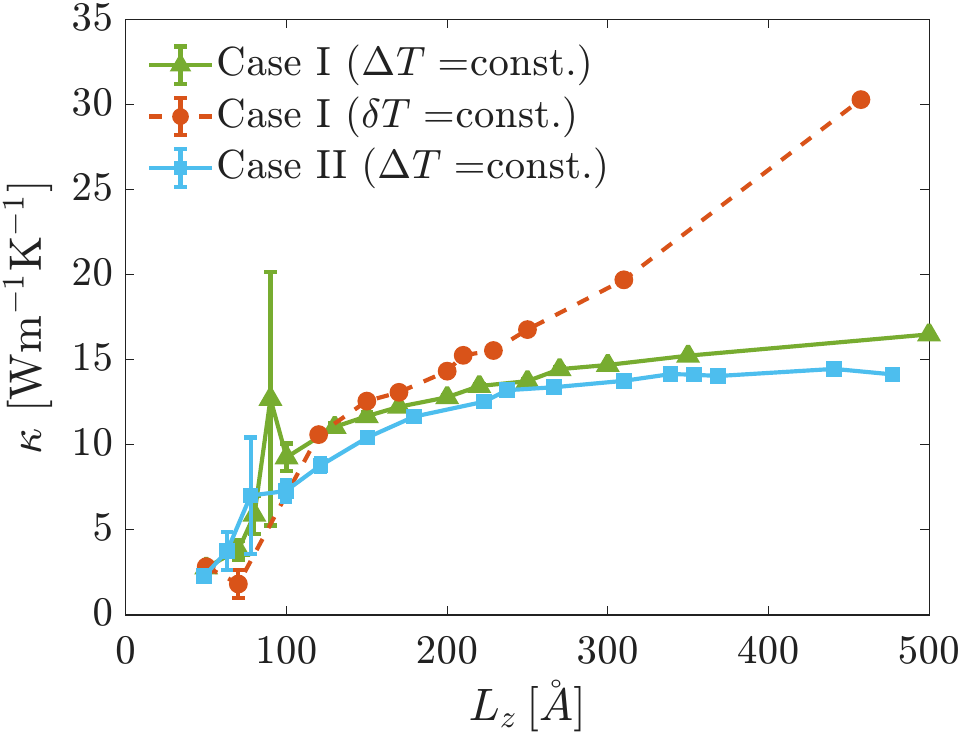}
  \end{subfigure}
    \caption{Conductivity $\kappa$ of the copper domains retrieved from NEMD simulations of the graphene-copper composites initialised through Case I, Case II, as a function of domain length $L_z$, where the temperatures at the boundaries are either kept constant $\Delta T=$constant, or extrapolated to keep the temperature gradient across the interface approximately constant $\delta T$=constant. For each composite NEMD simulation, two conductivities are plotted corresponding to the separate fits of the retrieved temperature profile on the hot side (left) and on the cold side of the interface (right).}
    \label{fig:Copper_kappa}
\end{figure}
Copper conductivities derived from the NEMD simulations for the domains on either side of the interface are plotted as a function of domain length in Figure \ref{fig:Copper_kappa}. Across all simulation settings, a pronounced increase in conductivity with domain length, $L_z$, is observed for domain lengths smaller than $200$ \AA, consistent with boundary-limited transport. Since the copper domains have a length of approximately half the domain size, and an estimated mean free path of $10$nm, the trend is attributed to a physical cutoff of the mean free path within each copper domain. In addition to the geometric restriction of the mean free path by the domain size, boundary scattering at these lengths may further reduce $\lambda$, and thereby suppress $\kappa$.\\
\\
For domains with a length larger than $200$\AA\, the simulations with constant boundary temperatures, ($\Delta T=$const.), show a weaker increase in conductivity with domain length, although a slight increase remains for both Cases I and II. This is consistent with reports of length-dependent conductivity approaching a plateau value asymptotically as the domain length increases, resolved for domain lengths up to 60 nm \cite{Saether2022}. Furthermore, it can be noted in Figure \ref{fig:VDOS_small_vs_big}, that the VDOS of different domain lengths with comparable conductivities of Case I are the same in terms of peak location and copper spectrum. \\
\\
Case I simulations with constant boundary temperatures show systematically higher conductivity than Case II. A higher lattice strain in Case I is consistent with higher lattice frequencies observed in the spectrum shift of copper of Case I in Figure \ref{fig:VDOS_set_vs_equil}. Strain-induced stiffening of dynamics is consistent with increased phonon velocities, which may contribute to a higher lattice thermal conductivity at the same temperatures as $\kappa\propto v_s$. Alternatively, the presence of a dampening region could suppress the higher frequencies in Case II, leading to the observed shift. In the same simulations, the conductivity is higher on the cold side than on the hot side of the interface. Since the boundary temperatures are $200$K and $400$K, near and above the Debye temperature of copper ($\Theta_D = 315$K), this trend can be attributed to stronger Umklapp scattering at higher temperature, reducing the mean free path \cite{ashcroft_mermin_ssp}.\\
\\
For simulations with constant temperature gradient, $(\nabla T_{Cu}=$const. for $\delta T=$const.), the interpretation of the conductivity trends is more subtle since the boundary temperatures vary with domain size. The typical temperature-dependent lattice thermal conductivity is sketched in Figure \ref{fig:cond_temp_sketch}. At the cold side, the temperature decreases from $226$ to $66$ K, well below the Debye temperature. Within this low-temperature regime, Umklapp scattering is increasingly suppressed by frozen states until scattering is governed by constant temperature-independent defect and boundary scattering \cite{Boer2023}. At these low temperatures, the $C_V$ decreases following quantum-mechanical reductions of phonon occupation. While phonon occupation is well resolved in the high-temperature limit by MD, in the low-temperature limit, MD overestimates the higher-frequency modes, following equipartition rather than Bose-Einstein statistics. Therefore, the low temperature decrease in conductivity is not observed in Figure \ref{fig:Copper_kappa}, and only the increase in conductivity due to scattering decrease is captured. On the hot side of the simulations, the boundary temperature increases from 374 to 535 K, well above the Debye temperature. In this temperature range, the Umklapp scattering is expected to cause a linear decay of the conductivity. However, we do not observe the expected decrease in conductivity at $T_{hot}$, but rather a faster approach to an asymptotic value as compared to the ($\Delta T=$constant) simulations, suggesting a competition between reduced boundary scattering at larger domain size and enhanced phonon scattering at higher temperature.\\
\\
Overall, the trends in copper thermal lattice conductivity are governed by the combined effects of mean free path and temperature. The pronounced increase in conductivity for small domains is consistent with boundary-limited transport, whereas the weaker increase for larger domain sizes is consistent with an asymptotic approach to bulk dynamics. The higher conductivity in Case I is plausibly related to the strain-induced stiffening of the lattice dynamics. The temperature dependence observed between the hot and cold sides is also consistent with phonon-scattering theory, although the low-temperature regime of the lattice conductivity is not resolved in these classical MD simulations, leading to a non-physical increase at low temperatures.

\section{Concluding Remarks}\label{sec:concl}
A study of the effects of initialisation choices and finite domain lengths on the interfacial thermal conductance at the copper-graphene interface has been conducted using molecular-dynamics simulations. Previous literature outcomes have been reproduced, and no trend in the ITC as a function of domain length or temperature definition has become apparent, suggesting that the ITC is robust to boundary artefacts. The only domain length dependence observed was due to a relation between the ITC and the density (number of atoms per unit volume), which is particularly pronounced for small domains due to initialisation choices. This dependency is related to the most striking effect observed in this study, namely, the differences in lattice constants due to initialisation choices. Two initialisation approaches were used: one initialised the lattice with lattice parameters consistent to experimentally measured properties (Case I), and the other initialised with minimal mismatch and residual strain (Case II). The difference in the resulting lattices, of a few per cent in the effective lattice parameters of the graphene and copper lattices, $(a_{Cu},a_G)$, results in an ITC difference of almost a factor of two. The lattice with higher residual strain showed a shift in the graphene spectrum to higher and lower frequencies in the in- and out-of-plane directions, respectively, relative to the equilibrated lattice spectrum, consistent with the literature. However, the change in overlap due to this shift does not correspond with the ITC difference. Instead, we argue that the additional interfacial resistance in the equilibrated case is due to wrinkling of the graphene layer and the emergence of a boundary layer of neighbouring Cu atoms that acts as a damping layer. This argument is supported by the broader phonon spectrum observed in the copper regions near the interface, which corresponds to increased scattering. In the copper-graphene system, the spectral overlap alone is insufficient to capture the interfacial heat-transfer dynamics. This study shows that although properties of NEMD simulations can be computed with small error estimates, the outcomes are sensitive to the domain configuration. We showed that the domain length and boundary definition in the current setting do not influence the ITC, although they did influence thermal conductivity in the copper domains. Therefore, the results suggest some robustness of the ITC to boundary artefacts. However, the ITC does not prove robust to small changes in lattice parameters due to initialisation procedures. Therefore, careful consideration of system choices is necessary to reliably draw conclusions from NEMD simulations. \\ 
\\
 In this study, the interface area was kept constant, which may be related to the graphene phonon mean free paths and warrants further investigation. Similarly, the thermostat settings were kept constant, which may highlight domain length effects and the propagation of boundary settings when varied. The interplay of other parameters with the domain length is not taken into account. A complete Uncertainty Quantification approach that simultaneously investigates all parameter dependencies to assess finite-size and domain artefacts fully is deferred to a subsequent study. Experimental confirmation of the strain-dependent interfacial thermal dynamics could be investigated through annealing techniques of manufactured composites.\\
 \\
Summarising, in order to ultimately control thermal conductivity in graphene-copper matrix composites for electronics cooling, we need a better understanding of interfacial dynamics, which can be achieved through robust molecular dynamics simulations. We demonstrated that the interfacial thermal conductance is robust to boundary temperature definitions considered in this study for sufficiently large domain lengths, but is highly sensitive to initialisation effects, including residual strain, which warrants further investigation.\\
\\
\textit{Acknowledgments.} This research has received funding from the EU grant agreement No 101046835 (Thermodust project) and is supported by an NWO grant 2024.0016.\\
\\
\textit{Data availability.}-All simulated data used in this study are available in
the 4TU.ResearchData database. 

\bibliographystyle{apsrev4-2}
\bibliography{ref_md}

@article{Stanley2023,
  author  = {Stanley, C. M.},
  title   = {Thermal boundary resistance: A review of molecular dynamics simulations and other computational methods},
  journal = {Phys. Status Solidi B},
  year    = {2023},
  volume  = {260},
  pages   = {2300095},
  doi     = {10.1002/pssb.202300095}
}

@article{wei2025,
    author = {Wei, Xiao-Ping and Li, Xue and Zhang, Ya-Ling},
    title = {Molecular dynamics simulations of the lattice thermal conductivity of Cu–Ni alloys},
    journal = {AIP Adv.},
    volume = {15},
    number = {6},
    pages = {065021},
    year = {2025},
    month = {06},
    issn = {2158-3226},
    doi = {10.1063/5.0259893},
}

@article{Zhu2022,
  author  = {Zhu, J. and Huang, S. and Xie, Z. and Guo, H. and Yang, H.},
  title   = {Thermal conductance of copper--graphene interface: A molecular simulation},
  journal = {Materials},
  year    = {2022},
  volume  = {15},
  number  = {21},
  pages   = {7588},
  doi     = {10.3390/ma15217588},
  note    = {PMID: 36363179; PMCID: PMC9654340}
}

@article{Landry2009,
  author  = {Landry, E. S. and McGaughey, A. J. H.},
  title   = {Thermal boundary resistance predictions from molecular dynamics simulations and theoretical calculations},
  journal = {Phys. Rev. B},
  year    = {2009},
  volume  = {80},
  pages   = {165304}
}

@article{Chen2022,
  author  = {Chen, J. and Xu, X. and Zhou, J. and Li, B.},
  title   = {Interfacial thermal resistance: Past, present, and future},
  journal = {Rev. Mod. Phys.},
  year    = {2022},
  volume  = {94},
  pages   = {025002},
  doi     = {10.1103/RevModPhys.94.025002}
}

@article{Pop2012,
  author  = {Pop, E. and Varshney, V. and Roy, A. K.},
  title   = {Thermal properties of graphene: Fundamentals and applications},
  journal = {MRS Bulletin},
  year    = {2012},
  volume  = {37},
  number  = {12},
  pages   = {1273--1281},
  doi     = {10.1557/mrs.2012.203}
}

@article{Saether2022,
  author  = {S{\ae}ther, S. and Erichsen, M. F. and Xiao, S. and Zhang, Z. and Lervik, A. and He, J.},
  title   = {Phonon thermal transport in copper: The effect of size, crystal orientation, and grain boundaries},
  journal = {AIP Adv.},
  year    = {2022},
  volume  = {12},
  pages   = {065301},
  doi     = {10.1063/5.0094170}
}

@article{Ghatage2020,
  author  = {Ghatage, D. and Tomar, G. and Shukla, R. K.},
  title   = {Thermostat-induced spurious interfacial resistance in non-equilibrium molecular dynamics simulations of solid--liquid and solid--solid systems},
  journal = {J. Chem. Phys.},
  year    = {2020},
  volume  = {153},
  number  = {16},
  pages   = {164110}
}

@article{Li2022,
  author  = {Li, Chao and Wang, Jie and Sheng, Yunhe and Yang, Lina and Su, Yu},
  title   = {The strain-dependent interfacial thermal resistance at graphene--silicon interface under various deformation conditions},
  journal = {Int. J. Heat Mass Transf.},
  year    = {2022},
  volume  = {198},
  pages   = {123383}
}

@article{Fang2025,
  author  = {Fang, Jiayuan and Zhang, Yang and Zhao, Pei},
  title   = {High-efficiency thermal transport in graphene-based composites via a copper interlayer},
  journal = {Cell. Rep. Phys. Sci.},
  year    = {2025},
  volume  = {6},
  number  = {11},
  doi     = {10.1016/j.xcrp.2025.102917},
  url     = {https://doi.org/10.1016/j.xcrp.2025.102917},
  issn    = {2666-3864},
  publisher = {Elsevier}
}

@article{Lee2025_RSER,
  author  = {Lee, Hyunjong and Ardeshiri Lordejani, Amir and van Goor, Leonore and Jurov, Andrea and Koutsioukis, Apostolos and Ruan, Siyuan and Santhosh, Neelakandan M. and Zarei, Fatemeh and Barreneche, Camila and Cvelbar, Uro{\v{s}} and Dosta, Sergi and Geurts, Bernard J. and Guagliano, Mario and Jafari, Davoud and Nicolosi, Valeria and Yin, Shuo and Zava{\v{s}}nik, Janez and Bagherifard, Sara and Lupoi, Rocco and Wits, Wessel W.},
  title   = {Review on properties, physics, and fabrication of two-dimensional material-based metal-matrix composites (2DMMCs) for heat transfer systems},
  journal = {Renew. Sustain. Energy Rev.},
  year    = {2025},
  volume  = {217},
  pages   = {115700},
  issn    = {1364-0321},
  doi     = {10.1016/j.rser.2025.115700}
}

@article{Dong2018,
  author  = {Dong, Haikuan and Fan, Zheyong and Shi, Libin and Harju, Ari and Ala-Nissila, Tapio},
  title   = {Equivalence of the equilibrium and the non-equilibrium molecular dynamics methods for thermal conductivity calculations: From bulk to nanowire silicon},
  journal = {Phys. Rev. B},
  year    = {2018},
  volume  = {97},
  pages   = {094305},
  doi     = {10.1103/PhysRevB.97.094305}
}

@article{Nejatolahi2021,
  author  = {Nejatolahi, M. and Golneshan, A. A. and Kamali, R. and others},
  title   = {Non-equilibrium versus equilibrium molecular dynamics for calculating the thermal conductivity of nanofluids},
  journal = {J. Therm. Anal. Calorim.},
  year    = {2021},
  volume  = {144},
  pages   = {1467--1481},
  doi     = {10.1007/s10973-020-09595-x}
}

@article{MuellerPlathe1997,
  author  = {M{\"u}ller-Plathe, F.},
  title   = {A simple non-equilibrium molecular dynamics method for calculating the thermal conductivity},
  journal = {J. Chem. Phys.},
  year    = {1997},
  volume  = {106},
  number  = {14},
  pages   = {6082--6085}
}

@article{Green1954,
  author  = {Green, M. S.},
  title   = {Markoff random processes and the statistical mechanics of time-dependent phenomena. {II}. Irreversible processes in fluids},
  journal = {J. Chem. Phys.},
  year    = {1954},
  volume  = {22},
  pages   = {398--413}
}

@article{Kubo1957,
  author  = {Kubo, R.},
  title   = {Statistical-Mechanical Theory of Irreversible Processes. {I}. General Theory and Simple Applications to Magnetic and Conduction Problems},
  journal = {J. Phys. Soc. Jpn.},
  year    = {1957},
  volume  = {12},
  pages   = {570--586}
}

@article{Khadem2013,
  author  = {Khadem, M. H. and Wemhoff, A. P.},
  title   = {Comparison of Green--Kubo and NEMD heat flux formulations for thermal conductivity prediction using the Tersoff potential},
  journal = {Comp. Mat. Sci.},
  year    = {2013},
  volume  = {69},
  pages   = {428--434}
}

@article{Matsubara2017,
  author  = {Matsubara, Hiroki and Kikugawa, Gota and Ishikiriyama, Mamoru and Yamashita, Seiji and Ohara, Taku},
  title   = {Equivalence of the EMD- and NEMD-based decomposition of thermal conductivity into microscopic building blocks},
  journal = {J. Chem. Phys.},
  year    = {2017},
  volume  = {147},
  number  = {11},
  pages   = {114104},
  doi     = {10.1063/1.4990593}
}

@article{Fugallo2014,
  author  = {Fugallo, G. and Cepellotti, A. and Paulatto, L. and Lazzeri, M. and Marzari, N. and Mauri, F.},
  title   = {Thermal conductivity of graphene and graphite: Collective excitations and mean free paths},
  journal = {Nano Lett.},
  year    = {2014},
  volume  = {14},
  number  = {11},
  pages   = {6109--6114}
}

@article{Xu2014,
  author  = {Xu, Xiangfan and Pereira, Luiz F. C. and Wang, Yu and Wu, Jing and Zhang, Kaiwen and Zhao, Xiangming and Bae, Sukang and Bui, Cong Tinh and Xie, Rongguo and Thong, John T. L. and Hong, Byung Hee and Loh, Kian Ping and Donadio, Davide and Li, Baowen and {\"O}zyilmaz, Barbaros},
  title   = {Length-dependent thermal conductivity in suspended single-layer graphene},
  journal = {Nature Comm.},
  year    = {2014},
  volume  = {5},
  number  = {1},
  pages   = {3689},
  doi     = {10.1038/ncomms4689},
  url     = {https://doi.org/10.1038/ncomms4689},
  issn    = {2041-1723},
}

@article{Klaver2015,
  author  = {Klaver, T. P. C. and Zhu, Shou-En and Sluiter, M. H. F. and Janssen, G. C. A. M.},
  title   = {Molecular dynamics simulation of graphene on {Cu} (100) and (111) surfaces},
  journal = {Carbon},
  year    = {2015},
  volume  = {82},
  pages   = {538--547}
}

@article{cai2010,
author = {Cai, Weiwei and Moore, Arden L. and Zhu, Yanwu and Li, Xuesong and Chen, Shanshan and Shi, Li and Ruoff, Rodney S.},
title = {Thermal Transport in Suspended and Supported Monolayer Graphene Grown by Chemical Vapor Deposition},
journal = {Nano Lett.},
volume = {10},
number = {5},
pages = {1645-1651},
year = {2010},
doi = {10.1021/nl9041966},
    note ={PMID: 20405895},

URL = { 
    
        https://doi.org/10.1021/nl9041966
    
    

},
}

@article{Mei2014,
  author  = {Mei, S. and Maurer, L. N. and Aksamija, Z. and Knezevic, I.},
  title   = {Full-dispersion Monte Carlo simulation of phonon transport in micron-sized graphene nanoribbons},
  journal = {J. Appl. Phys.},
  year    = {2014},
  volume  = {116},
  pages   = {164307},
  doi     = {10.1063/1.4899235}
}

@article{Lee2026_inprep,
  author = {Lee, Hyunjong and Ruan, Siyuan and Jafari, Davoud and Koutsioukis, Apostolos and Lupoi, Rocco and Yin, Shuo and {van Goor}, Leonore and Nicolosi, Valeria and Geurts, Bernard and Wits, Wessel},
  title = {Thermal Conductivity Characterization of Ball-Milled Copper-Graphene Composites: Insights from Microstructure-Informed Thermal Models},
  journal = {(Under review at) Mater. Des.},
  year   = {2026},
}

@book{Kubo1991_StatPhysII,
  author    = {Kubo, Ryogo and Toda, Morikazu and Hashitsume, Natsuki},
  title     = {Statistical Physics II},
  publisher = {Springer},
  address   = {Berlin, Heidelberg},
  year      = {1991},
}

@article{Kapitza1941,
  author  = {Kapitza, P. L.},
  title   = {To be completed},
  journal = {Zh. Eksp. Teor. Fiz.},
  year    = {1941},
  volume  = {11},
  pages   = {1},
  note    = {English transl.: J. Phys. U.S.S.R. 4, 181 (1941)}
}

@article{Onsager1931,
  author  = {Onsager, Lars},
  title   = {Reciprocal Relations in Irreversible Processes. {I}.},
  journal = {Phys. Rev.},
  year    = {1931},
  volume  = {37},
  pages   = {405--426},
  doi     = {10.1103/PhysRev.37.405}
}

@book{deGrootMazur1984,
  author    = {de Groot, Sybren R. and Mazur, Peter},
  title     = {Non-Equilibrium Thermodynamics},
  publisher = {Courier Corporation},
  year      = {1984}
}

@article{Gordiz2017,
  author  = {Gordiz, K. and Henry, A.},
  title   = {To be completed},
  journal = {J. Appl. Phys.},
  year    = {2017},
  volume  = {121},
  pages   = {025102}
}

@article{Thompson2022_LAMMPS,
  author  = {Thompson, A. P. and Aktulga, H. M. and Berger, R. and Bolintineanu, D. S. and Brown, W. M. and Crozier, P. S. and in 't Veld, P. J. and Kohlmeyer, A. and Moore, S. G. and Nguyen, T. D. and Shan, R. and Stevens, M. J. and Tranchida, J. and Trott, C. and Plimpton, S. J.},
  title   = {{LAMMPS}---a flexible simulation tool for particle-based materials modeling at the atomic, meso, and continuum scales},
  journal = {Comp. Phys. Comm.},
  year    = {2022},
  volume  = {271},
  pages   = {10817}
}

@article{Mishin2001,
  title = {Structural stability and lattice defects in copper: Ab initio, tight-binding, and embedded-atom calculations},
  author = {Mishin, Y. and Mehl, M. J. and Papaconstantopoulos, D. A. and Voter, A. F. and Kress, J. D.},
  journal = {Phys. Rev. B},
  volume = {63},
  issue = {22},
  pages = {224106},
  numpages = {16},
  year = {2001},
  month = {May},
  publisher = {American Physical Society},
  doi = {10.1103/PhysRevB.63.224106},
  url = {https://link.aps.org/doi/10.1103/PhysRevB.63.224106}
}

@article{DawBaskes1984,
  title = {Embedded-atom method: Derivation and application to impurities, surfaces, and other defects in metals},
  author = {Daw, Murray S. and Baskes, M. I.},
  journal = {Phys. Rev. B},
  volume = {29},
  issue = {12},
  pages = {6443--6453},
  numpages = {0},
  year = {1984},
  month = {Jun},
  publisher = {American Physical Society},
  doi = {10.1103/PhysRevB.29.6443},
  url = {https://link.aps.org/doi/10.1103/PhysRevB.29.6443}
}

@article{Stuart2001,
    author = {Stuart, Steven J. and Tutein, Alan B. and Harrison, Judith A.},
    title = {A reactive potential for hydrocarbons with intermolecular interactions},
    journal = {J. Chem. Phys.},
    volume = {112},
    number = {14},
    pages = {6472-6486},
    year = {2000},
    month = {04},
    issn = {0021-9606},
    doi = {10.1063/1.481208},
    url = {https://doi.org/10.1063/1.481208},
}

@article{Guo2006,
doi = {10.1088/0957-4484/17/18/033},
url = {https://doi.org/10.1088/0957-4484/17/18/033},
year = {2006},
month = {sep},
publisher = {},
volume = {17},
number = {18},
pages = {4726},
author = {Guo, Yufeng and Guo, Wanlin},
title = {Structural transformation of partially confined copper nanowires inside defected carbon
nanotubes},
journal = {Nanotechnology}
}

@article{Verlet1967,
  title = {Computer "Experiments" on Classical Fluids. I. Thermodynamical Properties of Lennard-Jones Molecules},
  author = {Verlet, Loup},
  journal = {Phys. Rev.},
  volume = {159},
  issue = {1},
  pages = {98--103},
  numpages = {0},
  year = {1967},
  month = {Jul},
  publisher = {American Physical Society},
  doi = {10.1103/PhysRev.159.98},
  url = {https://link.aps.org/doi/10.1103/PhysRev.159.98}
}

@book{Hairer1993,
    author = {Ernst Hairer , Gerhard Wanner , Syvert P. Nørsett},
    title = {Solving Ordinary Differential Equations I},
    publisher = {Springer Berlin, Heidelberg},
    year = {1993}
}

@book{chen2005nanoscale,
  title={Nanoscale Energy Transport and Conversion: A Parallel Treatment of Electrons, Molecules, Phonons, and Photons},
  author={Chen, G.},
  isbn={9780199774685},
  series={MIT-Pappalardo Series in Mechanical Engineering},
  url={https://books.google.nl/books?id=M3n3lUJpYDYC},
  year={2005},
  publisher={Oxford University Press}
}

@article{WANG2025,
title = {High thermal conductivity copper/graphene composites for efficient thermal management},
journal = {Carbon},
volume = {244},
pages = {120695},
year = {2025},
issn = {0008-6223},
doi = {https://doi.org/10.1016/j.carbon.2025.120695},
url = {https://www.sciencedirect.com/science/article/pii/S0008622325007110},
author = {Fei Wang and Chuanren Ye and Hao Yin and Shuaicheng Pan and Kanshuo Liu and Yanwu Zhu}
}

@article{MunGraphene,
author = {Mun, Sungkwang and Bowman, Andrew L. and Nouranian, Sasan and Gwaltney, Steven R. and Baskes, Michael I. and Horstemeyer, Mark F.},
title = {Interatomic Potential for Hydrocarbons on the Basis of the Modified Embedded-Atom Method with Bond Order (MEAM-BO)},
journal = {J. Phys. Chem. A},
volume = {121},
number = {7},
pages = {1502-1524},
year = {2017},
doi = {10.1021/acs.jpca.6b11343},
URL = { https://doi.org/10.1021/acs.jpca.6b11343
},
}

@article{Boden2014Nanoplatelet,
  author  = {Boden, Andreas and Boerner, B. and Kusch, P. and Firkowska, I. and Reich, Stephan},
  title   = {Nanoplatelet Size to Control the Alignment and Thermal Conductivity in Copper-Graphite Composites},
  journal = {Nano Lett.},
  year    = {2014},
  volume  = {14},
  number  = {6},
  pages   = {3640--3644},
  doi     = {10.1021/nl501411g},
  url     = {https://doi.org/10.1021/nl501411g}
}

@article{BokdamGraphenedistance,
  title = {Large potential steps at weakly interacting metal-insulator interfaces},
  author = {Bokdam, Menno and Brocks, Geert and Kelly, Paul J.},
  journal = {Phys. Rev. B},
  volume = {90},
  issue = {20},
  pages = {201411},
  numpages = {5},
  year = {2014},
  month = {Nov},
  publisher = {American Physical Society},
  doi = {10.1103/PhysRevB.90.201411},
  url = {https://link.aps.org/doi/10.1103/PhysRevB.90.201411}
}

@article{substrate_graphene_metal_Kelly,
  title = {First-principles study of the interaction and charge transfer between graphene and metals},
  author = {Khomyakov, P. A. and Giovannetti, G. and Rusu, P. C. and Brocks, G. and van den Brink, J. and Kelly, P. J.},
  journal = {Phys. Rev. B},
  volume = {79},
  issue = {19},
  pages = {195425},
  numpages = {12},
  year = {2009},
  month = {May},
  publisher = {American Physical Society},
  doi = {10.1103/PhysRevB.79.195425},
  url = {https://link.aps.org/doi/10.1103/PhysRevB.79.195425}
}

@article{MOORE2014163,
title = {Emerging challenges and materials for thermal management of electronics},
journal = {Mater. Today},
volume = {17},
number = {4},
pages = {163-174},
year = {2014},
issn = {1369-7021},
doi = {https://doi.org/10.1016/j.mattod.2014.04.003},
url = {https://www.sciencedirect.com/science/article/pii/S1369702114001138},
author = {Arden L. Moore and Li Shi}
}

@article{KHOSRAVI2024114834,
title = {Review of energy efficiency and technological advancements in data center power systems},
journal = {Energy Build.},
volume = {323},
pages = {114834},
year = {2024},
issn = {0378-7788},
doi = {https://doi.org/10.1016/j.enbuild.2024.114834},
url = {https://www.sciencedirect.com/science/article/pii/S0378778824009502},
author = {Ali Khosravi and Oscar R. Sandoval and Melika Sadat Taslimi and Tiia Sahrakorpi and Gessica Amorim and Juan Jose {Garcia Pabon}}
}

@article{grapheneproperties,
author = {Balandin, Alexander A. and Ghosh, Suchismita and Bao, Wenzhong and Calizo, Irene and Teweldebrhan, Desalegne and Miao, Feng and Lau, Chun Ning},
title = {Superior Thermal Conductivity of Single-Layer Graphene},
journal = {Nano Lett.},
volume = {8},
number = {3},
pages = {902-907},
year = {2008},
doi = {10.1021/nl0731872},
    note ={PMID: 18284217},

URL = { 
    
        https://doi.org/10.1021/nl0731872
    
    

},
}

@Article{qadirreview,
author ="Qadir, Akeel and Le, Top Khac and Malik, Muhammad and Amedome Min-Dianey, Kossi Aniya and Saeed, Imran and Yu, Yiting and Choi, Jeong Ryeol and Pham, Phuong V.",
title  ="Representative 2D-material-based nanocomposites and their emerging applications: a review",
journal  ="RSC Adv.",
year  ="2021",
volume  ="11",
issue  ="39",
pages  ="23860-23880",
publisher  ="The Royal Society of Chemistry",
doi  ="10.1039/D1RA03425A",
url  ="http://dx.doi.org/10.1039/D1RA03425A"}

@article{corona_graphene_nanocompostie,
    author = {Corona, Diego and Beatrici, Marco and Sbardella, Emanuele and Di Domenico, Gildo and Lucibello, Flavio and Zarcone, Mariano and Gaudio, Costantino Del},
    title = {3D printing copper – Graphene oxide nanocomposites},
    journal = {AIP Conf. Proc.},
    volume = {2416},
    number = {1},
    pages = {020007},
    year = {2021},
    month = {11},
    issn = {0094-243X},
    doi = {10.1063/5.0070350},
    url = {https://doi.org/10.1063/5.0070350},
}

@article{PhysRevB.81.245404,
  title = {Electron-phonon heat transfer in monolayer and bilayer graphene},
  author = {Viljas, J. K. and Heikkil\"a, T. T.},
  journal = {Phys. Rev. B},
  volume = {81},
  issue = {24},
  pages = {245404},
  numpages = {9},
  year = {2010},
  month = {Jun},
  publisher = {American Physical Society},
  doi = {10.1103/PhysRevB.81.245404},
  url = {https://link.aps.org/doi/10.1103/PhysRevB.81.245404}
}

@article{CASIMIR1938495,
title = {Note on the conduction of heat in crystals},
journal = {Physica},
volume = {5},
number = {6},
pages = {495-500},
year = {1938},
issn = {0031-8914},
doi = {https://doi.org/10.1016/S0031-8914(38)80162-2},
url = {https://www.sciencedirect.com/science/article/pii/S0031891438801622},
author = {H.B.G. Casimir}
}

@article{Park2013,
    author = {Park, Minkyu and Lee, Sun-Chul and Kim, Yong-Sung},
    title = {Length-dependent lattice thermal conductivity of graphene and its macroscopic limit},
    journal = {J. Appl. Phys.},
    volume = {114},
    number = {5},
    pages = {053506},
    year = {2013},
    month = {08},
    issn = {0021-8979},
    doi = {10.1063/1.4817175},
    url = {https://doi.org/10.1063/1.4817175},
}

@article{Thermalconductivities,
    author = {Ho, C. Y. and Powell, R. W. and Liley, P. E.},
    title = {Thermal Conductivity of the Elements},
    journal = {J. Phys. Chem. Ref. Data},
    volume = {1},
    number = {2},
    pages = {279-421},
    year = {1972},
    month = {04},
    issn = {0047-2689},
    doi = {10.1063/1.3253100},
    url = {https://doi.org/10.1063/1.3253100},
}

@article{MD_interface_length,
    author = {Park, Minkyu and Lee, Sun-Chul and Kim, Yong-Sung},
    title = {Length-dependent lattice thermal conductivity of graphene and its macroscopic limit},
    journal = {J. Appl. Phys.},
    volume = {114},
    number = {5},
    pages = {053506},
    year = {2013},
    month = {08},
    issn = {0021-8979},
    doi = {10.1063/1.4817175},
    url = {https://doi.org/10.1063/1.4817175},
}

@article{Zhan2015,
    author = {Zhan, Tianzhuo and Minamoto, Satoshi and Xu, Yibin and Tanaka, Yoshihisa and Kagawa, Yutaka},
    title = {Thermal boundary resistance at Si/Ge interfaces by molecular dynamics simulation},
    journal = {AIP Adv.},
    volume = {5},
    number = {4},
    pages = {047102},
    year = {2015},
    month = {04},
    issn = {2158-3226},
    doi = {10.1063/1.4916974},
    url = {https://doi.org/10.1063/1.4916974},
}

@article{Chalopin2012,
  title = {Thermal interface conductance in Si/Ge superlattices by equilibrium molecular dynamics},
  author = {Chalopin, Y. and Esfarjani, K. and Henry, A. and Volz, S. and Chen, G.},
  journal = {Phys. Rev. B},
  volume = {85},
  issue = {19},
  pages = {195302},
  numpages = {7},
  year = {2012},
  month = {May},
  publisher = {American Physical Society},
  doi = {10.1103/PhysRevB.85.195302},
  url = {https://link.aps.org/doi/10.1103/PhysRevB.85.195302}
}

@article{Pumarol2012,
author = {Pumarol, Manuel E. and Rosamond, Mark C. and Tovee, Peter and Petty, Michael C. and Zeze, Dagou A. and Falko, Vladimir and Kolosov, Oleg V.},
title = {Direct Nanoscale Imaging of Ballistic and Diffusive Thermal Transport in Graphene Nanostructures},
journal = {Nano Lett.},
volume = {12},
number = {6},
pages = {2906-2911},
year = {2012},
doi = {10.1021/nl3004946},
    note ={PMID: 22524441},

URL = { 
    
        https://doi.org/10.1021/nl3004946
    
    

},
}

@article{Tong2019,
  title = {Comprehensive first-principles analysis of phonon thermal conductivity and electron-phonon coupling in different metals},
  author = {Tong, Zhen and Li, Shouhang and Ruan, Xiulin and Bao, Hua},
  journal = {Phys. Rev. B},
  volume = {100},
  issue = {14},
  pages = {144306},
  numpages = {12},
  year = {2019},
  month = {Oct},
  publisher = {American Physical Society},
  doi = {10.1103/PhysRevB.100.144306},
  url = {https://link.aps.org/doi/10.1103/PhysRevB.100.144306}
}

@article{Schneider1978,
  title = {Molecular-dynamics study of a three-dimensional one-component model for distortive phase transitions},
  author = {Schneider, T. and Stoll, E.},
  journal = {Phys. Rev. B},
  volume = {17},
  issue = {3},
  pages = {1302--1322},
  numpages = {0},
  year = {1978},
  month = {Feb},
  publisher = {American Physical Society},
  doi = {10.1103/PhysRevB.17.1302},
  url = {https://link.aps.org/doi/10.1103/PhysRevB.17.1302}
}

@article{dunweg1991,
author = {Dunweg, Burkhard and Paul, Wolfgang},
title = {BROWNIAN DYNAMICS SIMULATIONS WITHOUT GAUSSIAN RANDOM NUMBERS},
journal = {Int. J. Mod. Phys. C},
volume = {02},
number = {03},
pages = {817-827},
year = {1991},
doi = {10.1142/S0129183191001037},

URL = { 
    
        https://doi.org/10.1142/S0129183191001037
    
    

},
}

@Inbook{Boer2023,
author="B{\"o}er, Karl W.
and Pohl, Udo W.",
title="Phonon-Induced Thermal Properties",
bookTitle="Semiconductor Physics",
year="2023",
publisher="Springer International Publishing",
address="Cham",
pages="157--190",
isbn="978-3-031-18286-0",
doi="10.1007/978-3-031-18286-0_5",
url="https://doi.org/10.1007/978-3-031-18286-0_5"
}

@article{Braun2019BestPractices,
  author    = {Efrem Braun and Justin Gilmer and Heather B. Mayes and David L. Mobley and Jacob I. Monroe and Samarjeet Prasad and Daniel M. Zuckerman},
  title     = {Best Practices for Foundations in Molecular Simulations [Article v1.0]},
  journal   = {Living J. Comput. Mol. Sci.},
  year      = {2019},
  volume    = {1},
  number    = {1},
  pages     = {5957},
  doi       = {10.33011/livecoms.1.1.5957},
  pmid      = {31788666},
  pmcid     = {PMC6884151},
  issn      = {2575-6524},
  publisher = {Living Journal of Computational Molecular Science},
  note      = {Epub 2018-11-29}
}

@book{ashcroft_mermin_ssp,
  author    = {Neil W. Ashcroft and N. David Mermin},
  title     = {Solid State Physics},
  publisher = {Cengage Learning, Inc.},
  year      = {1976},
  isbn      = {9780030839931}
}

@article{weights_report,
url = {https://doi.org/10.1515/pac-2019-0603},
title = {Standard atomic weights of the elements 2021 (IUPAC Technical Report)},
title = {},
author = {Thomas Prohaska and Johanna Irrgeher and Jacqueline Benefield and John K. Böhlke and Lesley A. Chesson and Tyler B. Coplen and Tiping Ding and Philip J. H. Dunn and Manfred Gröning and Norman E. Holden and Harro A. J. Meijer and Heiko Moossen and Antonio Possolo and Yoshio Takahashi and Jochen Vogl and Thomas Walczyk and Jun Wang and Michael E. Wieser and Shigekazu Yoneda and Xiang-Kun Zhu and Juris Meija},
pages = {573--600},
volume = {94},
number = {5},
journal = {Pure and Applied Chemistry},
doi = {doi:10.1515/pac-2019-0603},
year = {2022},
lastchecked = {2026-05-06}
}

@article{10.1063/1.328693,
    author = {Parrinello, M. and Rahman, A.},
    title = {Polymorphic transitions in single crystals: A new molecular dynamics method},
    journal = {J. Appl. Phys.},
    volume = {52},
    number = {12},
    pages = {7182-7190},
    year = {1981},
    month = {12},
    issn = {0021-8979},
    doi = {10.1063/1.328693},
}

@article{Cheng2021,
  author = {Cheng, Zhe and Li, Ruiyang and Yan, Xingxu and Jernigan, Glenn and Shi, Jingjing and Liao, Michael E. and Hines, Nicholas J. and Gadre, Chaitanya A. and Idrobo, Juan Carlos and Lee, Eungkyu and Hobart, Karl D. and Goorsky, Mark S. and Pan, Xiaoqing and Luo, Tengfei and Graham, Samuel},
  title = {Experimental observation of localized interfacial phonon modes},
  journal = {Nature Comm.},
  year = {2021},
  volume = {12},
  number = {1},
  pages = {6901},
  doi = {10.1038/s41467-021-27250-3},
  url = {https://doi.org/10.1038/s41467-021-27250-3},
  issn = {2041-1723}

  
}

\clearpage
\beginsupplement
\let\section\origsection
\let\subsection\origsubsection
\section*{Supplementary Information}

\section{Theory and Force Fields}\label{sec:add_theory}
Molecular Dynamics (MD) simulations provide a method for investigating phonon dynamics and their contributions to thermal transport across material interfaces \cite{Chen2022}. In these simulations, the lattice vibrations, known as phonons, are computed based on the collective dynamics of atoms under assumed force fields. Lattice vibrations account for over $99$\% of thermal transport in undoped substrated graphene, whereas in copper, electrons are the main heat carriers at room temperature \cite{Pop2012}\cite{substrate_graphene_metal_Kelly}. Although models that include electron interactions would be more accurate, due to challenges in time- and scale-separation in the respective dynamics, these models generally require additional treatment \cite{PhysRevB.81.245404}. MD simulations resolve nonlinear and anharmonic scattering at the interface. Using classical MD simulations, qualitative trends in interface thermal resistance have been identified, including responses to defect density, strain, and the number of graphene layers \cite{Zhu2022}\cite{Li2022}\cite{Fang2025}. MD simulations are a tool for constructing and further understanding interface dynamics at the graphene-copper interface, with the potential to provide qualitative insights into the underlying phenomena responsible for the wide range of measured properties of graphene-copper composites.\\
\\
The Molecular Dynamics simulations were performed using the Large-scale Atomic/Molecular Massively Parallel Simulator (LAMMPS) \cite{Thompson2022_LAMMPS}. The atoms are assumed to be point particles following Newton's laws under forces arising from prescribed potentials in the NVE ensemble. The velocities and positions of the atoms are updated through the St\o rmer-Verlet integration scheme, which conserves the symplectic form and bounds the total energy of the system without drift. The scheme resolves the dynamics with an accuracy of $O(\Delta t^2)$, with $\Delta t$ the step size in case of no thermostatting \cite{Verlet1967}\cite{Hairer1993}.
\begin{equation}   \label{eq:Hamiltonian}
H(p,q) = \frac{1}{2}\sum_{i=1}^N \frac{1}{m_i}p_i^Tp_i + V(q_1,q_2,...,q_N),  
\end{equation} 
\begin{equation}  \label{eq:Verlet2}
q_{n+1} -2q_n +q_{n-1} = \Delta t^2M^{-1}F(q_n)
\end{equation} 
with $H$ the Hamiltonian in $(p,q)$ the atomic momenta and positions, where $N$ is the total number of atoms, and $M = \text{diag}(m_1,m_2,...,m_N)$ is the mass matrix, $F$ the sum of forces in the system arising from interatomic potentials as $F_{\text{potential}} = -\nabla V$, and from thermostat forcing, $F_{\text{Langevin}}$.

Thermostatted regions within the domain are endowed with stochastic noise generated using a Langevin thermostat \cite{Schneider1978}. 
\begin{equation}\label{eq:Lang}
    F_{\text{Langevin}} = F_f+F_r
\end{equation}  
\begin{equation}
    F_f = -\frac{mv}{t_\text{damp}}, \quad F_r = \sqrt{\frac{k_BT m}{\Delta t  \cdot t_\text{damp}}}\eta
\end{equation}
with $t_{\text{damp}}$ the damping time, $T$ the target temperature, and $\eta$ the stochastic noise generated from a uniform distribution. Although the distribution is not Gaussian, the uniform distribution defines a valid short-time propagator for the same Fokker-Planck dynamics in the limit $\Delta t \to 0$ \cite{dunweg1991}. The assigned atomic masses are 63.546 amu for copper and 12.011 amu for carbon. 

Since a single force field for a graphene-copper embedding is currently unavailable, a combination of force fields is set \cite{Zhu2022} \cite{Li2022}\cite{Fang2025}. The interactions between Cu atoms are calculated following the Embedded Atom Method (EAM) potential for copper atom interactions as developed by Mishin et al. (2001) using ab initio calculations \cite{Mishin2001}. The EAM potential considers both a pairwise potential function $\phi$ and an embedding contribution $f_{\text{emb}}$ from the electron density $\rho$ \cite{DawBaskes1984}: 
\begin{equation}  
V_{\text{EAM}} = \sum_i \left[f_{\text{emb}}\left( \sum_{j\neq i} \rho(r_{ij})\right) +\frac{1}{2}\sum_{j\neq i}\phi(r_{ij}) \right]
\end{equation}  
where $r_{ij}$ is the distance between the atoms labeled $i$ and $j$ and the embedding $f_{\text{emb}}$ and potential functions $\phi$ are calculated for all $i\neq j$ within a cutoff radius of $\sim 5.5$ \AA\, \cite{Mishin2001}.

For the C-C interactions, the Adaptive Intermolecular Reactive Bond Order  (AIREBO) potential is used \cite{Stuart2001}. The AIREBO potential is known for representing both chemical and van der Waals bonding, enabling bond formation and breaking dynamics, which are particularly relevant for graphene, where the short-range intralayer $sp^2$ bonds are much stronger and chemically bonded compared to interlayer bonds, which are held only by van der Waals forces, and are prone to wrinkling, folding, and defects \cite{Pop2012}. 
\begin{equation*} V_{\mathrm{AIREBO}}
=
\frac{1}{2}\sum_i \sum_{j\neq i}
\left[
V^{\mathrm{REBO}}_{ij}
+
V^{\mathrm{LJ}}_{ij}
+
\sum_{k\neq i,j}\sum_{l\neq i,j,k}
V^{\mathrm{TORS}}_{ijkl}
\right] \label{eq:AIREBO}
\end{equation*} 
\begin{alignat*}{2}
&V^{\mathrm{REBO}}_{ij}
&&=
S'(t_c(r_{ij}))\left(V^{R}_{ij}(r_{ij}) + b_{ij}\,V^{A}_{ij}(r_{ij})\right), \\
&V^{\mathrm{LJ}}_{ij}
&&=
f_{\mathrm{out}}^{\mathrm{LJ}}(r_{ij})\,
C_{ij}\,
\Big[
S\!\big(t_r(r_{ij})\big)\,S\!\big(t_b(b_{ij}^{*})\big)
+
1-S\!\big(t_r(r_{ij})\big)
\Big]\,\\ & && \quad
\times 4\epsilon_{ij}
\left[
\left(\frac{\sigma_{ij}}{r_{ij}}\right)^{12}
-
\left(\frac{\sigma_{ij}}{r_{ij}}\right)^6
\right], \\
&V^{\mathrm{TORS}}_{ijkl}
&&=
w_{ij}(r_{ij})\,w_{jk}(r_{jk})\,w_{kl}(r_{kl})\,
V_{\mathrm{tors}}(\omega_{ijkl}) .
\end{alignat*}
with $r$ the distance between atoms, $V^R, V^A$ the repulsive and attractive (exponential) potentials, $b$ the bond order, $S'$ the switching function $S'(t) = \Theta(-t)+\Theta(t)\Theta(1-t)\frac{1}{2}[1+\cos(\pi t)]$ , with with scaling function $t_a (a_{ij}) =  \frac{a_{ij}-a_{ij}^{min}}{a_{ij}^{max}-a_{ij}^{min}}$ and $\Theta$ the Heavyside step-function, $S(t) = \Theta(-t)+\Theta(t)\Theta(1-t)[1-t^2(3-2t)]$ the switching function for the LJ contribution, $\epsilon$ the depth of the LJ potential well, $\sigma$ the distance at which the LJ potential is zero. Additionally, we have $\omega$ the torsion angle, $V_{torsion}$ the torsion function \cite{Stuart2001}. The cutoff radius for the LJ term is $10.2$\AA\,, as implemented through a cutoff scaling factor of $3.0$ in LAMMPS.

The Cu-C atoms are assumed to be only bonded by van der Waals forces. The long-range attractive van der Waals force and a short-range repulsion of overlapping electron orbitals are captured by the Lennard-Jones (LJ) potential. Here, the coefficients in the LJ potential for Cu-C are based on a mixing rule with $\epsilon=0.02578$ eV, $\sigma=3.0825$ \AA\,  and a cutoff radius of $r_c=10.0$ \AA\, \cite{Guo2006}.

\begin{equation}
V_{\mathrm{LJ}}
=
\sum
4\varepsilon
\left[
    \left( \frac{\sigma}{r_{ij}} \right)^{12}
    -
    \left( \frac{\sigma}{r_{ij}} \right)^{6}
\right]\Theta(r_c)
\end{equation}

\section{Uncertainties}\label{sec:uncertainties}
Uncertainty of the Kapitza resistance as calculated from equations \ref{eq:Kapitza} and \ref{eq:T_Cu}, obtained from a single simulation of the Cu-G-Cu interface with a temperature gradient along $\mathbf{e}_z$ is estimated by standard error propagation of (i) the uncertainty in the linear fits of the temperature profiles $\sigma_{T_{Cu_i}}$, and (ii) the uncertainty in the linear fit of the flux output in time $\sigma_{J_z}$:
    \begin{equation}
        \sigma_{R_K} = R_K\sqrt{\left(\frac{\sigma_{J_z}}{J_z}\right)^2+\left(\frac{\sigma_{\delta T}}{\delta T}\right)^2 }, \end{equation}\begin{equation} \sigma_{\delta T} =\sqrt{ \sigma_{T_{Cu_1}}^2+\sigma_{T_{Cu_2}}^2}
    \end{equation} 
A similar standard error propagation is applied to the standard error in the calculated copper conductivities $\kappa = -J_z/a$.

\section{Velocities: Equilibration, Autocorrelation and Distributions}\label{sec:details_velocities}
  \begin{figure}
  \centering
  \begin{subfigure}[c]{\linewidth}
    \centering
    \includegraphics[width=\linewidth]{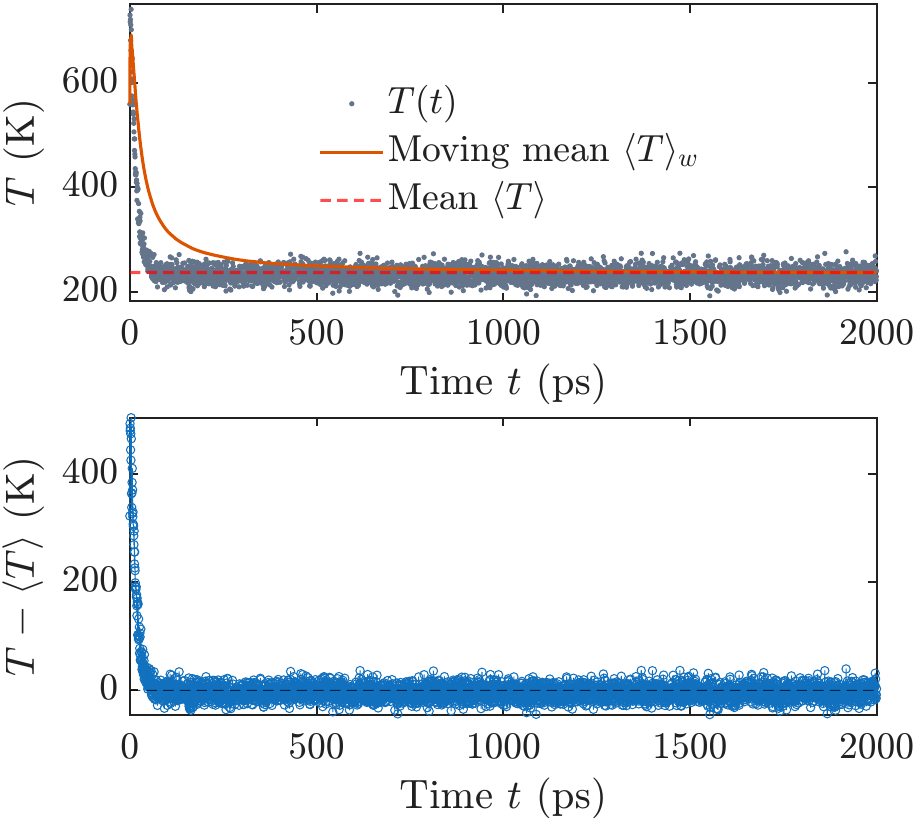}
  \end{subfigure}\hfill
  \begin{subfigure}[c]{\linewidth}
    \centering
\includegraphics[width=1\linewidth]{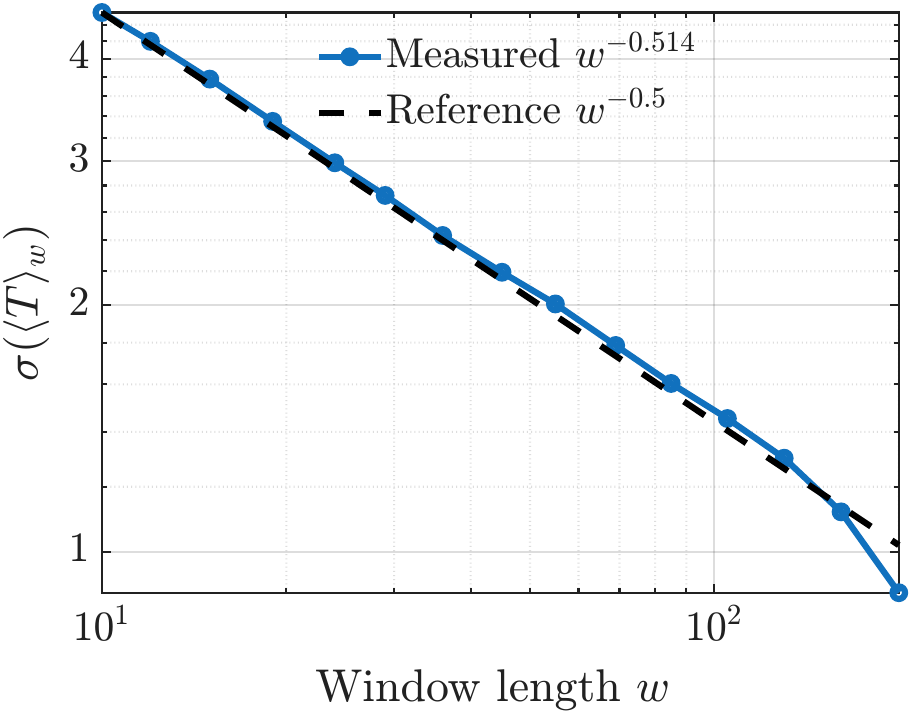}
  \end{subfigure}
  \caption{Equilibration and deviation of a single bin temperature per $0.5$ ps within a copper block in the Cu-G-Cu composite during an equilibration and production run (2 ns total) (left) and the standard deviation over window length (right) of samples taken per $1$ ps during a production run (1 ns).}
\label{fig:time_covergence}
\end{figure}
 \begin{figure}
    \centering
    \includegraphics[width=\linewidth]{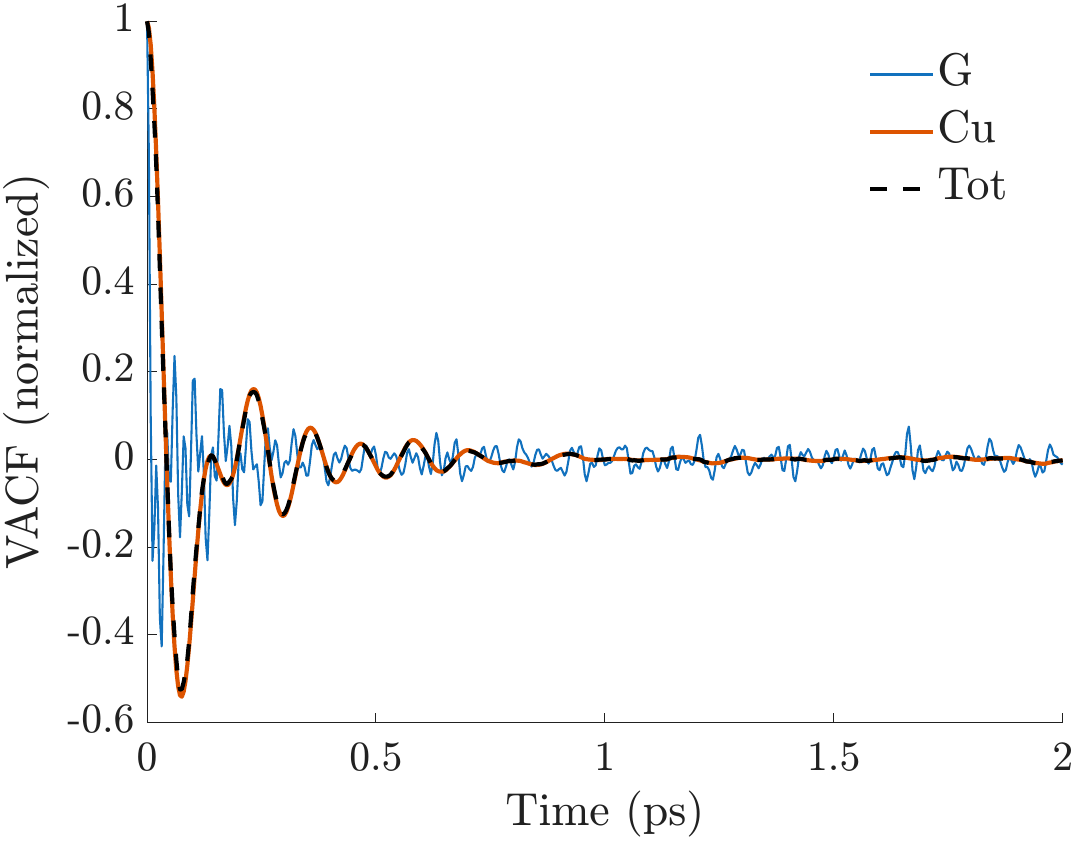}
    \caption{Normalised VACF of the graphene atoms (G), the copper atoms (Cu), and all atoms combined (Tot).}
    \label{fig:VACF1}
\end{figure}
\begin{figure}
    \centering
\includegraphics[width=\linewidth]{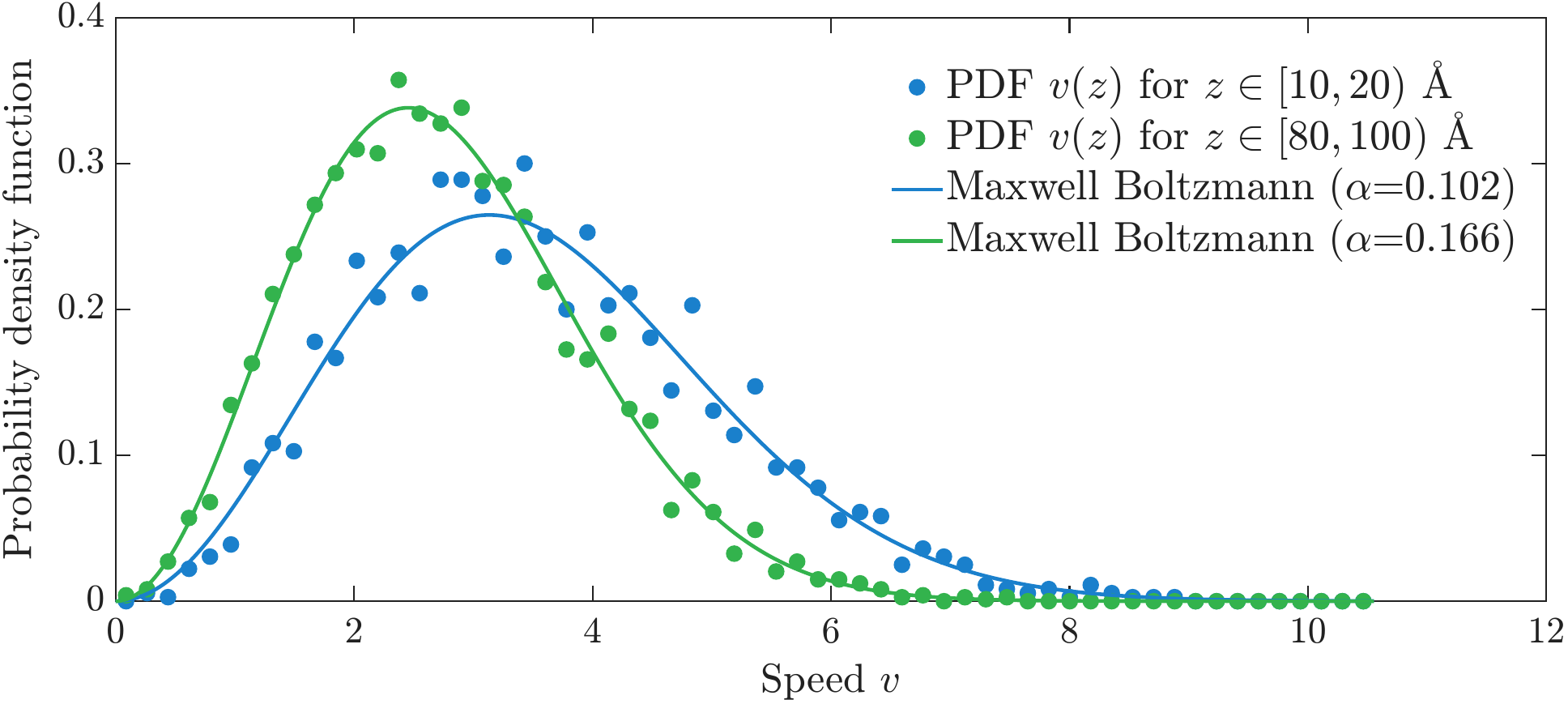}
    \caption{Maxwell-Boltmann velocity distributions $f(v) = 4\pi \left(\frac{\alpha}{\pi}\right)^{3/2}v^2 e^{-\alpha v^2}$ fitted to the velocity distribution of atoms in slabs of $20$\AA\, during the production NEMD run.}
    \label{fig:Maxwell-Boltzman}
\end{figure}

Each NEMD simulation is equilibrated for 1 ns. After the equilibration run, the system is assumed to be in a statistically steady state, and a production simulation is performed for 1 ns unless otherwise specified. The timestep size is set to 1 fs. In Figure \ref{fig:time_covergence}, the temperature samples of a single slab $\Omega_k = \left\{ \mathbf{x} \in \Omega : z_k-\frac{\Delta z}{2} \leq \mathbf{x}\cdot\mathbf{e}_z < z_k+\frac{\Delta z}{2} \right\}$ along $z$ as computed from equation \ref{eq:temperature} show no trend in the local mean after 1 ns, and the statistically steady-state assumption is confirmed. Figure \ref{fig:time_covergence} shows the standard deviation of the time-averaged temperature as a function of averaging window length. The near $0.5$ scaling of the standard deviation indicates the expected $1/\sqrt{n}$ reduction in uncertainty, consistent with sufficiently decorrelated samples. This confirms the Velocity Autocorrelation function decay observed in Figure \ref{fig:VACF1}.\\
\\
Additionally, to assess the local equilibrium and equipartition assumption, the distribution functions of the velocities in different arbitrary slabs, along $z$ are shown in Figure \ref{fig:Maxwell-Boltzman}. Maxwell-Boltzmann distributions are overlaid and show reasonable agreement, indicating that the local velocity statistics are approximately Gaussian even for intervals of these sizes.
\section{Lattice initialisation}\label{sec:lattice_initilization_caseII}
In this section, the steps of Case II initialisation are described. The initialisation is started by relaxing the copper and graphene lattices separately at 0K, followed by equilibration at 300K. Then, matching the number of lattice repetitions to minimise the lattice, the equilibrated graphene layer is combined with the copper lattice, and subsequently the whole structure is allowed to equilibrate and relax at 300K. The confirmed structures, details on the equilibration steps and retrieved domain sizes are given in the following subsections.
\subsection{Copper initialisation}\label{subsub:Copperinitialise}
 The copper lattice is initialised in an FCC crystal structure with an initial lattice constant of $a_{Cu}=3.6$  \AA\,. A full energy minimisation at 0K of the lattice in the domain as well as the atomic positions is performed. After minimisation, there is a remaining pressure of $-0.0003$ bar and a remaining maximum force of $\max_{i} \max(|F_{x,i}|, |F_{y,i}|,|F_{z,i}|) = 3.10\cdot 10^{-4}$ eV/\AA\, which is small compared to characteristic interatomic forces (cohesive energy at typical length scales of Cu in the EAM potential to $10^0$ eV/\AA), indicating near fully relaxed state. The relaxation at 0K yields a lattice constant of $3.615$\AA\, and energy of $E_C = 3.54$ eV/atom. This corresponds exactly to the known values of the EAM potential for fcc Cu by \cite{Mishin2001}.
 
 \begin{figure}
  \centering
  \begin{subfigure}[c]{\linewidth}
    \centering
    \includegraphics[width=0.7\linewidth]{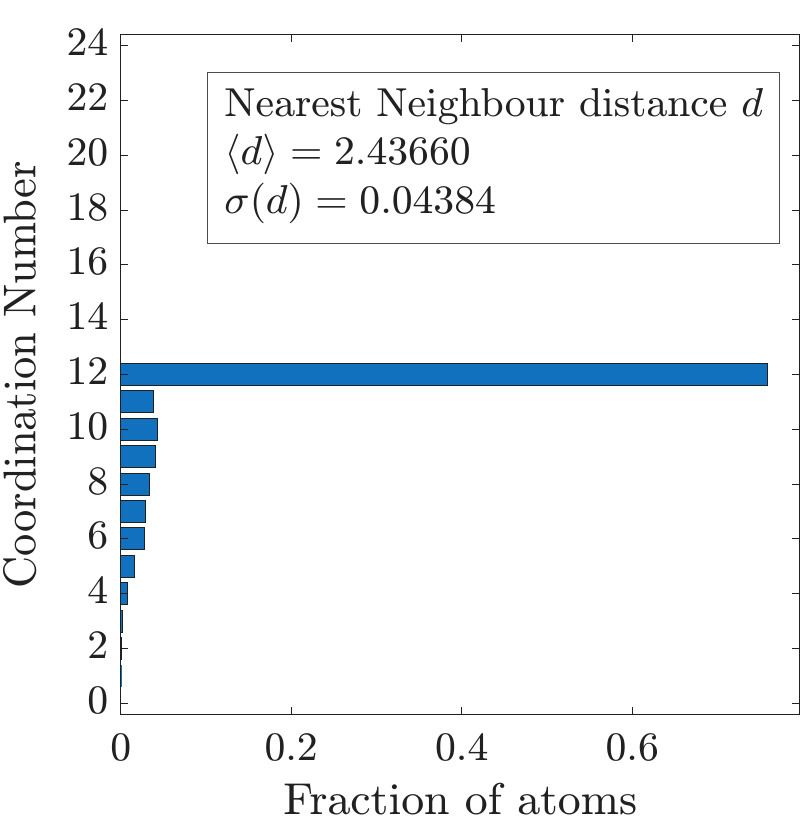}
  \end{subfigure}\vspace{0.4cm}
  \begin{subfigure}[c]{\linewidth}
    \centering
    \includegraphics[width=0.7\linewidth]{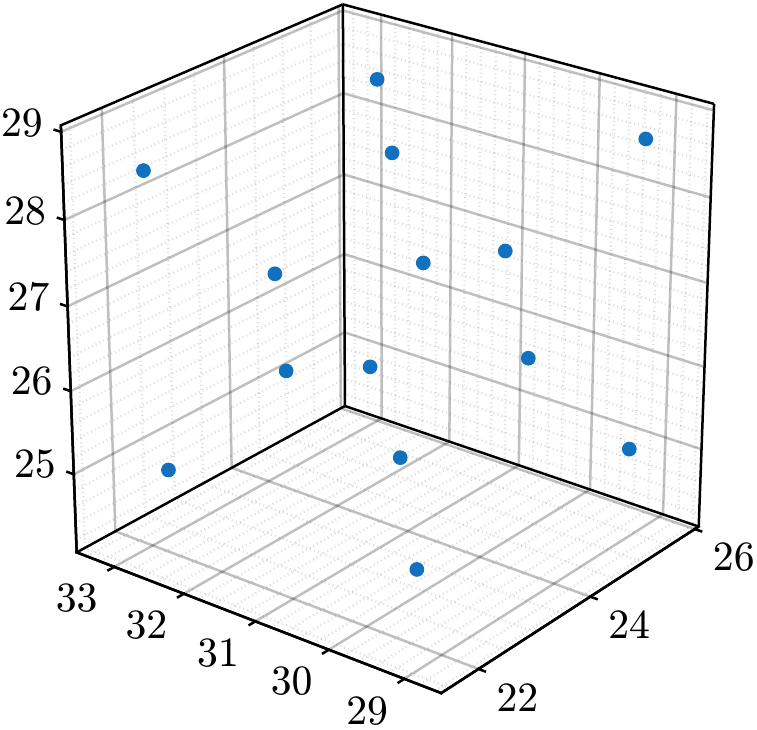}
  \end{subfigure}
  \caption{Structure of copper equilibrated at 300K. The coordination number of nearest neighbours within $r_c=3$\AA\, and their interdistances (left). Atomic positions (dots) of a center domain within the copper lattice (right).}
\label{fig:structure_Copper_300K}
\end{figure}
\begin{figure}
  \centering
  \begin{subfigure}[c]{\linewidth}
    \centering
    \includegraphics[width=\linewidth]{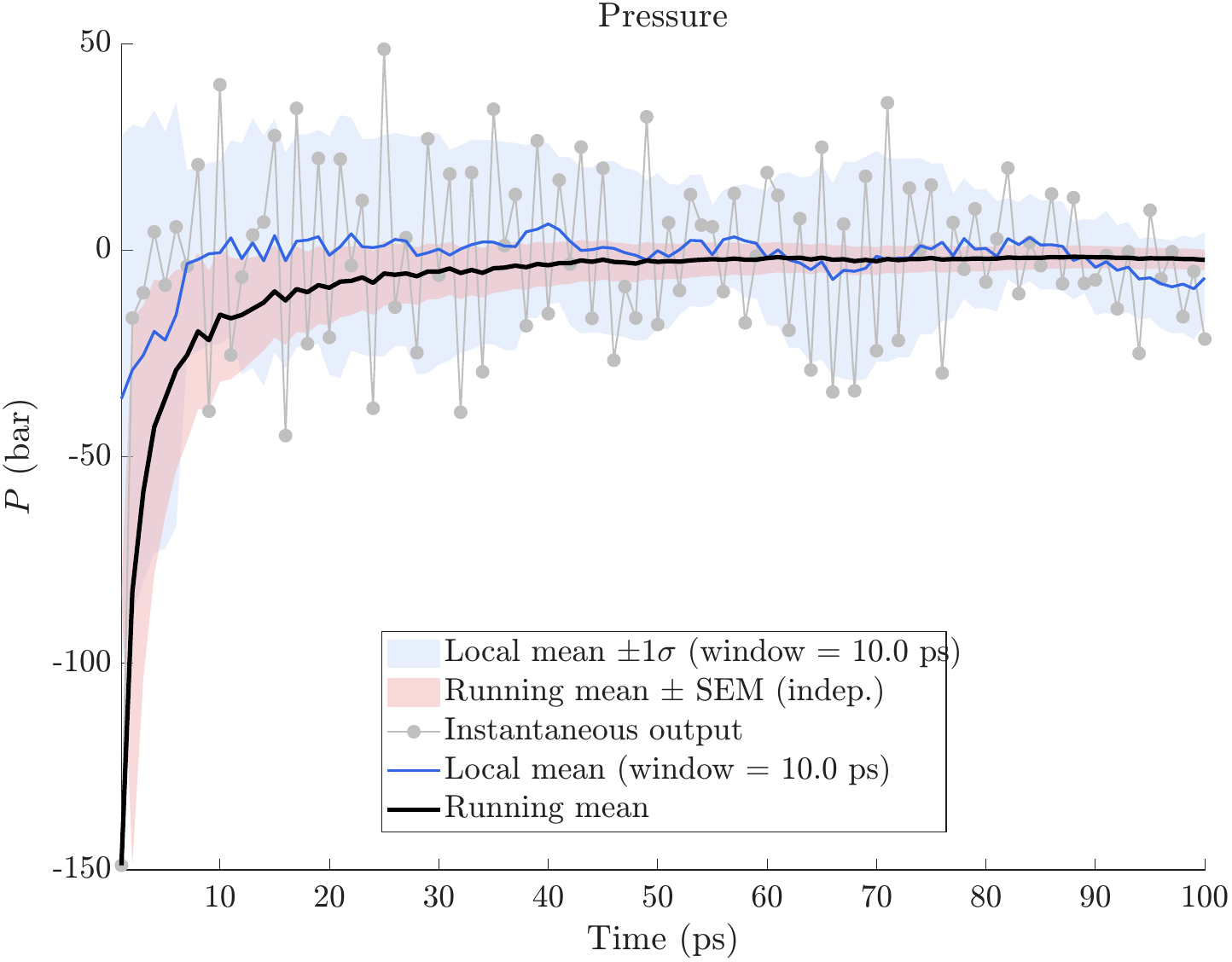}
  \end{subfigure}\hfill
  \begin{subfigure}[c]{\linewidth}
    \centering
    \includegraphics[width=\linewidth]{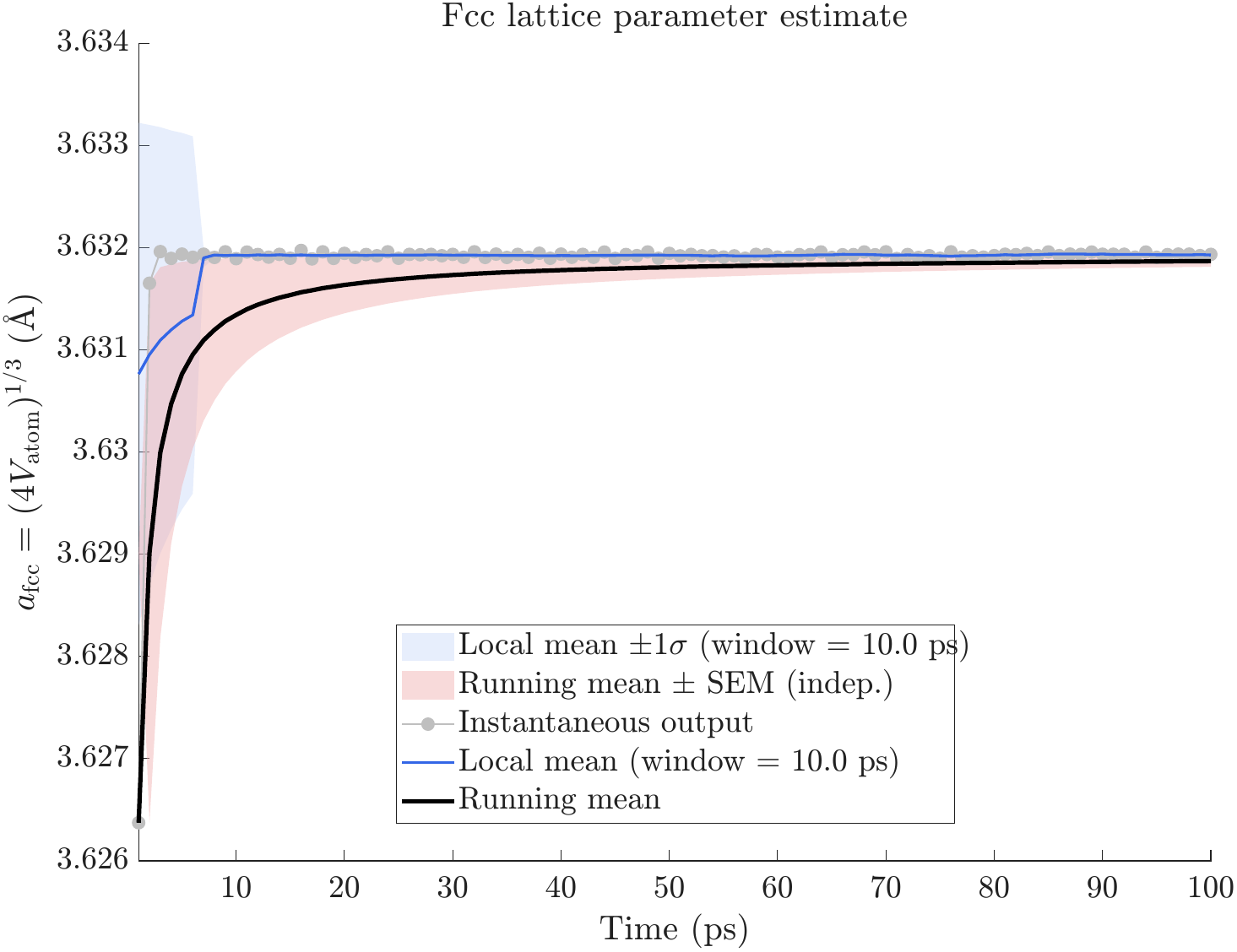}
  \end{subfigure}
  \caption{Equilibration process copper at 300K. The pressure (left) and the estimated lattice constant from the domain size (right) as a function of time during the NPT equilibration run with $\Delta t = 10^{-3}$ps.}
  \label{fig:relaxation300KCopper_pressure_lattice}
\end{figure}

 The relaxation at 0K is followed by an equilibration at 300K for 100 ps through an NPT ensemble with periodic boundary conditions. This results in an equilibrated lattice constant of $a=3.632$ \AA\, and cohesive energy of $3.50$ eV/atom. The relaxation and equilibration can be inspected in Figures \ref{fig:relaxation300KCopper_pressure_lattice}.  The structure remains a stable FCC after equilibration at 300K, confirmed by the atomic positions and number of nearest neighbours consistent with FCC in Figure \ref{fig:structure_Copper_300K}.  
 \subsection{Graphene initialisation}\label{subsub:grapheneinitial}
 The carbon atoms are initialised in a honeycomb structure from a 2D hexagonal lattice with lattice constant $a_G=2.46$ \AA\,. The lattice is set up such that the periodic boundaries align with periodicity in the honeycomb structure by initialising a rotated honeycomb structure accordingly. The domain is assumed to be periodic in the graphene xy-plane to mimic an infinite sheet. In the z-direction, wall boundaries are imposed at $z=\pm3$\AA\, with a potential of: \begin{equation} E_{wall} = \begin{cases}
     4\epsilon \left[\left(\frac{\sigma}{r}\right)^{12}-\left(\frac{\sigma}{r}\right)^{6} \right], \, &r<r_c \\
     0, \, &r\geq r_c
 \end{cases} \end{equation} with $r$ begin the distance of the atom to the wall, $r_c= 8$\AA\, the cutoff distance, $\sigma =3.0825$\AA\, and $\epsilon=0.02578$ eV resembling the LJ potential imposed by a copper atom. 
 
 By imposing a wall with a potential similar to that of the copper blocks, the relaxation and equilibration of the sheet will be closer to that of embedded graphene than to free or suspended graphene, and less sensitive to wrinkling and defects. This way, the relaxation will not be dominated by the relaxation of any neighbouring lattice, and the energy imposed by the walls is directly excluded from the NPT relaxation energy and pressure, such that these do not dominate the relaxation either.
 
 The graphene sheet is relaxed in three steps. First, by minimising the potential energy by changing the atomic positions, followed by changing both positions as well as the domain to minimise the potential energy, the hydrostatic pressure, and a strain energy expression as proposed by Parrinello and Rahman \cite{10.1063/1.328693}\cite{Thompson2022_LAMMPS}. Lastly, at a set box size, the atomic positions are once more relaxed to minimise the potential energy at 0K. The relaxed lattice is found to have an approximate lattice parameter $a_0 = 2.419$ \AA\, and an energy $E_C = 7.4$ eV/atom. From the atomic positions a nearest neighbour distance of $\langle d\rangle = 1.40$\AA\, with $\sigma(d)= 2\cdot 10^{-4}$ is found for neighbours within a radius of $r=2$\AA\,. Furthermore, the honeycomb structure is confirmed by investigating the number of nearest neighbours, which is three, consistent with the honeycomb structure. The obtained values match reported values from \cite{MunGraphene} for the graphene structure under the REBO potential of $a=2.419$ \AA\, and $E_C =7.243 $ eV/atom.
 
\begin{figure}
  \centering
  \begin{subfigure}[c]{\linewidth}
    \centering
    \includegraphics[width=0.6\linewidth]{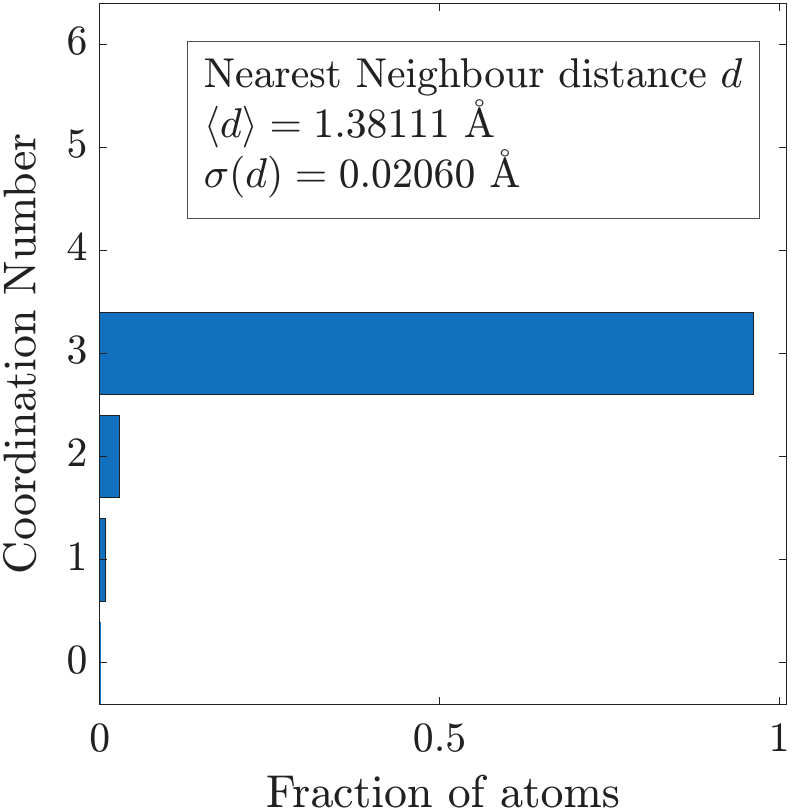}
  \end{subfigure}\vspace{0.4cm}
  \begin{subfigure}[c]{\linewidth}
    \centering
    \includegraphics[width=0.9\linewidth]{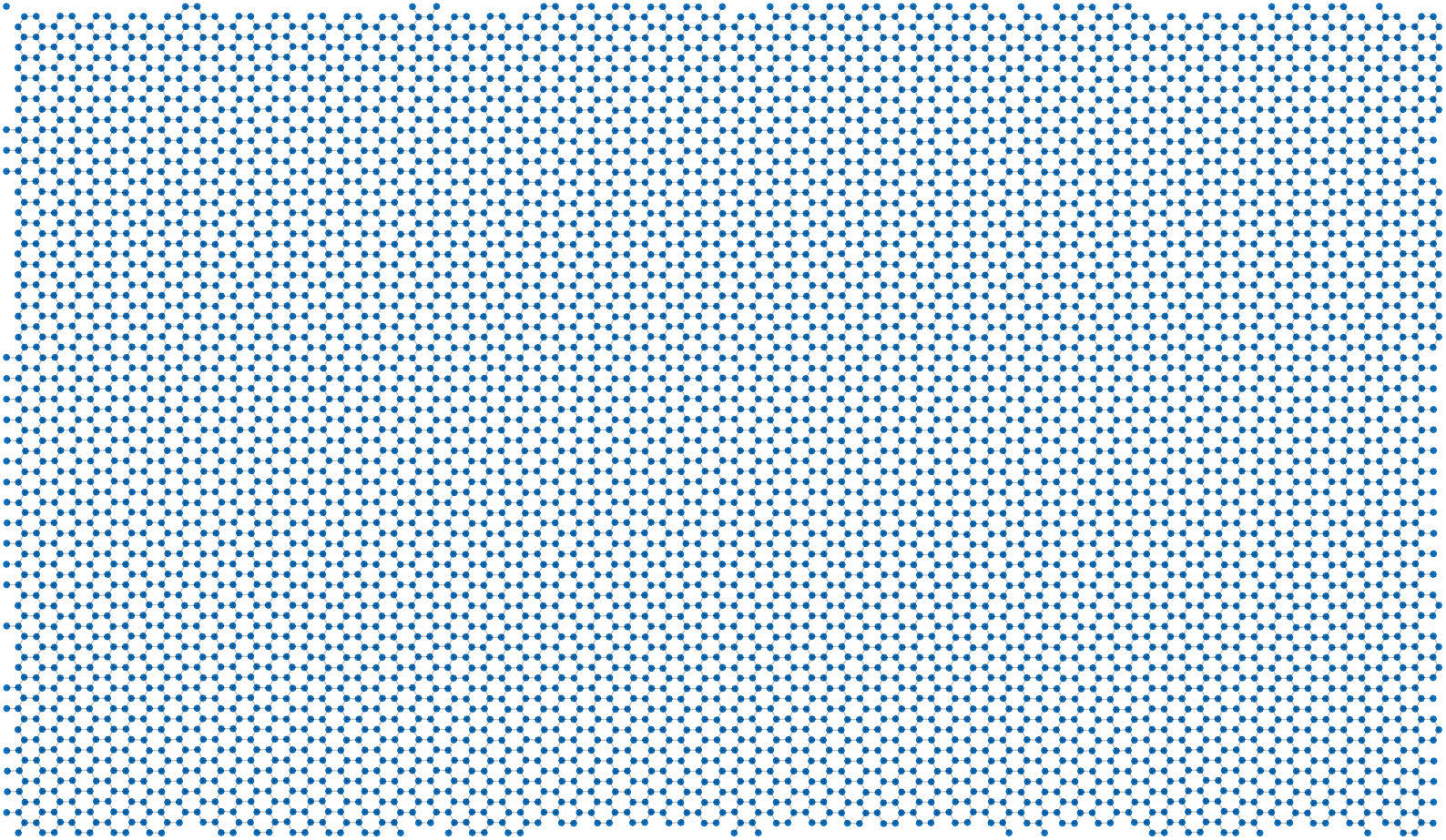}
  \end{subfigure}
  \caption{Structure of graphene equilibrated at 300K. The coordination number of nearest neighbours within $r_c=2$\AA\, and their interdistances (left). Atomic positions (dots) and bonding (lines) within $r_c=2$\AA\, in the x-y plane (right).}
  \label{fig:structuregraphene300K}
\end{figure}

\begin{figure}[t]\vspace{1cm}
  \begin{subfigure}{\linewidth}
    \centering
    \includegraphics[width=\linewidth]{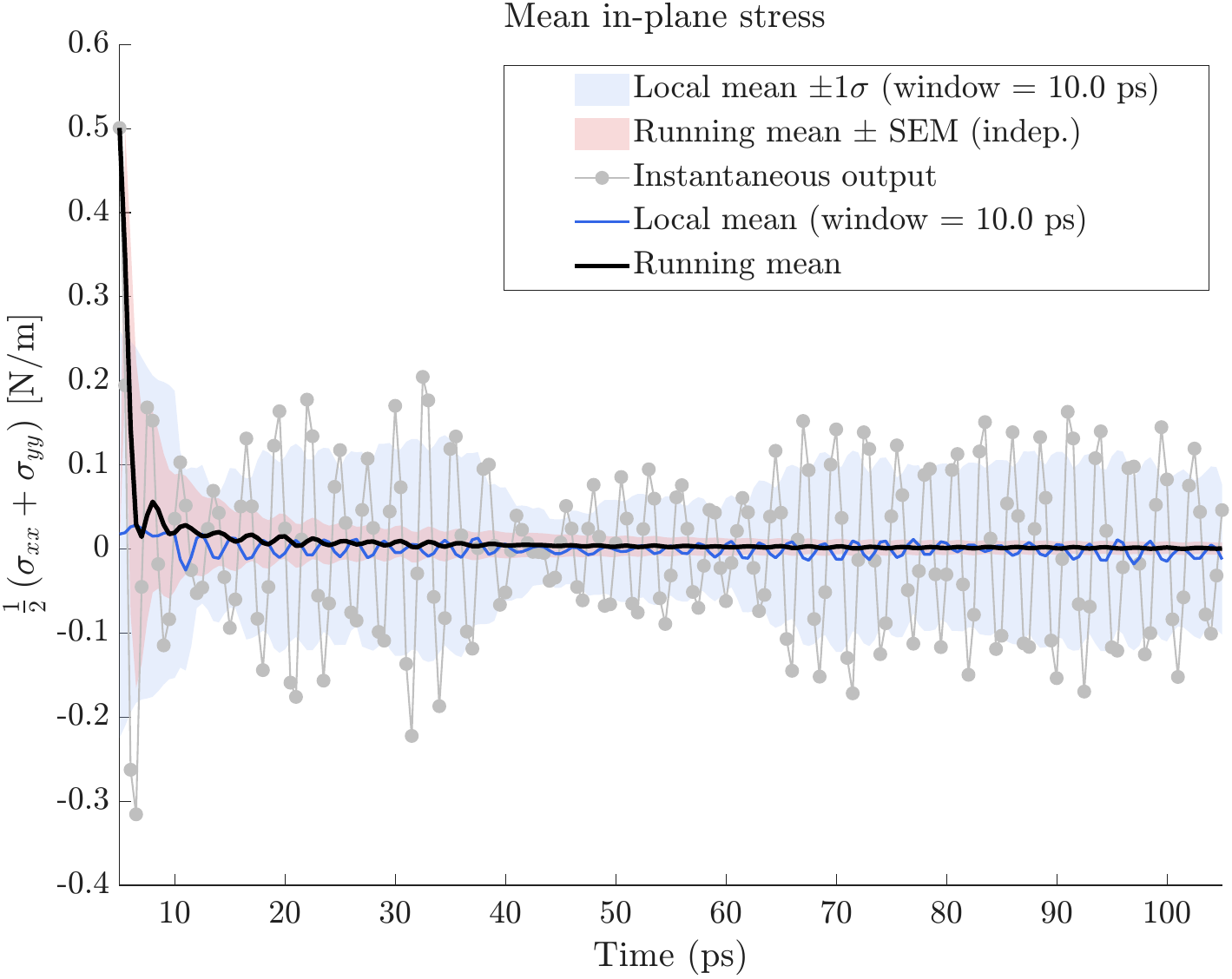}
  \end{subfigure}\vspace{0.4cm}
  \begin{subfigure}{\linewidth}
    \centering
    \includegraphics[width=\linewidth]{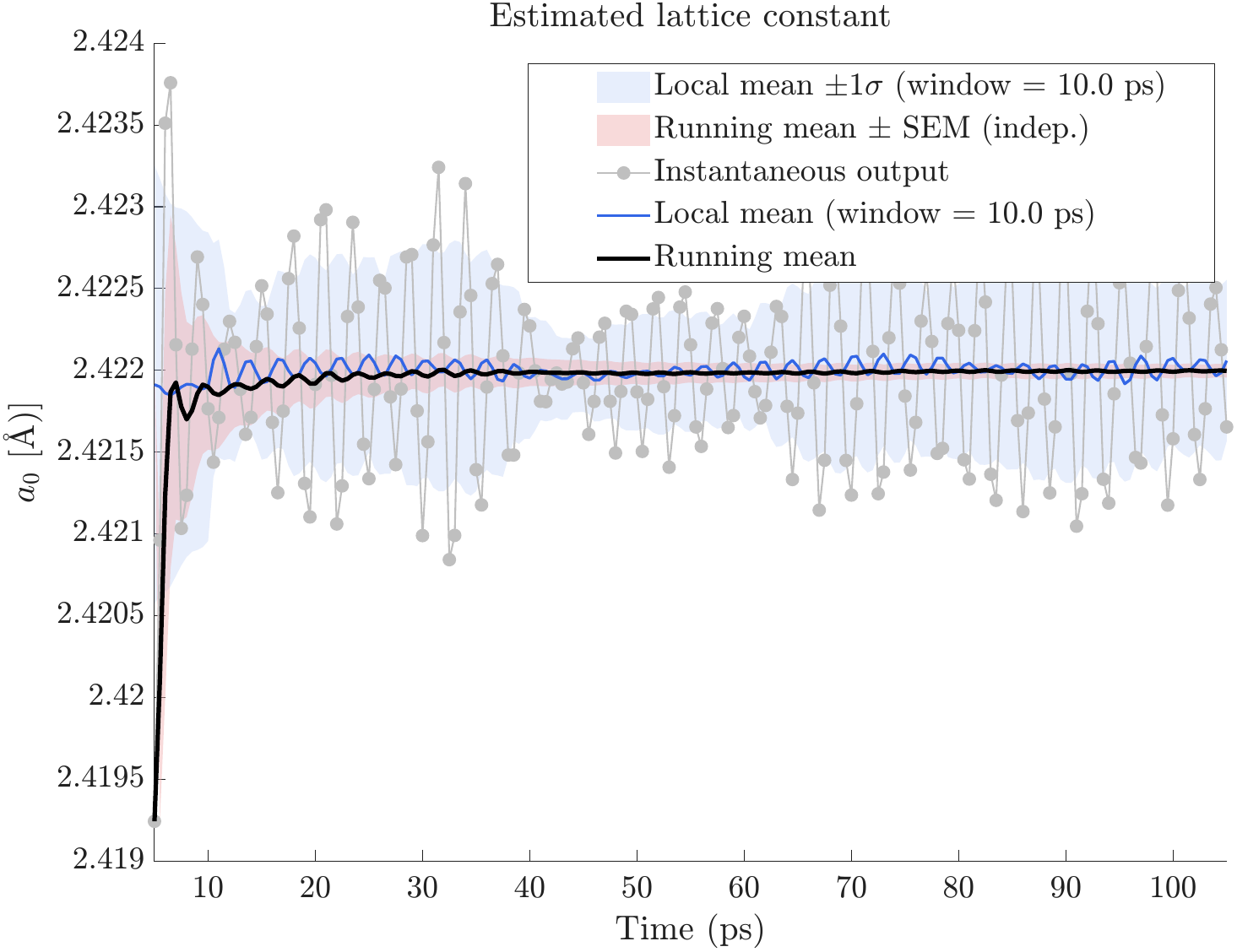}
  \end{subfigure}
  \caption{Equilibration process graphene at 300K. The mean in-plane stress (left) and the estimated lattice constant from the domain size (right) as a function of time during the NPT equilibration run with $\Delta t =2.5\cdot 10^{-4}$ps.}
  \label{fig:relaxation300Kgraphene_stress_lattice}
\end{figure}

 After the relaxation at 0K, the sheet is heated gradually to 300K using a sequence of NVT runs. The gradual heating up to 300K is followed by an NPT run of 50 ps with a timestep of $0.25$ fs. During the NPT run, the domain in the xy plane is relaxed such that the energy and pressure are minimised (excluding the wall energy). The equilibration process is displayed in Figure \ref{fig:relaxation300Kgraphene_stress_lattice}, and no trend in the running means is observed after about $60$ ps. The in-plane stress averages at $0$ N/m as desired, with remaining fluctuations of $10^{-1}$ N/m. Similarly, the approximate lattice constant derived from the domain sizes fluctuates with $10^{-3}$\AA\, around a mean of $a = 2.422$ \AA. The fluctuations in the mean in-plane stress follow the fluctuations in the lattice constant. The equilibrated structure displays a stable honeycomb as shown in Figures \ref{fig:structuregraphene300K},  with a nearest neighbour distance of $1.38$\AA\..
 
\begin{figure}\centering
  \begin{subfigure}{\linewidth} 
  \centering
\includegraphics[width=0.6\linewidth]{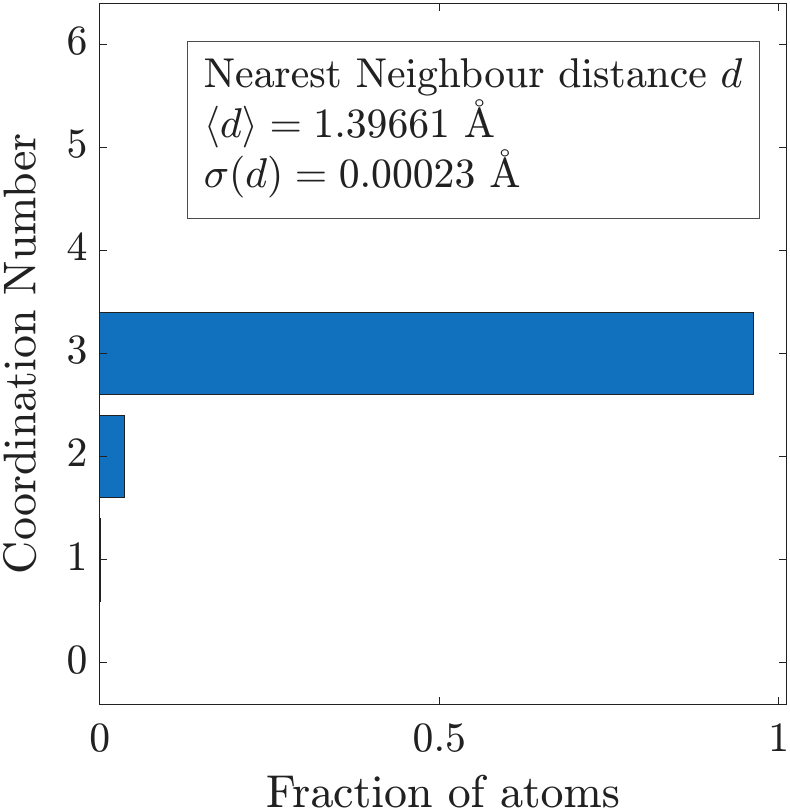} 
\end{subfigure}
  \begin{subfigure}{\linewidth}
  \includegraphics[width=0.9\linewidth]{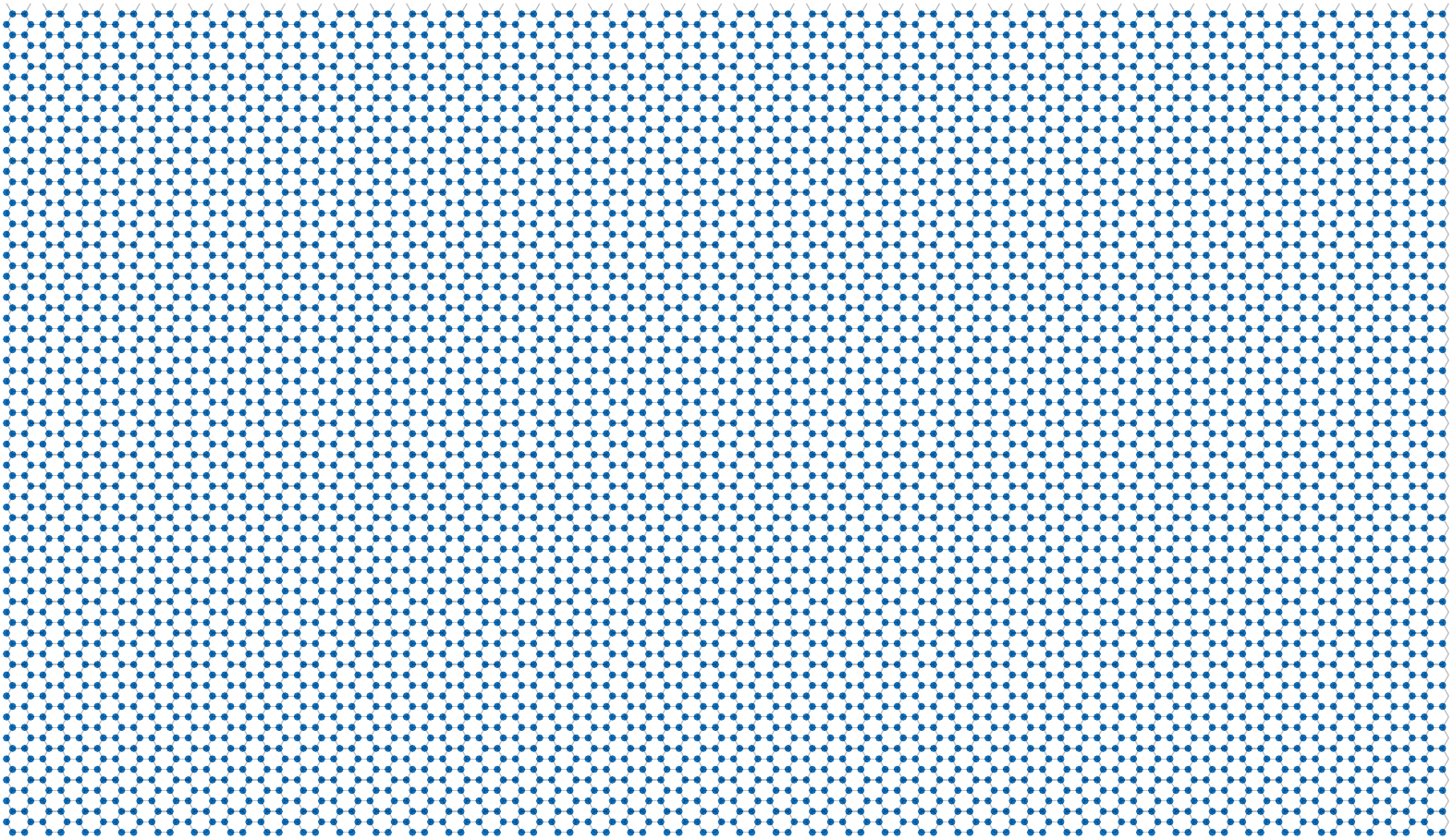} \end{subfigure} 
\caption{Structure of graphene relaxed at 0K. The coordination number of nearest neighbours within $r_c=2$\AA\, and their interdistances (left). Atomic positions (dots) and bonding (lines) within $r_c=2$\AA\, in the x-y plane (right).} \label{fig:structure_graphene_0K} \end{figure}

\subsection{Copper-Graphene construction}\label{subsub:Coppergraphene}
 The atomic positions of copper and graphene, and their domains, are separately equilibrated as described above. The atomic positions are used to initialise the copper blocks on the right and left of the graphene sheet within a single domain, and the system is subsequently equilibrated. The choices of the graphene and copper lattices used and their initial mismatch at 300K are given in Table \ref{tab:mismatchlattice}. 
 The distance between the copper blocks and the graphene sheet is taken to be $3$ \AA\, on either side, corresponding to an equilibrium distance between copper and single-layer graphene of roughly $3$ \AA\, \cite{BokdamGraphenedistance}. 
 
 The copper atoms interface with the graphene sheet on the (001) plane. \cite{Saether2022} concluded that the crystal orientation did not influence the phonon heat transport in copper significantly, and \cite{Zhu2022} showed that the crystal orientation also has a minimal effect of 33 MW/m2K on the ITC with graphene, as well as minimal changes in the VDOS. Therefore, we choose to investigate domain properties for only the (001) plane and expect the results to be representative of the dynamics of all crystal orientations. 
 
\begin{figure*}
    \centering   \includegraphics[width=0.8\linewidth]{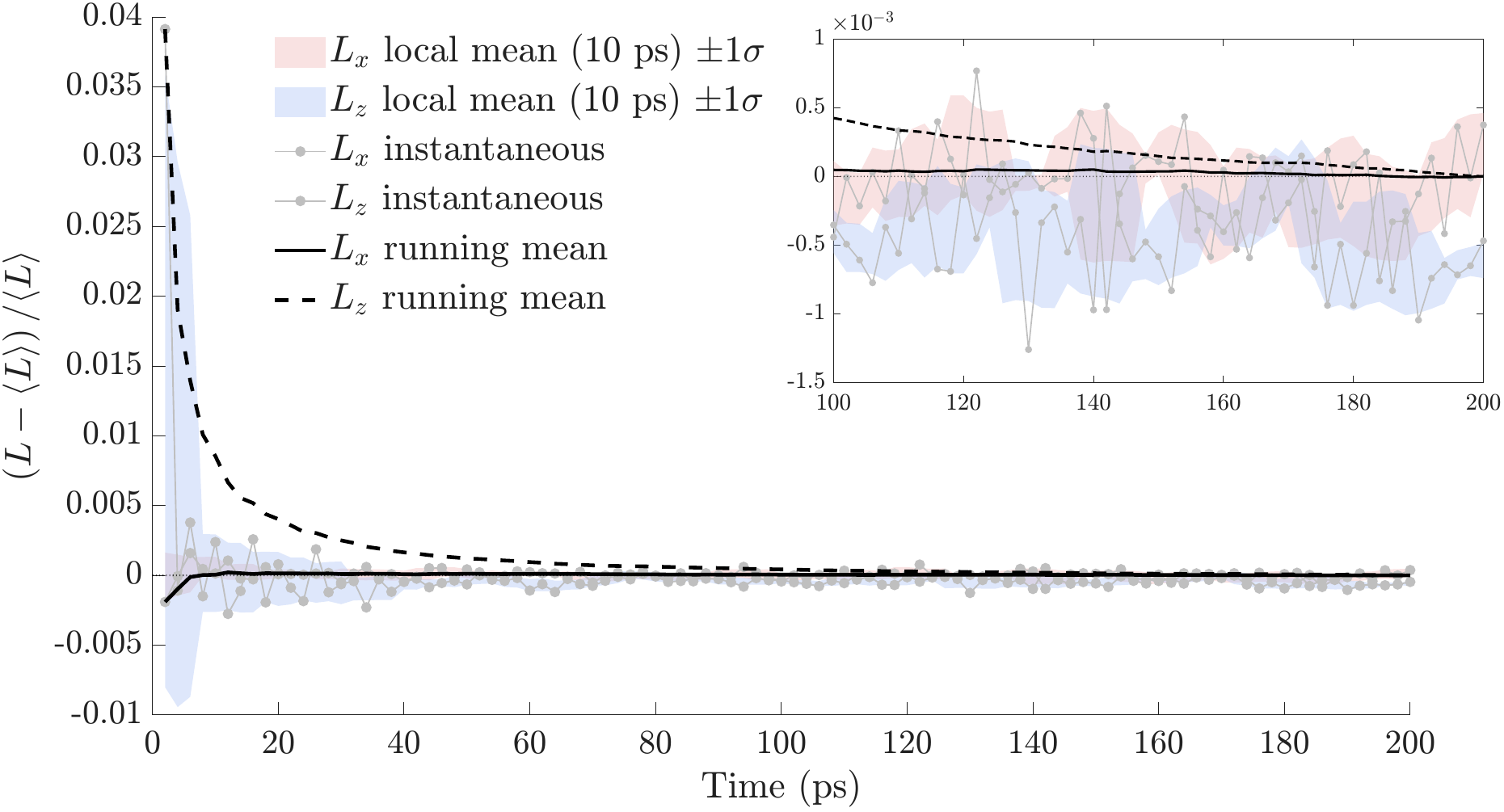}
  \caption{Case II Equilibration of relative domain sizes $\frac{L-\langle L\rangle}{\langle L\rangle }$ a function of time during the NPT run of 200 ps with $\Delta t =10^{-3}$ps and zoomed into the tail (top, right). This shows no trend in the running mean value and a remaining oscillation in $L$ of $<0.2\%$, with no trend in the running mean.}
  \label{fig:relaxation300Kgraphene_cu__lattice}
\end{figure*}

 The combined structure is further equilibrated using an NPT run of 200 ps. During this run, the atomic positions and the domain size in the xyz plane are anisotropically changed to minimise pressure and temperature, yielding a final domain given in Table \ref{tab:mismatchlattice}. The NPT equilibration for a single case is plotted in Figure \ref{fig:relaxation300Kgraphene_cu__lattice}, from which it is clear that the final domain is less than $0.2\%$ from the mean equilibrium domain. With this procedure, we obtain the lattice structure as visualised for one of the cases in Figure \ref{fig:relaxation300Kgraphene_cu__lattice}. There is clearly wrinkling of the graphene sheet, which is not present for the other initialisation method. The average distance of the graphene atoms to the copper atoms is also much smaller. Wrinkling of substrated graphene on Cu surfaces in free vacuum has been observed in literature \cite{Klaver2015}.
\FloatBarrier
\begin{table*}
\centering
\begin{tabular}{|l|l|l|l|l|l|l|l|l|l|l|}
\hline
 $L_x^G$ [\AA] & $L_y^G$ [\AA] &  $L_x^{Cu}$ [\AA] & $L_y^{Cu}$ [\AA] & $L_z^{Cu}$[\AA] &$\epsilon_x$ & $\epsilon_y$  &$L_x^{Cu-G}$& $L_y^{Cu-G}$ & $L_z^{Cu-G}$ \\
\hline 
 $54.5454$ & $58.1387$ &  $54.4769$& $58.1087$ & $21.7908$&$-0.13\%$ & $-0.052\%$ &$53.2920$ &$57.6794$ &$48.7643$\\
\hline 
 $54.5454$ & $58.1387$ &  $54.4517$& $58.0818$ & $29.0409$&$-0.17\%$ & $-0.094\%$ &$54.3147$ &$56.8201$ &$63.0374$ \\
 \hline
  $54.5454$ & $58.1387$ &  $54.4941$& $59.1270$ & $36.3294$&$-0.10\%$ & $+1.67\%$ &$54.2237$ &$57.7836$ &$77.6569$ \\
 \hline
   $54.5454$ & $58.1387$ &  $54.4694$& $58.1007$ & $47.2068$&$-0.14\%$ & $-0.065\%$ &$54.2190$ &$57.5388$ &$99.5149$ \\
 \hline
   $54.5454$ & $58.1387$ &  $54.4811$& $58.1131$ & $58.1131$&$-0.12\%$ & $-0.044\%$ &$54.2992$ &$ 57.9377$ &$121.1998$ \\
 \hline
   $54.5454$ & $58.1387$ &  $54.4807$& $58.1128$ & $72.6410$&$-0.12\%$ & $-0.045\%$ &$54.3781$ &$57.9225$ &$150.3269$ \\
 \hline
   $54.5454$ & $58.1387$ &  $54.4855$& $58.1179$ & $87.1769$&$-0.11\%$ & $-0.036\%$ &$54.3249$ &$57.9483$ &$179.2682$ \\
 \hline
   $54.5454$ & $58.1387$ &  $54.4770$& $58.1088$ & $94.4268$&$-0.13\%$ & $-0.051\%$ &$53.9215$ &$57.9763$ &$193.8497$ \\
 \hline
   $54.5454$ & $58.1387$ &  $54.4623$& $58.0931$ & $108.9245$&$-0.15\%$ & $-0.079\%$ &$53.9664$ &$57.9216$ &$223.0062$ \\
 \hline
   $54.5454$ & $58.1387$ &  $54.4735$& $58.1051$ & $116.2101$&$-0.13\%$ & $-0.058\%$ &$54.3071$ &$57.6344$ &$237.3817$ \\
 \hline
   $54.5454$ & $58.1387$ &  $54.4851$& $58.1174$ & $130.7642$&$-0.11\%$ & $-0.037\%$ &$54.3446$ &$57.9011$ &$266.3766$ \\
 \hline
   $54.5454$ & $58.1387$ &  $54.4840$& $58.1163$ & $152.5553$&$-0.11\%$ & $-0.039\%$ &$54.4166$ &$57.4022$ &$310.0778$ \\
 \hline
   $54.5454$ & $58.1387$ &  $54.4861$& $58.1185$ & $167.0906$&$-0.11\%$ & $-0.035\%$ &$54.2688$ &$57.2537$ &$339.1457$ \\
 \hline
   $54.5454$ & $58.1387$ &  $54.4682$& $58.0994$ & $174.2983$&$-0.14\%$ & $-0.068\%$ &$ 53.9449$ &$58.0169$ &$353.7044$ \\
 \hline
\end{tabular}\caption{Case II: Domain sizes after equilibration of the graphene lattice $L_{x,y}^G$, the copper lattice ${L_{x,y,z}^{Cu}}$, and initial lattice mismatch $\epsilon_{x,y} = 100 \cdot(1-L_{x,y}^G/L_{x,y}^{Cu})$ from the difference in the initial domain sizes in the xy plane, $L_{x,y}$, of the relaxed copper and relaxed graphene structures, and the total equilibrated domain sizes $L_{x,y,z}^{Cu-G}$.} \label{tab:mismatchlattice}
\end{table*}

\clearpage

\section{Additional Supporting Figures}
\begin{figure}[H]
\includegraphics[width=0.9\linewidth]{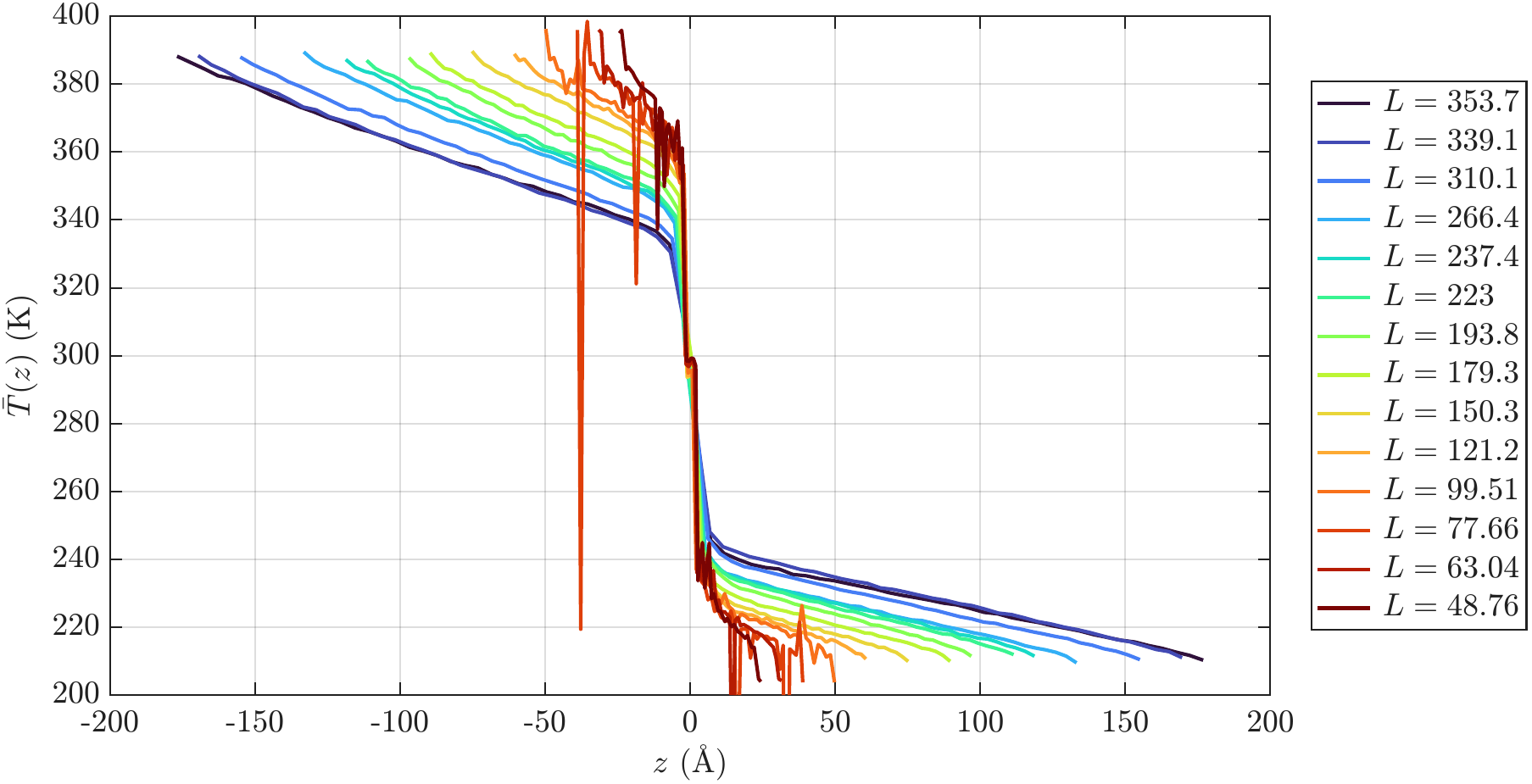}
\caption{Temperature profiles with $L/2$ centered at $z=0$ for the set lattice with contant $\Delta T$ for Case II.}
\label{fig:temp_profiles_case2}
\end{figure}

\begin{figure}[H]
\begin{subfigure} {0.9\linewidth}
\includegraphics[width=0.9\linewidth]{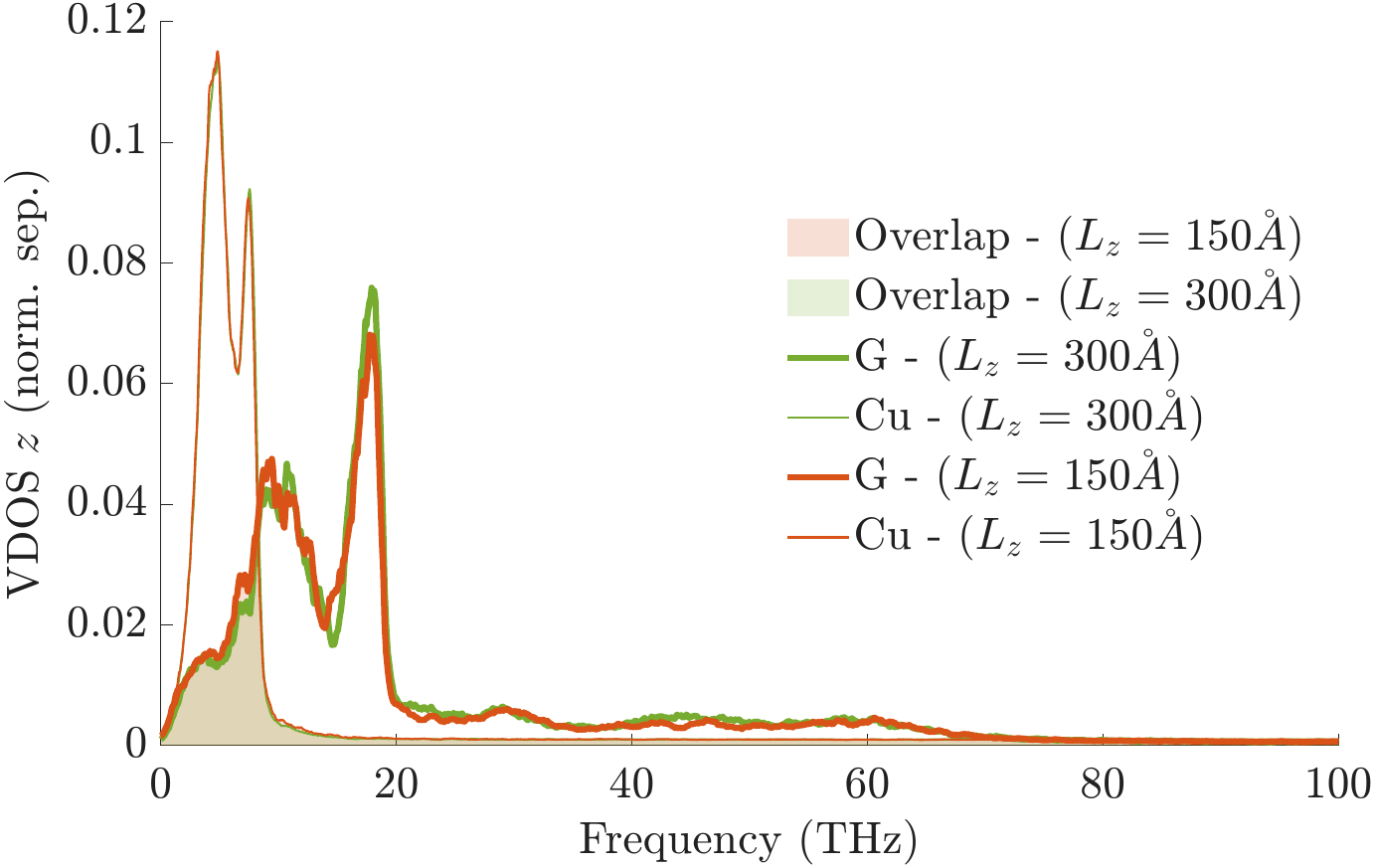}
  \end{subfigure}
  \begin{subfigure}{0.9\linewidth}
    \includegraphics[width=0.9\linewidth]{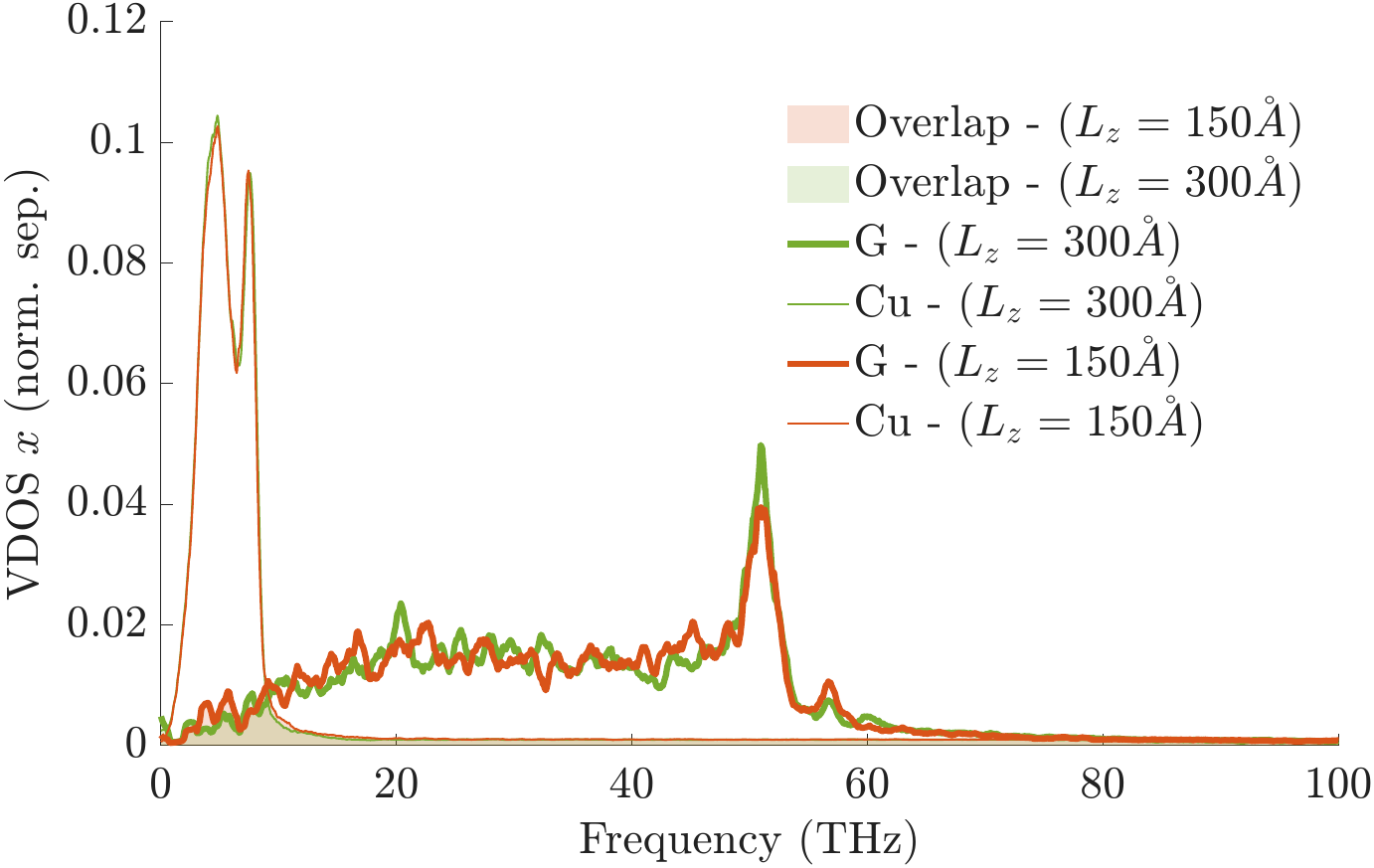}
  \end{subfigure}
  \caption{VDOS set for Case I with $(L_x,L_y,L_z)=(50.0, 50.0, 150) $\AA\, (top) and Case I with $(L_x,L_y,L_z)=(50.0, 50.0, 300.0)$\AA\, (bottom), with ITC$=526.0\pm2$[MW/m$^2$K]  and  ITC$=557.3\pm3$[MW/m$^2$K], respectively. }
\label{fig:VDOS_small_vs_big}
\end{figure}

In Figure \ref{fig:temp_profiles_case2}, the temperature profiles obtained for the $\Delta T$ kept constant for Case II are plotted. The same $T_{cold,hot}$ temperatures are used for varying domain lengths. Consequently, the temperature jump at the interface is not constant. \\ 
Figure \ref{fig:VDOS_small_vs_big} displays the VDOS of two Case I simulations with varying domain lengths and comparable ITC values. The VDOS show the same peak locations for both graphene and copper spectra, unlike the VDOS shifts observed between Case I and Case II configurations as visualised for two simulations in Figure \ref{fig:VDOS_set_vs_equil}.

\begin{figure}[H]
  \centering
    \begin{subfigure}{0.49\linewidth}
    \centering
    \includegraphics[width=\linewidth]{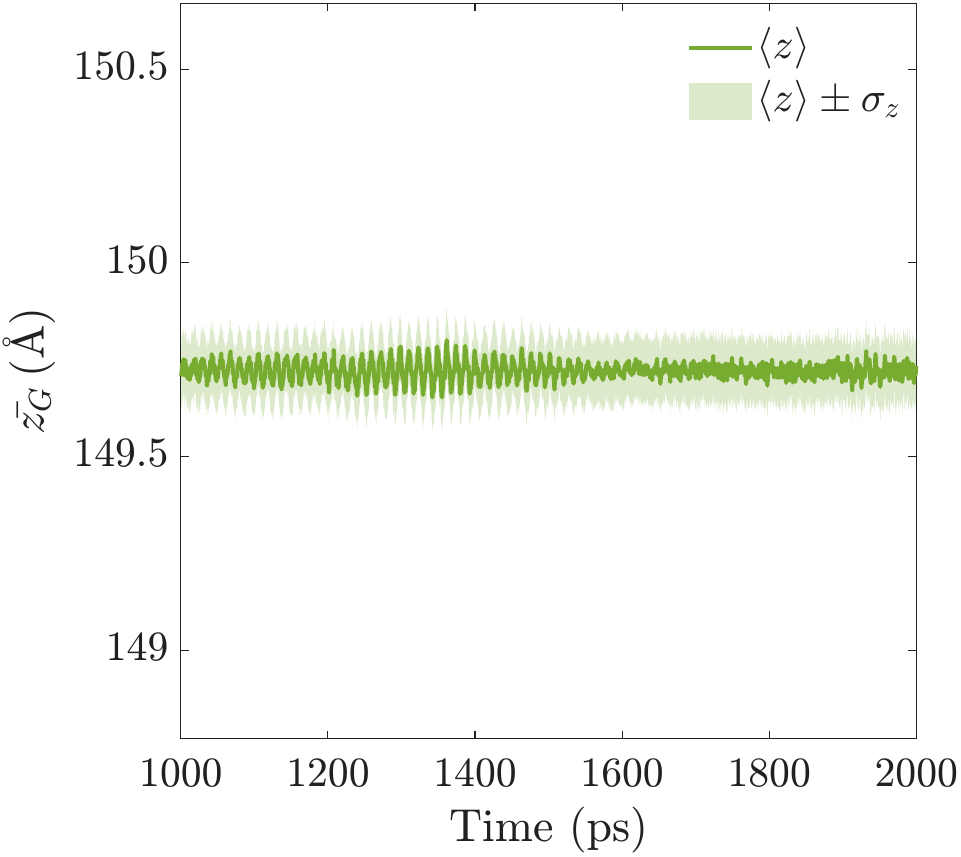}
  \end{subfigure}
  \begin{subfigure}{0.49\linewidth}
    \centering
\includegraphics[width=\linewidth]{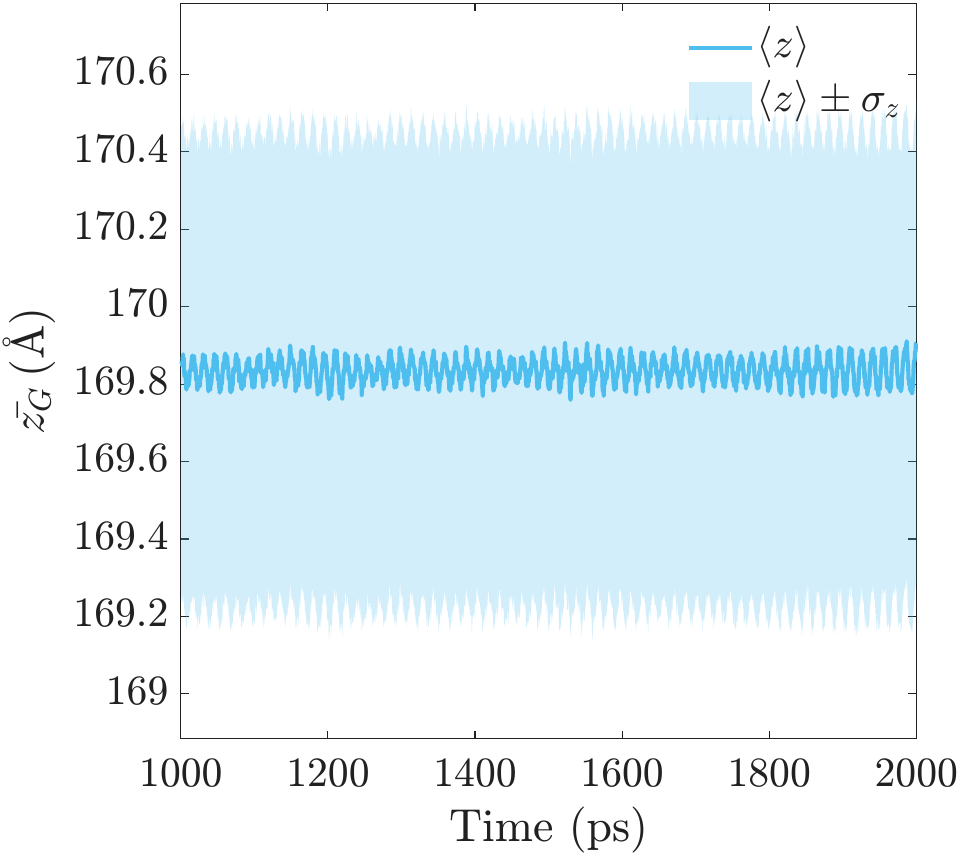}
  \end{subfigure}
  \caption{Average position along $z$ of the graphene atoms at every 0.5 ps including the standard deviation $\sigma_z$, for Case I (left)  with $(L_x,L_y,L_z)=(50.0, 50.0, 300.0) $\AA\,,  ITC$=557.3\pm3$[MW/m$^2$K] and Case II (right) with $(L_x,L_y,L_z)=(54.3,  57.3,  339.1)$\AA\,, ITC$=297.8 \pm 1$[MW/m$^2$K].}
\label{fig:average_Z_supp}
\end{figure}

In Figure \ref{fig:average_Z_supp}, the average position along $z$ as a function of time of Case I and Case II are plotted with the same axis magnitudes, demonstrating that the Case I and Case II do not on average change in position, and that the deviation from the average position in larger in Case II, with a constant standard deviation throughout the simulation time.

\end{document}